\documentclass[12pt]{article}
\pdfoutput=1
\usepackage{amsfonts}
\usepackage{url}
\usepackage{hyperref}
\usepackage{bbold}
\usepackage{slashed}
\usepackage{graphicx}
\usepackage{pgfplots}
\usepackage{color}
\usepackage{mathrsfs}
\usepackage{fancybox}
\usepackage{lipsum}
\usepackage{eurosym}
\usepackage{tcolorbox}
\usepackage{lmodern}
\usepackage{tikz}
\usepackage{tikz,pgfplots}
\usepackage{pgfplots}
\usepackage{amsmath,amsthm,amssymb}
\usepgfplotslibrary{fillbetween}
\usepackage{feynmp}
\usepackage{slashed}
\usepackage{graphicx}
\usepackage{epstopdf}
\usepackage{subfigure}
\usepackage{color}
\usepackage{pgfplots}
\usetikzlibrary{decorations.pathmorphing}
\tikzset{snake it/.style={decorate, decoration=snake}}
\usetikzlibrary{shapes}
\usepackage{empheq}
\usepackage{mathrsfs}

\def\({\left (}
\def\){\right )}
\def\[{\left [}
\def\]{\right ]}
\def\d{\mathrm{d}}

\numberwithin{equation}{section}

\usepackage{environ}
\NewEnviron{myequation}{%
\begin{eqnarray}
\scalebox{1.1}{$\BODY$}
\end{eqnarray}
}

\newcommand{\beq}{\begin{equation}}
\newcommand{\eeq}{\end{equation}}
\newcommand{\nn}{\nonumber\\} 
\newcommand{\bea}{\begin{eqnarray}}
\newcommand{\ea}{\end{eqnarray}}
\newcommand{\barr}{\!\begin{array}}
\newcommand{\earr}{\end{array}\!}
\newcommand{\lb}{{\langle}}
\newcommand{\rb}{{\rangle}}

\def\d{{\partial}}
\def\n{{\bf \widehat n}}
\def\k{{\bf k}}

\begin{document}
\begin{titlepage}

\setcounter{page}{1} \baselineskip=15.5pt \thispagestyle{empty}

\vfil

${}$
\vspace{1cm}

\begin{center}

\def\thefootnote{\fnsymbol{footnote}}
\begin{changemargin}{0.05cm}{0.05cm} 
\begin{center}
{\Large \bf  Expanding the Black Hole Interior:\\[5mm]
Partially Entangled Thermal States in SYK}
\end{center} 
\end{changemargin}

~\\[1cm]
{Akash Goel\footnote{\href{mailto:akashg@princeton.edu}{\protect\path{akashg@princeton.edu}}}, Ho Tat Lam\footnote{\href{mailto:htlam@princeton.edu}{\protect\path{htlam@princeton.edu}}}, Gustavo J. Turiaci\footnote{\href{mailto:turiaci@princeton.edu}{\protect\path{turiaci@princeton.edu}}} and Herman Verlinde\footnote{\href{mailto:verlinde@princeton.edu}{\protect\path{verlinde@princeton.edu}}}}
\\[0.3cm]

{\normalsize { \sl Physics Department,  
Princeton University, Princeton, NJ 08544, USA}} \\[3mm]

\end{center}


 \vspace{0.2cm}
\begin{changemargin}{01cm}{1cm} 
{\small  \noindent 
\begin{center} 
\textbf{Abstract}
\end{center} }
We introduce a family of partially entangled thermal states in the SYK model that interpolates between the thermo-field double state and a pure (product) state. The states are prepared by a euclidean path integral describing the evolution over two euclidean time segments separated by a local scaling operator ${\cal O}$. We argue that the holographic dual of this class of states consists of two black holes with their interior regions connected via a domain wall, described by the worldline of a massive particle. We compute the size of the interior region and the entanglement entropy as a function of the scale dimension of ${\cal O}$ and the temperature of each black hole. We argue that the one-sided bulk reconstruction can access the interior region of the black hole.

\end{changemargin}
 \vspace{0.3cm}
\vfil
\begin{flushleft}
\today
\end{flushleft}

\end{titlepage}

\newpage
\tableofcontents
\newpage

\addtolength{\abovedisplayskip}{.5mm}
\addtolength{\belowdisplayskip}{.5mm}

\def\plus{\raisebox{.5pt}{\tiny$+$\smpc}}

\addtolength{\parskip}{.6mm}
\def\spc{\hspace{1pt}}

\def\nspc{{\hspace{-2pt}}}
\def\ff{\rm\smpc f\smpc} 
\def\fff{\mbox{Y}}
\def\ww{{\rm w}}
\def\smpc{{\hspace{.5pt}}}

\def\zz{{\spc \rm z}}
\def\xx{{\rm x\smpc}}
\def\xxi{\mbox{\footnotesize \spc $\xi$}}
\def\jj{{\rm j}}
 \addtolength{\baselineskip}{-.1mm}

\renewcommand{\Large}{\large}

\def\calO{{b}}
\def\be{\begin{equation}}
\def\ee{\end{equation}}




\def\mathbi#1{\textbf{\em #1}} 
\def\som{{ \textit{\textbf s}}} 
\def\tom{{ \textit{\textbf t}}} 
\def\nom{n} 
\def\mom{m} 
\def\la{\langle}
\def\bea{\begin{eqnarray}}
\def\eea{\end{eqnarray}}
\def\is{\!  & \!  = \!  &  \!}
\def\ra{\rangle}
\def\half{{\textstyle{\frac 12}}}
\def\cL{{\cal L}}
\def\halfi{{\textstyle{\frac i 2}}}
\def\ba{\bea}
\def\ea{\eea}
\def\lb{\langle}
\def\rb{\rangle}
\newcommand{\rep}[1]{\mathbf{#1}}

\def\uU{\bf U}
\def\be{\bea}
\def\ee{\eea}
\def\delbar{\overline{\partial}}
\def\ra{\rangle}
\def\la{\langle}
\def\ccdot{\!\spc\cdot\!\spc}
\def\nspc{\!\spc\smpc}
\def\tr{{\rm tr}}
\def\li{|\spc}
\def\ri{|\spc}

\def\hf{\textstyle \frac 1 2}

\def\bfcdot{\raisebox{-1.5pt}{\bf \LARGE $\spc \cdot\spc $}}
\def\spc{\hspace{1pt}}
\def\is{\! &  \! = \! & \!}
\def\d{{\partial}}
\def\n{{\bf \widehat n}}
\def\k{{\bf k}}
\def\GO{{\cal O}}

\def\pp{{\mbox{\tiny$+$}}}
\def\mm{{\mbox{\tiny$-$}}}

\setcounter{tocdepth}{2}
\addtolength{\baselineskip}{0mm}
\addtolength{\parskip}{.6mm}
\addtolength{\abovedisplayskip}{1mm}
\addtolength{\belowdisplayskip}{1mm}

\def\sss{{\bf s}}

\def\fff{e}
\def\lL{{{}_{{}^{L}}}}

\def\rR{{{}_{{}^{R}}}}

\def\llL{{{\! \spc}_{{}^{L}}}\!\spc}
\def\rrR{{{\!\spc}_{{}^{R}}}\!\spc}

\section{Introduction}
\vspace{-1mm}
It is generally believed that black holes must admit a self-consistent quantum description. In AdS/CFT, this microscopic theory takes the form of a finite temperature CFT on the asymptotic AdS boundary. While the rules of quantum mechanics are manifestly obeyed in this holographic setting, it has proven to be a non-trivial task to extract local bulk physics inside the black hole horizon from the dual quantum theory. The logical tension between QM and the semi-classical bulk description is most directly underlined by the firewall argument \cite{Almheiri:2012rt, Almheiri:2013hfa}.

An often studied finite temperature state in AdS/CFT is the (unnormalized) thermo-field double state
\bea
|\mbox{\sc TFD} \rb \! \is \! \sum_n e^{-\beta E_n /2} |n\rb_\lL |n\rb_{\! \smpc \rR}
\eea
 living in the tensor product Hilbert space of a left- and right CFT.
It defines the purification of the thermal density matrix, and can be thought of as obtained by performing a CFT path integral describing the euclidean time evolution over half a thermal circle with period $\beta$. The TFD state of a holographic CFT is believed to be dual to a maximally extended black hole space-time with two asymptotic regions separated by a bifurcate horizon \cite{Maldacena:2001kr}.

Another type of thermal states are typical pure states of some given total energy. Assuming that the ETH applies, these states will look thermal relative to the set of local bulk observables. Alternatively, one can consider pure states of the form
\bea
|\spc \Psi \spc \rangle_{\!\smpc \rrR} \is e^{-\frac \beta 2  H} \spc | \spc B\spc \ra_{\! \smpc \rrR} \; = \; {}_\llL\!\spc\langle \spc B\spc | \mbox{\sc TFD} \rangle
\eea
with $|\spc B\spc \rangle$ some typical CFT boundary state. Assuming $|\spc B\spc \rangle$ is uncorrelated with the local bulk observables, this state also looks thermal from the outside. We will call them ``thermal pure states". They are believed to describe a one-sided black hole geometry.
 
Our interest is to learn more about the holographic reconstruction of the black hole interior. 
For thermal pure states, one can use the mirror operator construction of \cite{Papadodimas:2013wnh}, or more generally, the quantum error correction procedure of \cite{Verlinde:2012cy}, to reconstruct the interior operators from a single CFT. This situation must be contrasted with the thermo-field double case. For the TFD state, the one-sided quantum state is a thermal density matrix and the firewall argument of \cite{Almheiri:2012rt, Almheiri:2013hfa} implies that the one-sided bulk reconstruction is limited to the region outside the horizon.

\subsection{Partially entangled thermal states in SYK}
\vspace{-2mm} 

An attractive dual perspective on the thermal pure states was recently suggested in \cite{Kourkoulou:2017zaj}, within the context of the SYK model \cite{KitaevTalks}. The SYK model is a quantum theory of $N$ Majorana variables $\{ \psi_i, \psi_j \} = \delta_{ij}$ dynamically coupled via the Hamiltonian
\bea
\label{hsyk}
 H =  {\rm i}^{\frac q 2}\!\! \sum_{i_1, ..,i_q} \spc j_{i_1 \dots i_q} \spc \psi_{i_1} \ldots \psi_{i_q}
\eea
with $j_{i_1 .. i_q}$ random couplings picked from a gaussian distribution with variance given by $\la j_{i_1 .. i_q}^2\ra~=~\frac{(q-1)! J^2}{N^{q-1}}$.

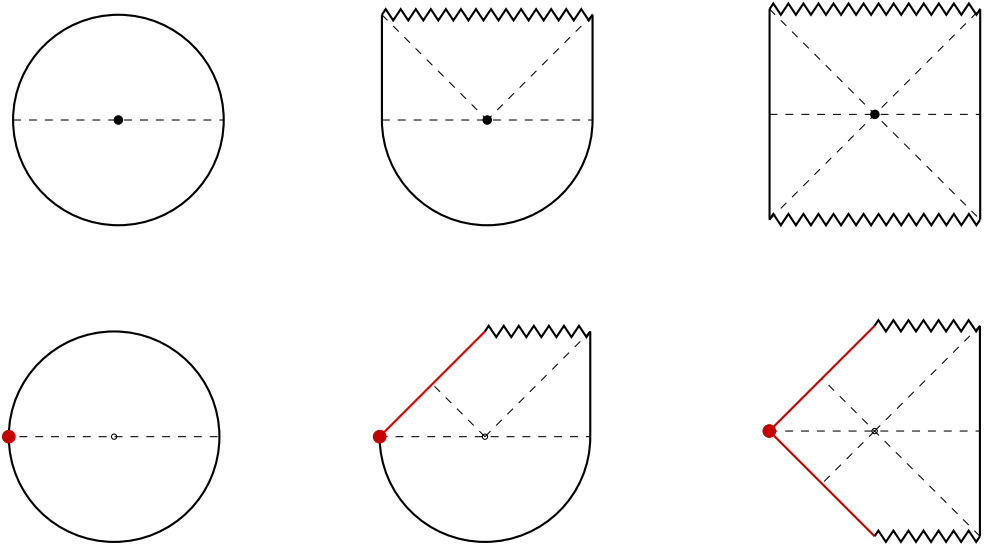
\begin{figure}[t!]
\begin{center}
\begin{tikzpicture}[scale=0.7]
\draw[dashed] (-2,0) -- (2,0);
\draw[thick] (2,0) arc (360:0:2);
\draw[fill=black] (0,0) circle (0.08);
\end{tikzpicture}
\hspace{1.8cm}
\begin{tikzpicture}[scale=0.7]
\draw[dashed] (-2,0) -- (2,0);
\draw[thick] (-2,0) -- (-2,2);
\draw[thick] (2,0) -- (2,2);
\draw[dashed] (-2,2) -- (0,0);
\draw[dashed] (2,2) -- (0,0);
\draw[fill=black] (0,0) circle (0.08);
\draw[thick,decoration = {zigzag,segment length = 2mm, amplitude = 0.75mm},decorate] (-2,2)--(2,2);
\draw[thick] (-2,0) arc (-180:0:2);
\end{tikzpicture}
\hspace{2.05cm}
\begin{tikzpicture}[scale=0.7]
\draw[dashed] (-2,0) -- (2,0);
\draw[thick] (-2,-2) -- (-2,2);
\draw[thick] (2,-2) -- (2,2);
\draw[dashed] (-2,2) -- (2,-2);
\draw[dashed] (-2,-2) -- (2,2);
\draw[fill=black] (0,0) circle (0.08);
\draw[thick,decoration = {zigzag,segment length = 2mm, amplitude = 0.75mm},decorate] (-2,2)--(2,2);
\draw[thick,decoration = {zigzag,segment length = 2mm, amplitude = 0.75mm},decorate] (-2,-2)--(2,-2);
\end{tikzpicture}\ \  \  \\[12mm]
\begin{tikzpicture}[scale=0.7]
\draw[dashed] (-2.1,0) -- (2,0);
\draw[thick] (2,0) arc (360:0:2);
\draw[thin] (0,0) circle (0.05);
\draw[color={rgb:red,10; black,3}, fill={rgb:red,10; black,3}] (-2,0) circle (0.12);
\end{tikzpicture}
\hspace{1.75cm}
\begin{tikzpicture}[scale=0.7]
\draw[dashed] (-4,0) -- (0,0);
\draw[thick] (0,0) arc (360:180:2);
\draw[dashed] (-2,0) -- (0,2);
\draw[dashed] (-2,0) -- (-3,1);
\draw[thick] (0,0) -- (0,2);
\draw[thick,color={rgb:red,20; black,3}] (-4,0) -- (-2,2);
\draw[thin] (-2,0) circle (0.05);
\draw[thick,decoration = {zigzag,segment length = 2mm, amplitude = 0.75mm},decorate] (-2,2)--(0,2);
\draw[color={rgb:red,10; black,3}, fill={rgb:red,10; black,3}] (-4,0) circle (0.12);
\end{tikzpicture}
\hspace{2cm}
\begin{tikzpicture}[scale=0.7]
\draw[dashed] (-2,0) -- (0,2);
\draw[dashed] (0,-2) -- (-3,1);
\draw[dashed] (-2,0) -- (-3,-1);
\draw[dashed] (-4,0) -- (0,0);
\draw[thick] (0,-2) -- (0,2);
\draw[thick,color={rgb:red,10; black,3}] (-2,-2) -- (-4,0);
\draw[thick,color={rgb:red,10; black,3}] (-4,0) -- (-2,2);
\draw[thin] (-2,0) circle (0.05);
\draw[thick,decoration = {zigzag,segment length = 2mm, amplitude = 0.75mm},decorate] (-2,2)--(0,2);
\draw[thick,decoration = {zigzag,segment length = 2mm, amplitude = 0.75mm},decorate] (-2,-2)--(0,-2);
\draw[color={rgb:red,10; black,3}, fill={rgb:red,10; black,3}] (-4,0) circle (0.12);
\end{tikzpicture}\ \ \ \ 
\end{center}
\caption{\small The euclidean (left), lorentzian (right) space-time associated with the thermo-field double state (top) and thermal pure state (bottom). The middle column shows the total geometry describing the state preparation and real time evolution, obtained by gluing the euclidean and lorentzian geometry together along the equator of the disc }
\label{PETSgeomfig1}
\end{figure}

The SYK Hilbert space contains a natural basis of $2^{N/2}$ boundary states labeled as $|\spc \sss \spc \rangle\equiv|s_1,s_2,...,s_{N/2}\rangle$, with $s_i$ taking values in $\{-1,1\}$, defined by arranging the Majorana variables into pairs (we assume $N$ is even) and requiring that
\bea
\label{sdef}
\(\psi^{2k-1}-is_k\psi^{2k}\)|\spc \sss \spc \rangle\! \is \! 0  \,.
\eea
The thermal pure state defined via
\bea
\label{tps}
|\Psi\rangle \is |\spc \sss,\beta\ra \, \equiv \, e^{-\frac \beta 2  H} |\spc \sss\spc\rangle \, = \, {}_{\llL}\!\smpc \langle \sss|\mbox{\sc TFD}\rangle
\eea
 looks like a thermal state relative to the class of flip invariant operators, that do not depend on the sign of the individual Majorana variables \cite{Kourkoulou:2017zaj}: all n-point correlation functions $\tr(\rho_s\, {\cal O}_1 ...{\cal O}_n) = \langle \Psi |{\cal O}_1 ...{\cal O}_n |\Psi  \rangle$ of collections of operators that are invariant under the `flip group' 
are equal to the thermal expectation values with inverse temperature $\beta$ (as long as $n\ll N$).  

As suggested in \cite{Kourkoulou:2017zaj}\cite{Almheiri:2018ijj}, the projection ${}_{\llL}\!\smpc \langle \sss|\mbox{\sc TFD}\rangle$ of the thermo-field double state onto a particular boundary state can be holographically represented as an `end-of-the-world particle' that removes the left asymptotic region of the two-sided black hole geometry, but keeps part of the left region in place. The left region thus becomes identified with the black hole interior as seen from~the~right.

A qualitative description of the proposed dual geometry corresponding to the thermo-field double and the thermal pure states is indicated in figure \ref{PETSgeomfig1}, c.f. \cite{Kourkoulou:2017zaj}. The top row indicates the thermal circle (left), which in the holographic setting, constitutes the boundary of a Poincar\'e disc, the euclidean AdS$_2$ bulk space-time. The corresponding two-sided black hole geometry is shown on the top right. The middle column shows the total geometry that includes the state preparation and the real time evolution is obtained by gluing the euclidean and lorentzian geometry together along the equator of the disc. The bifurcate horizon is situated at the center of the disk. The bottom row indicates the geometry of the thermal pure state. The trajectory of the end of the world-particle starts at the intersection point between the left boundary and the equator, where the lorentzian geometry is glued onto the euclidean half-circle. This geometric argument that pure black hole state has a smooth interior geometry provides support for the aforementioned (state-dependent) QEC procedure for constructing interior operators \cite{Papadodimas:2013wnh, Verlinde:2012cy}.

Thermal pure states and the thermo-field double are both mathematical idealizations. Generic states are somewhere in between: macroscopic systems are never in a pure state nor in a perfect thermally mixed state, since typically we know somewhere between everything or nothing about a system.\footnote{ By the same token, observables that measure properties of a macroscopic quantum system are usually defined with reference to some classical environment or restriction. So they are typically neither purely state-dependent nor completely state-independent.}
In current terminology, a class of states compatible with a classical background is called a code subspace \cite{Verlinde:2012cy}\cite{Almheiri:2014lwa}. 
It thus becomes natural to look for a practical generalization of the thermo-field double or thermal pure states, in the form of an interpolating family of partially mixed thermal states.

In the context of the SYK model, there are two natural ways to define such an interpolating family of states. The first method is a straightforward modification of the above construction of the thermal pure states. We will describe this method first. Then we introduce a second class of partially mixed thermal states with a better understood holographic description. This second type of states will be the main focus of this paper.

Consider the $2^N$ dimensional Hilbert space ${\cal H}$ spanned by $2N$ Majorana variables $\psi^i$. We assume $N$ is even. Introduce the basis of $2^N$ states $|\sss\rangle$ defined in eqn \eqref{sdef}.
Next we partition the $2N$ Majorana fermions into two groups of $N$ Majorana fermions $\{\psi_{\llL,\rrR}\}$, each spanning sub-Hilbert spaces $\mathcal{H}_{\llL,\rrR}$ of dimension $2^{N/2}$. Let $H_\llL$ and $H_\rrR$ denote two identical SYK Hamiltonians acting on each subsystem. Note that the choice of the Hamiltonian depends on the choice of partition.\footnote{Alternatively, we could have picked a fixed partition into left- and right variables $\{\psi_{\llL,\rrR}\}$, but allowed ourselves the freedom to chose an arbitrary partition into even and odd variables $\psi_{\rm even}$ and $\psi_{\rm odd}$. In this case, the Hamiltonians $H_\llL$ and $H_\rrR$ would be held fixed, and the state $|\sss\rb$ would depend on the choice of partition.} Now consider the following class of $2^N$ entangled states
\bea
\label{petsone}
|\Psi\rangle\is |\sss;  \beta_\llL, \beta_\rrR \rangle=e^{-\frac 1 2 \beta_\llL H_\llL-\frac 1 2 \beta_\rrR H_\rrR}|\spc \sss \spc \rangle\,.
\eea
By choosing different partitions, we obtain a large class of states with different degrees of entanglement between the left-sector ${\cal H}_\llL$ and right-sector ${\cal H}_\rrR$.
The thermo-field double is a state for which the partition into $\{\psi_\llL, \psi_{\rrR}\}$ precisely coincides with the division into $\{\psi_{\rm even},\psi_{\rm odd}\}$, and for which all $s_k = 1$. On the other end of the spectrum, the thermal pure states correspond to the case for which the $\{\psi_\llL\}$ consists of $N/2$ Majorana pairs $(\psi_{2k},\psi_{2k+1})$, so that the boundary state \eqref{sdef} factorizes into left and right boundary state $|\sss\rb = |\sss\rb_\llL \otimes |\sss\rb_\rrR$. The state (\ref{petsone}) then factorizes into a product of two thermal pure states. For the generic choice of partition, the states \eqref{petsone} are partially entangled thermal states with an entanglement entropy somewhere in between zero (for the pure product states) and the thermal entropy (for the TFD state). We discuss some further properties of the class of state \eqref{petsone} in Appendix \ref{app:SYKPETS}.

In the rest of this paper, we will study the properties and holographic dual geometry of another class of partially entangled states of the form
\bea\label{ostate}
| \Psi \rb \is \sum_{m,n} e^{{-\frac 1 2 \beta_\llL} E_m{-\frac 1 2 \beta_\rrR}E_n}\, \mathcal{O}_{n,m} \, |m\rb_\lL |n\rb_\rR
\eea
where $\mathcal{O}_{m,n}=\lb m | \mathcal{O} |n \rb$ are the matrix elements of some arbitrarily chosen local scaling operator ${\cal O}$. This state satisfies the property 
\bea
\label{dostate}
{}_\lL\!\smpc\lb \psi_1 | {}_\rR\!\smpc\lb \psi_2 | \Psi \rb \is \lb \psi_1 | e^{-\frac 1 2 \beta_\llL H} \spc \mathcal{O}\spc e^{-\frac 1 2 \beta_\rrR H} | \psi^\star_2 \rb,
\eea
with $\psi_1$ and $\psi_2$ labeling generic states in the left and right Hilbert spaces. From \eqref{dostate} we recover the expression \eqref{ostate} by inserting a complete basis of $\mathcal{H}_\lL \otimes \mathcal{H}_\rR$. As seen from the second representation, the state $|\Psi\rangle$ can be thought of as produced by evolution of a single QM system over a euclidean time interval $\beta_\rrR$, acting with a local operator ${\cal O}$, and then evolving over a second euclidean time interval $\beta_\llL$. It is tempting to identify $\beta_{\llL,\rrR}$ with the effective temperature of the left- and right QM system, but as we will see, this identification is in general not correct.

The class of states \eqref{ostate} includes the TFD and thermal pure states as special limits.
If we chose ${\cal O} = \mathbb{1}$ with $\la n | \mathbb{1} | m \ra = \delta_{nm}$, the state \eqref{ostate} reduces to the thermo-field double with inverse temperature $\beta = \beta_\llL + \beta_\rrR$. On the other end, if we send $\beta_\llL \to \infty$, we project the left CFT onto the vacuum state. For a sufficiently random choice of the operator ${\cal O}$, the state \eqref{ostate} then takes the form of a thermal pure state \eqref{tps}. We will call the above general class of states `partially entangled thermal states' (PETS)\footnote{When $\mathcal{O}_{n,m}$ is constant, the state factorizes into two thermal pure states of inverse temperature $\beta_L$ and $\beta_R$. This can happen if $\ell \to \infty$ for our setup.}. 

The reduced density matrix for QM$_\rR$ after tracing over the left Hilbert space $\rho_\rR = {\rm Tr}_\lL\spc |\Psi\rb \lb \Psi |$ is given by  
\bea
\rho \is \sum_{m,n,n'} e^{-{\frac 1  2 \beta_\rrR} E_n}\, \mathcal{O}_{n,m}\, e^{-\beta_\llL E_{m}}\, \mathcal{O}_{m,n'}\,e^{-{\frac 1 2 \beta_\rrR} E_{n'}}|\spc n\spc \rb\spc \lb\spc n'\spc |
\label{rhoom}
\eea
or more succinctly
\bea
\rho \is e^{-\beta_\rrR H/2 } \, {\cal O}\, 
e^{- \beta_\llL  H} {\cal O} \,  
e^{-\beta_\rrR H/2}
\eea
In the following, we will usually choose ${\cal O}$ to be a scaling operator ${\cal O}_\ell$, with scaling dimension $\ell$.
We will be mostly interested in large scaling dimensions of order $N/\beta J$.
More generally, we will also consider the generalization of PETS in which we replace the single operator ${\cal O}$ by an incoherent sum of operators ${\cal O}_i$ with all approximately the same conformal dimension. The PETS then becomes a partially entangled mixed state. 
\bigskip

\subsection{A useful graphical notation}\label{sec:graphicalnot}
\vspace{-2mm}
We are interested in determining the entanglement and thermal properties of the partially entangled thermal state and of their holographic dual. For this purpose, we briefly pause to introduce a helpful graphical notation for the three types of states.

\medskip

\noindent{\bf The thermofield double state} 
represents the purification of the (unnormalized) thermal density matrix  
\bea \rho \, = \, 
 \spc e^{-{\beta H}}
\is \, \raisebox{2pt}{\begin{tikzpicture}[scale=.5, baseline={([yshift=0cm]current bounding box.center)}]
\draw[thick] (-0.2,0.5) arc (10:350:1.52);
\draw[fill,black] (-0.2,0.5) circle (0.05);
\draw[fill,black] (-0.2,-0.0) circle (0.05);
\draw (-2.8,0.35) node {\scriptsize\bf $\beta$};;\end{tikzpicture}}\eea 
The TFD state can be thought of as being prepared by a euclidean path-integral of a single QM system, evolved over half of the thermal circle
\bea
\label{tfd}
| \mbox{\sc TFD} \rangle\nspc \;\; \cong \;\; e^{-\frac \beta 2  H} \is \ \begin{tikzpicture}[scale=.52, baseline={([yshift=0cm]current bounding box.center)}]
\draw[thick] (-0.2,0) arc (190:350:1.52);
\draw[fill,black] (-0.2,0.0375) circle (0.05); 
\draw[fill,black] (2.8,0.0375) circle (0.05);
\draw (1.2,-.6) node {\scriptsize\bf $\frac \beta 2$};;
\end{tikzpicture} 
\eea
Here for later convenience we adopted the congruence $ \mbox{\small $\sum$\,\footnotesize$e^{-\frac \beta 2\nspc E_n} | n \rangle_{\mbox{\tiny \sc l}} | n \rangle_{\mbox{\tiny \sc r}}$}\, \cong\,  \nspc \mbox{\small $\sum$\,\footnotesize $e^{-\frac \beta 2\nspc E_n}  | n\rangle \langle n |$} $ between entangled states of two identical systems and linear operators.

\medskip

\noindent{\bf Thermal pure states} are states of the form \cite{Kourkoulou:2017zaj} 
\bea
\label{pure}
\li  \spc \sss, \beta\spc \ra \is e^{-\frac{\beta}{2} H} \li\spc {\bf s} \spc \ra
 \; = \; \, \begin{tikzpicture}[scale=.52, baseline={([yshift=0cm]current bounding box.center)}]
\draw[thick] (-0.2,-0.1) arc (188:355:1.52);
\draw[fill,black] (2.825,0.0375) circle (0.05);
\draw[color={rgb:red,10; black,3}, fill={rgb:red,10; black,3}]   (-.22,0) circle (0.13); 
\draw (1.2,-.7) node {\footnotesize\bf $\frac \beta 2$};;
\end{tikzpicture}
\eea
The half circle indicates the euclidean time evolution over $\beta/2$ and the red dot indicates the projection onto the state $\li {\bf s} \ra$ defined in eqn \eqref{sdef}. The corresponding density matrix is denoted~by 
\bea
\rho_s\; \equiv \;  \li \sss,\beta \ra \la \sss,\beta \ri  \, = \, e^{-\frac{\beta}{2} H}\spc {\rm P}_{\! \sss}\,\spc e^{-\frac{\beta}{2} H}  \, \is \; \raisebox{2.5pt}{\begin{tikzpicture}[scale=.52, baseline={([yshift=0cm]current bounding box.center)}]);
\draw[thick] (-0.2,0.5) arc (10:350:1.52);
\draw[fill,black] (-0.2,0.5) circle (0.05);
\draw[fill,black] (-0.2,-0.0) circle (0.05);
\draw[color={rgb:red,10; black,3}, fill={rgb:red,10; black,3}]  (-3.2,0.25) circle (0.13); 
\draw (-1.75,1.25) node {\scriptsize\bf $\beta/2$};;
\draw (-1.75,-.85) node {\scriptsize\bf $\beta/2$};;
\end{tikzpicture}}
\eea
with P$_{\! \sss} = | \sss \rb\lb \sss|$ the projection on the state $|\sss\ra$.

\medskip

\noindent
{\bf Partially entangled thermal states} are represented in this notation as
\bea
\label{psio}
\li\spc \Psi\spc \ra & \cong & \ 
\, e^{-\frac 1 2 \beta_\llL H} \, {\cal O}_\ell\,\spc 
e^{-\frac 1 2 \beta_\rrR H} \  \;\; = \;\; \raisebox{1pt}{\begin{tikzpicture}[scale=.52, baseline={([yshift=0cm]current bounding box.center)}]
\draw[thick] (-0.65,-1) arc (245:355:1.6);
\draw[thick] (-3.75,0.3) arc (190:300:1.9);
\draw[fill,black] (1.6,0.25) circle (0.05);
\draw[fill,black] (-3.75,0.3) circle (0.05);
\draw[color={rgb:red,10; black,3}, fill={rgb:red,10; black,3}]   (-.8,-.95) circle (0.14); 
\draw (-2.4,-.4) node {\scriptsize $\frac 1 2 \beta_\llL$};
\draw (.3,-.4) node {\scriptsize $\frac 1 2 \beta_\rrR$};;
\end{tikzpicture}} \
\eea
As indicated by the figure, this class of PETS is prepared by performing a path integral over two segments of a thermal circle separated by the insertion of a local scaling 
operator ${\cal O}_\ell$. This insertion has a number of non-trivial effects.

If the dimension $\ell$ of the operator is small, the operator insertion produces a small perturbation of the TFD state. The dual space-time will just look like the two-sided black hole with a single particle excitation propagating in the bulk. For this paper, we will instead be interested in the case in which the scale dimension of $\mathcal{O}_\ell$ is of order $\ell \sim N/\beta J$. As we will see, in this regime the insertion of the operator ${\cal O}_\ell$ leads to a non-trivial modification of the dual geometry. This backreaction is indicated graphically in eqn \eqref{psio} via the kink connecting the two arcs. Due to the presence of the kink, the two arcs each span an angle bigger than $\pi/2$, reflecting the physical difference between the quantities $\beta_{\llL,\rrR}$, that specify the left- and right-euclidean time lapse, and the effective temperature as seen by the corresponding one-sided observer. The ratio between the two is parameterized by an angle $\theta_{\llL, \rrR}$ via
\bea
\label{thetaeff}
 \beta_\llL^{\rm \!\!\!\smpc eff} \! \is\! \frac{2 \pi \beta_\llL}{2\pi - \theta_\llL} \, \qquad \qquad 
\beta_\rrR^{\rm \!\!\!\smpc eff}\, = \,  \frac{2 \pi \beta_\rrR}{2\pi - \theta_\rrR}\, \qquad \qquad \theta_{\llL,\rrR} \in [0,\pi]
\eea
One of our tasks is to compute how these angles $\theta_{\llL,\rrR}$ depend on the scale dimension $\ell$ of the local operator and on $\beta_\llL$ and $\beta_\rrR$.

The density matrix in graphical notation reads
\bea
\label{rhoom}
\rho \is e^{-\frac 12 \beta_\rrR H } \, {\cal O}\, 
e^{- \beta_\llL  H} {\cal O} \, 
e^{-\frac 1 2 \beta_\rrR H}
\;\; = \; \; 
\raisebox{4pt}{\begin{tikzpicture}[scale=.48, baseline={([yshift=0cm]current bounding box.center)}]
\draw[thick] (-0.65,-1.05) arc (245:350:1.6);
\draw[thick] (-0.65,1.85) arc (115:10:1.6);
\draw[thick] (-.95,1.85) arc (50:310:1.9);
\draw[fill,black] (1.6,0.1) circle (0.05);
\draw[fill,black] (1.6,0.7) circle (0.05);
\draw[dotted,thick,red] (-0.8,-0.7) -- (-.8,1.6);
\draw[color={rgb:red,10; black,3}, fill={rgb:red,10; black,3}]   (-.8,1.75) circle (0.14); 
\draw[color={rgb:red,10; black,3}, fill={rgb:red,10; black,3}]   (-.8,-.95) circle (0.14); 
\draw (-3.5,.45) node {\footnotesize\bf $\beta_\llL$};
\draw (.3,1.4) node {\scriptsize $\frac 1 2 \beta_\rrR$};
\draw (.3,-.65) node {\scriptsize $\frac 1 2 \beta_\rrR$};
\end{tikzpicture}}
\eea
Its partition sum $Z_\ell = \tr \rho$
reduces to the thermal SYK two-point function, with inverse temperature $\beta = \beta_\llL+\beta_\rrR$ of two local scaling operators ${\cal O}_\ell$
\bea
\label{zmb}
Z_\ell(\beta, \tau)\, \equiv \,  \tr(\spc \rho \spc )
\! \is \!
\bigl\la {\cal O}_\ell
(0) \, {\cal O}_\ell 
(\tau ) 
\bigr\rangle_{\!\beta}   \;\;= \; \;
\raisebox{4pt}{
\begin{tikzpicture}[scale=.48, baseline={([yshift=0cm]current bounding box.center)}]
\draw[thick] (-0.65,-1.05) arc (245:475:1.6);
\draw[thick] (-.95,1.85) arc (50:310:1.9);
\draw[dotted,thick,red] (-0.8,-0.7) -- (-.8,1.6);
\draw[color={rgb:red,10; black,3}, fill={rgb:red,10; black,3}]   (-.8,1.75) circle (0.14); 
\draw[color={rgb:red,10; black,3}, fill={rgb:red,10; black,3}]   (-.8,-.95) circle (0.14); 
\draw (-3.15,.45) node {\footnotesize\bf $\beta\!-\! \tau$};;
\draw (1.15,.45) node {\footnotesize\bf $\tau$};;
\end{tikzpicture}}
\eea
with $\tau \equiv \beta_\rrR$.
This thermal two-point function has been analyzed and can be explicitly computed in the Schwarzian limit of the SYK model. This is the appropriate limit for our purpose of extracting the holographic dual interpretation of this class of partially entangled states.

\subsection{Overview of results}
\vspace{-2mm}
In this paper we will determine the 2D space-time dual to the partially entangled states \eqref{ostate} in the SYK model \eqref{hsyk}, for $\ell \sim N/\beta J$, and compute the entanglement entropy between the two sides. We will work in the low energy approximation of the SYK model, described by Schwarzian quantum mechanics. This is the appropriate regime for comparison with AdS${}_2$ gravity.

\begin{figure}[t!]
\begin{center}
\begin{tikzpicture}[scale=0.78]
\draw[dashed] (-4.15,-1.72) -- (3.03,-1.72);
\draw[thin] (1,-1.72) circle (0.05);
\draw[fill=black] (-1.75,-1.72) circle (0.09);
\draw[thick,color={rgb:red,10; black,3}] (0,-3.46) -- (0,0);
\draw[thick] (0,0) arc (120:-120:2);
\draw[thick] (0,0) arc (45:315:2.45);
\draw[thin] (-1.55,-1.52) arc (53:-53:0.24);
\draw[thin] (.9,-1.88) arc (245:115:0.2);
\draw (-1.05,-2.02) node {\scriptsize $\theta_\llL$};
\draw (0.532,-2.02) node {\scriptsize $\theta_\rrR$};
\draw[thin][dashed] (1,-1.72) -- (0,0);
\draw[thin][dashed] (1,-1.72) -- (0,-3.45);
\draw[thin][dashed] (-1.7,-1.72) -- (0,0);
\draw[thin][dashed] (-1.7,-1.72) -- (0,-3.45);
\draw (2.1,-1.37) node[color=blue] {\footnotesize $b$};
\draw (.4,-1.39) node[color=blue] {\footnotesize $a$};
\draw (-0.75,.-1.39) node[color=blue] {\footnotesize $c$};
\draw (-3.1,-1.39) node[color=blue] {\footnotesize $d$};
\draw[color={rgb:red,10; black,3}, fill={rgb:red,10; black,3}] (0,-3.48) circle (0.11);
\draw[color={rgb:red,10; black,3}, fill={rgb:red,10; black,3}] (0,0) circle (0.11);
\end{tikzpicture}
\hspace{2cm}
\begin{tikzpicture}[scale=0.8]
\draw[dashed] (-5,-2.2) -- (-.5,2.2);
\draw[dashed] (-5,2.2) -- (-.5,-2.2);
\draw[thick] (-5,-2.2) -- (-5,2.2);
\draw[thick] (2,-2.2) -- (2,2.2);
\draw[thick,color={rgb:red,10; black,3}] (-1.0,-2.2) -- (-1.0,2.2);
\draw[dashed] (-2,2.2) -- (2,-2.2);
\draw[dashed] (2,2.2) -- (-2,-2.2);
\draw[dashed] (-5,0) -- (2,0);
\draw[fill=black] (-2.75,0) circle (0.08);
\draw[thin] (0,0) circle (0.05);
\draw[thick,decoration = {zigzag,segment length = 2mm, amplitude = 0.75mm},decorate] (-5,2.2)--(2,2.2);
\draw[thick,decoration = {zigzag,segment length = 2mm, amplitude = 0.75mm},decorate] (-5,-2.2)--(2,-2.2);
\draw (1.1,.26) node[color=blue] {\footnotesize $b$};
\draw (-.6,.22) node[color=blue] {\footnotesize $a$};
\draw (-1.8,.22) node[color=blue] {\footnotesize $c$};
\draw (-4.1,.26) node[color=blue] {\footnotesize $d$};
\end{tikzpicture}
\end{center}
\caption{\small The euclidean and lorentzian space-time dual to the partially entangled states \eqref{ostate}. The worldline of the massive bulk particle created by the
operator insertion ${\cal O}_\ell$ is indicated by the red line. It divides the space-time into two AdS${}_2$ regions. For a sufficiently massive particle, the worldline is hidden behind two horizons. In this figure, the left-horizon is the true `extremal surface' with minimal value $\Phi_\llL$ of the dilaton.}
\label{PETSgeom2}
\end{figure}
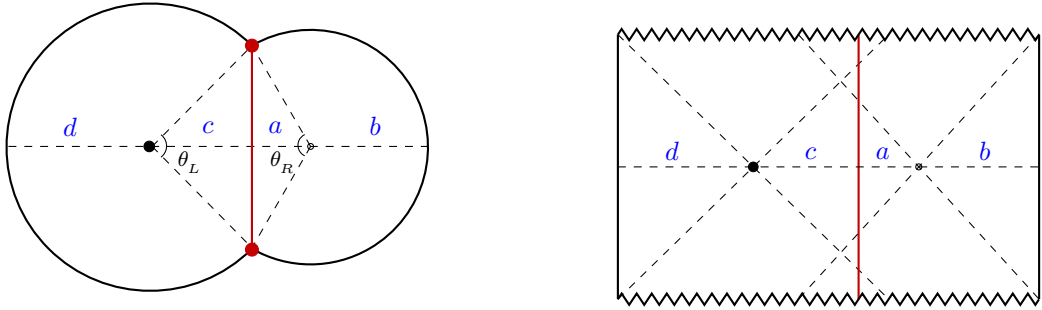

There exists an elegant and for our purpose very useful reformulation of the Schwarzian theory in terms of the motion of a charged particle on AdS${}_2$ in a large constant magnetic (euclidean) or electric (lorentzian) field \cite{Kitaev16} \cite{Maldacena:2017axo}. The classical action of the 1D effective theory is proportional to the area of AdS${}_2$ enclosed by the worldline of this charged particle. For the euclidean finite temperature partition function, this worldline follows a circular path, which we identify with the thermal circle.
In this description, the two-point function \eqref{zmb} is obtained by adding an extra term to the 1D effective action equal to $\ell$ times the length of a bulk geodesic connects the two points $t=0$ and $t=\tau$ along the worldline of the charged boundary particle. The semi-classical path of the charged particle then looks like the squeezed thermal circle shown on the left in figure \ref{PETSgeom2}. The two thermal circle segments represent the piece-wise-circular trajectory of the charged particle, and the red line represents the geodesic worldline of a massive bulk particle with mass $\ell$.\footnote{A very similar geometric set up has been considered previously in \cite{Kourkoulou:2017zaj} and \cite{Gu:2017njx}.}
The holographic dual geometry consists of two AdS${}_2$ regions glued together along the path of the massive particle. As shown in figure \ref{PETSgeom2}, each AdS${}_2$ region contains a center point, that after analytic continuation to lorentzian signature, corresponds to a bifurcate horizon of a two-sided black hole. 

For sufficiently large $\ell$ above some critical value, determined by $\beta_{\llL,\rrR}$, the worldline of the bulk particle is hidden in the region behind two horizons. In this regime, the state will look thermal relative to the observables that probe the left and right exterior region. The effective left and right temperature and the opening angles $\theta_{\llL,\rrR}$ are determined via the effective Schwarzian dynamics. We will compute these effective temperatures in section \ref{sec:JTback}.

In figure \ref{PETSgeom2}, the left-horizon is the true `extremal surface' with minimal value $\Phi_\llL$ of the dilaton. In section \ref{sec:EEPETS}, we will show that its value governs the entanglement entropy between the two sides via $S_{\rm ent} = S_0 + \Phi_\llL/{4G_N}$
with $S_0$ the microscopic ground state entropy of the SYK model.
In section \ref{sec:bulkrec} we will argue that this extremal surface separates the regions accessible through one-sided bulk reconstruction from each side. In particular, the right-sided entanglement wedge includes the regions $a$ and $c$ behind the horizon shown in figure \ref{PETSgeom2}. Finally, in section \ref{sec:genmultins} we discuss some generalizations of PETS with more than one operator insertion. In the Appendix we collect some useful formulas for determining and reconstructing the classical bulk geometry.

\section{Space Time Geometry of PETS}\label{sec:JTback}
\vspace{-1mm}
In this section our interest is to determine the holographic dual geometry described by the partially entangled thermal states, in the semiclassical regime. One approach would be to start from the Jackiw-Teitelboim model \cite{JT, Almheiri:2014cka, Maldacena:2016upp, Engelsoy:2016xyb}. As mentioned above, this JT model can be recast as the mechanics of a charge boundary particle in a magnetic field \cite{Kitaev16} (see also \cite{Maldacena:2017axo} and \cite{Gu:2017njx}). Here we will follow a somewhat different route: we will start from the exact correlators of the low energy effective theory of the SYK model, given by Schwarzian quantum mechanics, computed in \cite{Mertens:2017mtv}. We then take their semiclassical limit \cite{Lam:2018pvp} and derive the semi-classical space-time geometry from the resulting expression. As we will see, this procedure is remarkably efficient. 

We will denote the JT dilaton by $\Phi$. The coupling constant that appears in the Schwarzian action is $C=\frac{\Phi_r}{8\pi G_N}$, with $\Phi_r=\epsilon \Phi_b$ the renormalized boundary dilaton value. In SYK, the coupling $C$ corresponds to the heat capacity $C = \alpha_S N/J$, with $\alpha_S$ an order one constant \cite{KitaevTalks, Maldacena:2016hyu}. For the Schwarzian action we follow the notation in \cite{Mertens:2017mtv}. We summarize the coordinates and our conventions in Appendix \ref{app:conv}. In the following, we will parametrize the energy $E$ and thermal entropy $S$ of a finite energy state by means of a dimensionless `momentum' variable $k$ via
\bea
E(k) \! \is \! \frac{k^2}{2C} , \qquad \qquad S\, =\, S_0 + 2 \pi k,
\eea
where $S_0$ denotes the microscopic SYK ground state entropy.

As explained in the Introduction, the partition function associated with a PETS is given by the two-point function of two operators of dimension $\ell$. The exact two-point function obtained in \cite{Mertens:2017mtv} can be written as \footnote{The notation $\pm$ inside the Gamma function means one should take a product over all signs combinations. See \cite{Mertens:2017mtv} for more details.}
\bea
\lb \mathcal{O}(\tau)\mathcal{O}(0)\rb_\beta &=& \int \prod_{i=1,2}dk_i\rho(k_i)~ e^{-\frac{k_1^2}{2C}\tau - \frac{k_2^2}{2C}(\beta-\tau)}\frac{\Gamma(\ell \pm ik_1\pm i k_2)}{\Gamma(2\ell)},\\
&=& \int \prod_{i=1,2} dk_id\theta_i ~e^{-I(k_i, \theta_i, \tau, \ell)},
\ea
where the `action' appearing in the exponent is given by 
\beq
I(k_i, \theta_i, \tau, \ell) =   \sum_{i=1,2} \left(\frac{k_i^2}{2C}\tau_i + \theta_i k_i - \log \rho(k_i)\right) +\ell \log \left(\cos \frac{\theta_1}{2}+ \cos \frac{\theta_2}{2} \right)^2 + I_0(\ell),
\eeq
and we defined $\tau_1=\tau$, $\tau_2=\beta-\tau$ and the density of states $\rho(k)= 2k\sinh 2 \pi k$. This second way of expressing the two-point function will be very useful below. We will refer to $I(k_i,\theta_i)$ as the action associated to the two-point function with values $k_i$ and $\theta_i$. At this point this gives an exact expression computing the two-point function, up to an unimportant normalization factor $I_0$ which appears as a constant term in the action. 

We now take a semiclassical limit, $C$ and $\ell$ both large with $\ell/C$ fixed. Since $\ell$ is a dimensionless number it should be compared with a dimensionless ratio such as $2 \pi C/\beta$. Since we will take $\beta$ to be of order one we will simply compare $\ell$ directly with $C$. In this case the integrals over $k_i$ and $\theta_i$ become dominated by their saddle point. The saddle point scaling is such that $k_i\sim C$ and $\theta_i \sim 1$. This approximation is reliable since the action scales as $I \sim C$. We define the (order one) semiclassical action $I_{\rm s.c.}$ as 
\beq
I(k_i,\theta_i) = C I_{\rm s.c.}(k_i,\theta_i).
\eeq
In this limit the action simplifies to 
\beq\label{eq:spactionsc}
C I_{\rm s.c.}(k_i, \theta_i, \tau, \ell) =   \sum_{i=1,2} \left(\frac{k_i^2}{2C}\tau_i +( \theta_i-2\pi) k_i \right) +\ell \log \left(\cos \frac{\theta_1}{2}+ \cos \frac{\theta_2}{2} \right)^2 + I_0.
\eeq
We see that when $\ell \sim \mathcal{O}(C)$, the saddle point will depend on the value of $\ell$. We can interpret this as the result of backreaction of the space-time geometry. The saddle-point equations $\partial_{k_i} I_{\rm s.c.} = \partial_{\theta_i} I_{\rm s.c.} =0$ simplify to 
\bea\label{eq:sp2pt}
\frac{k_i}{C}\tau_i = 2 \pi - \theta_i,\\
\frac{k_i}{\sin \frac{\theta_i}{2}} = \frac{\ell}{\cos \frac{\theta_1}{2}+ \cos \frac{\theta_2}{2}},
\ea
for $i=1,2$. Using the first equations one can eliminate the angles $\theta_i$. This gives the equivalent system of equations
\bea
\label{sad2}
\frac{ k_1\tau_1 }{C} + 2 \arctan \frac{k_1+k_2}{\ell} +2 \arctan \frac{k_1-k_2}{\ell} &=&2 \pi ,\\
 \frac{k_2\tau_2 }{C} + 2 \arctan \frac{k_1+k_2}{\ell} - 2 \arctan \frac{k_1-k_2}{\ell}&=& 2\pi.
\ea
The geometric meaning of the above equations will be explained below. \footnote{In the small $\ell$ limit, $\ell/C \ll 1$, the backreaction is turned off. The solution then becomes 
$k_i \approx {2 \pi C}/{\beta}$ and $\theta_i \approx {2 \pi \tau_i}/{\beta}$
Keeping track of the subleading $\mathcal{O}(\ell/C)$ terms, one finds the expected form of a thermal two-point function in a 1D CFT 
\beq
\label{eq:sl2r2pt}
\lb \mathcal{O}(\tau) \mathcal{O}(0) \rb \sim \Bigl(\frac{\pi}{\beta \sin \frac{\pi}{\beta} \tau} \Bigr)^{2\ell}.
\eeq}
\begin{figure}
\begin{center}
\begin{tikzpicture}[scale=1.2]
\draw[thick] (0,0) coordinate (a) arc (30:330:2) coordinate (d);
\draw (a)+(0.1,0.4) node {$X_1$};
\draw (d)+(0.1,-0.4) node {$X_2$};
\draw[thick,dotted] (0,0) arc (30:-30:2);
\draw[thick] (0,0) arc (120:-120:1.15);
\draw[thick,dotted] (0,0) arc (120:240:1.15);
\draw[fill=black] (-1.732,-1) coordinate (b) circle (0.08);
\draw[fill=black] (0.577,-1) coordinate (c) circle (0.08);
\draw[dashed] (-3.732,-1) -- (1.757,-1);
\draw[dashed] (a) -- (b);
\draw[dashed] (a) -- (c);
\draw[dashed] (d) -- (c);
\draw[dashed] (d) -- (b);
\draw (-4.232,-1) node {$\tau_2$};
\draw (1.977,-1) node {$\tau_1$};
\draw (-1.132,-0.4) node {$\rho_2$};
\draw (0.537,-0.4) node {$\rho_1$};
\draw (-1.232,-1.5) node {$\theta_2$};
\draw (0.537,-1.5) node {$\theta_1$};
\draw (a)+(-.2,-0.5) node {$\phi$};
\draw[thick] (b)+(0.4,0) arc (0:-30:0.4);
\draw[thick] (b)+(0.4,0) arc (0:30:0.4);
\draw[thick] (c)+(-0.25,0) arc (180:240:0.25);
\draw[thick] (c)+(-0.25,0) arc (180:120:0.25);
\draw[thick] (a)+(0,-0.25) arc (270:300:0.25);
\draw[thick] (a)+(0,-0.25) arc (270:210:0.25);
\draw[color={rgb:red,5; black,2}, fill={rgb:red,5; black,2}] (a) circle (0.08);
\draw[color={rgb:red,5; black,2}, fill={rgb:red,5; black,2}] (d) circle (0.08);
\draw[color={rgb:red,5; black,2}] (a) -- (d);
\end{tikzpicture}
\end{center}
\vspace{-0.7cm}
\caption{\small The curve that maximizes the action with two operator insertion (red dots) at $\tau_1=\tau$ and $\tau_2=\beta-\tau$. The horizons of each side are located at the black dots. }
\label{fig:curve1}
\end{figure}
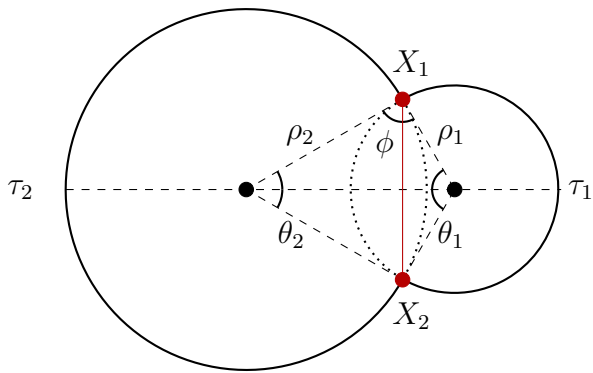

\subsection{Backreaction}
\vspace{-2mm}
In this subsection we will extract the geometric interpretation of our saddle-point equations \eqref{eq:sp2pt} and compare our results with the action described in \cite{Gu:2017njx}. 
In \cite{Gu:2017njx}, the authors exploit the fact (\cite{Kitaev16} and \cite{Maldacena:2017axo}) that the Schwarzian action associated to the reparametrization mode $f(u)$, $u\in(0,\beta)$ is proportional to the area enclosed by the curve $(\rho(u),\theta(u)=\frac{2\pi}{\beta} f(u))$ in a hyperbolic space with metric
\beq
ds^2=d\rho^2+\sinh^2\rho \hspace{0.1cm}d\theta^2\,,
\eeq
where $\rho(u)$ is determined from the constraint that the induced metric is $g_{uu}=1/\epsilon^2$, with $\epsilon$ a small cut-off scale. This describes a cut-off version of the Poincare disk in Euclidean signature. The Schwarzian action can be recast as a geometric problem regarding the boundary particle as
\bea\label{Gu:2017njxaction}
-S&=&-C\int du ~ \text{Sch}\left( \tan \frac{\pi}{\beta}f(\tau),\tau \right)+\ell \log\frac{f'(u_1)f'(u_2)}{\left(\sin\frac{\pi(f(u_1)-f(u_2))}{\beta}\right)^2} \\[2mm]
&\simeq& -\frac{C}{\epsilon}\left[(A-L+2\pi)+\frac{\epsilon\ell}{C}\log\left(2\epsilon^2\cosh D(X_1,X_2)\right)\right]
\ea
where $D(X_1,X_2)$ is the geodesic distance between the location of the insertions $X_1,X_2$. The approximation is valid when the cut-off $\epsilon\ll1$ so that $\rho(u)$ is large. $A$ denotes the area enclosed by the trajectory of the boundary particle, and $L \sim \beta$ its length. 

For $\ell \ll C$ we can neglect the term in the action depending on the geodesic distance between $X_1$ and $X_2$. Then the curve that minimizes the area with a fixed length is given by a circle inside the Poincare disk. In Lorenzian signature this maps to a black hole with the horizon located at the center of the disk. The location of this circle as a function of the length (temperature) is 
\beq
k= \frac{2\pi C}{\beta},~~~\sinh \rho = \frac{C}{k \epsilon}.
\eeq
In this case it is easy to see that the action in \eqref{Gu:2017njxaction} matches the first terms of our action \eqref{eq:spactionsc} using the explicit expression for the area $A=2 \pi\cosh \rho $ and length $L= 2 \pi \sinh \rho $ in for the Poincare disk.
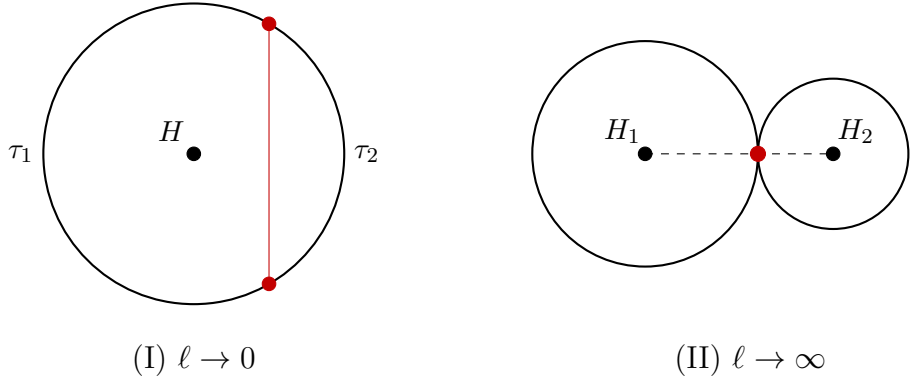
\begin{figure}[t!]
\begin{center}
\begin{tikzpicture}[scale=1]
\draw[thick] (0,0) circle (2);
\draw[color={rgb:red,10; black,3}, fill={rgb:red,10; black,3}] (1, 1.73) circle (0.09);
\draw[color={rgb:red,10; black,3}, fill={rgb:red,10; black,3}] (1, -1.73) circle (0.09);
\draw[color={rgb:red,10; black,3}] (1, 1.73) -- (1, -1.73);
\draw (-2.3,0) node {\small $\tau_1$};
\draw (-0.3,0.3) node {\small $H$};
\draw (2.3,0) node {\small $\tau_2$};
\draw (0,-2.75) node {(I) $\ell \to 0$};
\draw[fill=black] (0,0) circle (0.09);
\draw[dashed] (6, 0) -- (8.5,0);
\draw[thick] (6,0) circle (1.5);
\draw[thick] (6+2.5,0) circle (1);
\draw[color={rgb:red,10; black,3}, fill={rgb:red,10; black,3}] (6+1.5,0) circle (0.1);
\draw[fill=black] (6+0,0) circle (0.09);
\draw[fill=black] (6+2.5,0) circle (0.09);
\draw (6-0.3,0.3) node {\small $H_1$};
\draw (8.5+0.3,0.3) node {\small $H_2$};
\draw (6+1.4,-2.75) node {(II) $\ell \to \infty$};
\end{tikzpicture}
\vspace{-0.6cm}
\end{center}
\caption{\small Backreaction generated by an operator insertion (red dots) of dimension $\ell$ when $\ell\to0$ and $\ell\to\infty$. We indicate the backreaction by depicting the deformations of the boundary curve in the Euclidean Poincare disk. We indicate the (local) horizons by a black dot.}
\label{fig:backreaction}
\end{figure}

When a heavy operator is inserted we also need to minimize the distance between the insertion points $X_1$ and $X_2$. First one can approximate each side of the boundary by circles as shown in figure \ref{fig:curve1}. Each has a radius given by 
\beq\label{distcirc}
\sinh \rho_i = \frac{C}{k_i \epsilon}.
\eeq
If we define the opening angle of each circle by $\theta_i$ as in figure \ref{fig:curve1} then the length on each side is related to the time insertions as 
\beq
\sinh \rho_i (2\pi -\theta_i) = \frac{\tau_i}{\epsilon}.
\eeq
By using this equation and \eqref{distcirc} one gets precisely the first relation of our saddle-point equations \eqref{eq:sp2pt}. With these identifications, the geodesic distance between $X_1$ and $X_2$ is given by as
\bea
\cosh D_{12} &=& 1 + 2 \sinh^2 \rho_1 \sin^2 \frac{\theta_1}{2} =1 + 2 \sinh^2 \rho_2 \sin^2 \frac{\theta_2}{2}, \nn
&\approx& \frac{C^2}{\epsilon^2} \left( \frac{\sin \frac{\theta_1}{2}}{k_1} \right)^2 = \frac{C^2}{\epsilon^2} \left( \frac{\sin \frac{\theta_2}{2}}{k_2} \right)^2.
\ea
The first line of this equation is a purely geometric result. In the second line we have used our proposal to identify our variables with geometry. A first observation is that the two equivalent geometric expressions become consistent when one takes into account the second set of saddle point equations \eqref{eq:sp2pt} since it implies $k_1^{-1}\sin \frac{\theta_1}{2}=k_2^{-1}\sin \frac{\theta_2}{2}$. From the geometry of figure \ref{fig:curve1} this is simply the hyperbolic version of the sine rule. A second observation is that the second equation in \eqref{eq:sp2pt} allows us to write 
\bea
\cosh D_{12} \is \frac{C^2}{\epsilon^2 \ell^2} \left( \cos \frac{\theta_1}{2} + \cos \frac{\theta_2}{2} \right)^2.
\eea
 Inserting this relation in the action proposed by \cite{Gu:2017njx} we find a match with our on-shell action \eqref{eq:spactionsc}. The same is true for the $\ell$ independent terms in \eqref{eq:spactionsc}. This connection allows us to extract the backreaction due to operator insertions in terms of our variables $\theta$ and $k$.

One important parameter of the geometry is the angle $\phi$ defined in figure \ref{fig:curve1} which can be shown to be equal to $\phi = \ell \epsilon/C$. If we take the cut-off to be $\epsilon \sim 1/\beta J$ in terms of SYK variables, then $\phi \sim \ell/N$. 
Another interesting parameter of the geometry is the distance between the horizons (specified by the center of each circle segment). This distance $D_H$ can be expressed as
\bea\label{distancehorizons}
\cosh D_H \is \frac{1 + \cos \frac{\theta_1}{2}\cos \frac{\theta_2}{2}}{\sin \frac{\theta_1}{2}\sin \frac{\theta_2}{2}}= \frac{k_1^2+k_2^2+\ell^2}{2k_1k_2}.
\eea
We can separate this into a minimal geodesic distance $D_1$ ($D_2$) between the left (right) horizon and the world line of the bulk particle as 
\bea\label{eq:distHB}
\sinh D_1 \is \frac{\ell^2 +k_2^2 - k_1^2}{2 \ell k_1},~~~~\sinh D_2 = \frac{\ell^2 +k_1^2 - k_2^2}{2 \ell k_2}.
\eea
One can verify they add up to the distance between horizons $D_1 + D_2 = D_H$. 

Assume $\tau_1 \neq \tau_2$. Then call $k_{\rm max}={\rm max}(k_1,k_2)$ and $k_{\rm min}={\rm min}(k_1,k_2)$. For any $\ell$, the horizon associated to $k_{\rm min}$ is always visible from the outside, and it is always part of the geometry. This is not true for the other horizon associated with $k_{\rm max}$. When $\ell$ takes values between $\ell=0$ and a critical $\ell_*$, the right horizon is not part of the geometry, it is removed by the gluing procedure across the worldline of the bulk particle. The second horizon becomes visible for $\ell>\ell_*$. From the formulas above (in particular the one for $D_1$ and $D_2$) one can write a condition that determines the critical scaling dimension $\ell_*$ as
\bea
\ell_*^2 \is k^2_{\rm max}(\ell_*)-k^2_{\rm min}(\ell_*),
\eea
where the momenta in the right-hand side are functions of $\ell_*$, $\beta$ and $\tau$, determined by solving the saddle point equations \eqref{sad2}. From the geometry of figure \ref{fig:curve1} one can see that this is equivalent to the condition $\theta=\pi$ (for which the bulk particle worldline crosses the horizon associated to $k_{\rm max}$). 

Another way of writing the condition of both horizons being part of the geometry is 
\bea
|\Delta E| \is |E_2- E_1| \, < \, \frac{\ell^2}{2C}.
\eea
In this regime, the trajectory of the bulk particle lies between the two horizons. The corresponding state of the SYK system will look thermal from the perspective of simple observables that can only measure the state outside each horizon. We will elaborate more on bulk reconstruction for these geometries in section \ref{sec:bulkrec}.

Using this knowledge about the backreaction we can analyze the cases $\ell \to 0$ and $\ell \to \infty$. We show both cases in figure \ref{fig:backreaction} and we explain below how these simple figures allow us to find approximate solutions to the saddle-point equations.

 When $\ell \to 0$ backreaction is negligible, $k_1 \approx k_2 \approx 2 \pi C/\beta$. Then the geometry is a circle in the Poincare disk with length $\propto \beta$ and the evaluation of the geodesic distance reproduces equation \eqref{eq:sl2r2pt}. We will study the leading correction to this limit in section \ref{sec:eepetssummary}.
 
On the other hand when $\ell \to \infty$ points $X_1$ and $X_2$ want to be as close as possible. The two arcs become full circles touching at a point. The renormalized length of each circle is fixed to be $\tau_1=\tau$ and $\tau_2=\beta-\tau$. We can anticipate then $k_1 \approx 2 \pi C/\tau_1$ and $k_2 \approx 2 \pi C/\tau_2$ with $\theta_i \approx 0$. From equation \eqref{distancehorizons} we can deduce the distance between horizons as a function of the dimension in this limit as $D_H \approx 2\log \frac{\beta \ell}{C}$. 

\subsection{Dilaton Profile}
The Schwarzian dynamics fixes the backreaction and therefore fixes the boundary curve as explained above. From the boundary curve, one can easily find the dilaton profile inside the bulk. The detailed formulas are left for Appendix \ref{app:conv}. Here we point out the relevant qualitative features of the euclidean configuration and its continuation to Lorenzian signature. 
 
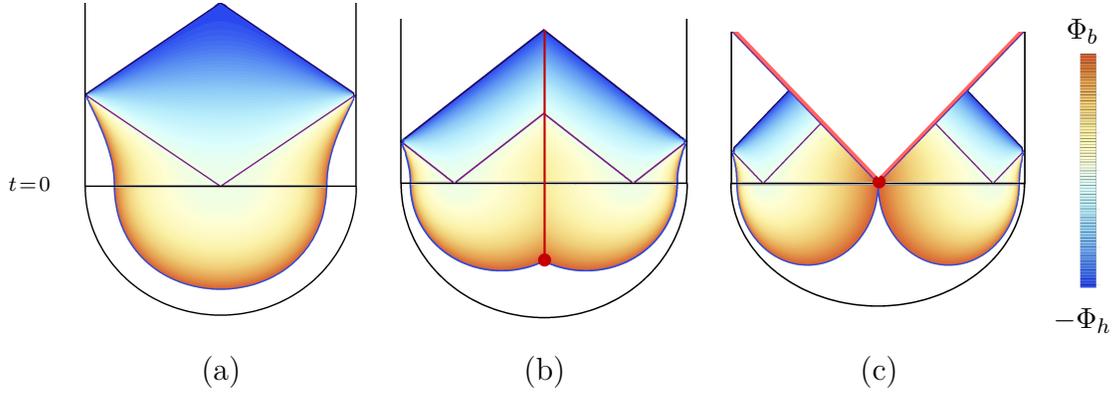
\begin{figure}
\begin{center}
\begin{tikzpicture}
\draw (-2.4,-0.3) node {\scriptsize$t\!=\!0\ \ \ $};
\node[inner sep=0pt] (russell) at (0,0){\includegraphics[scale=0.32]{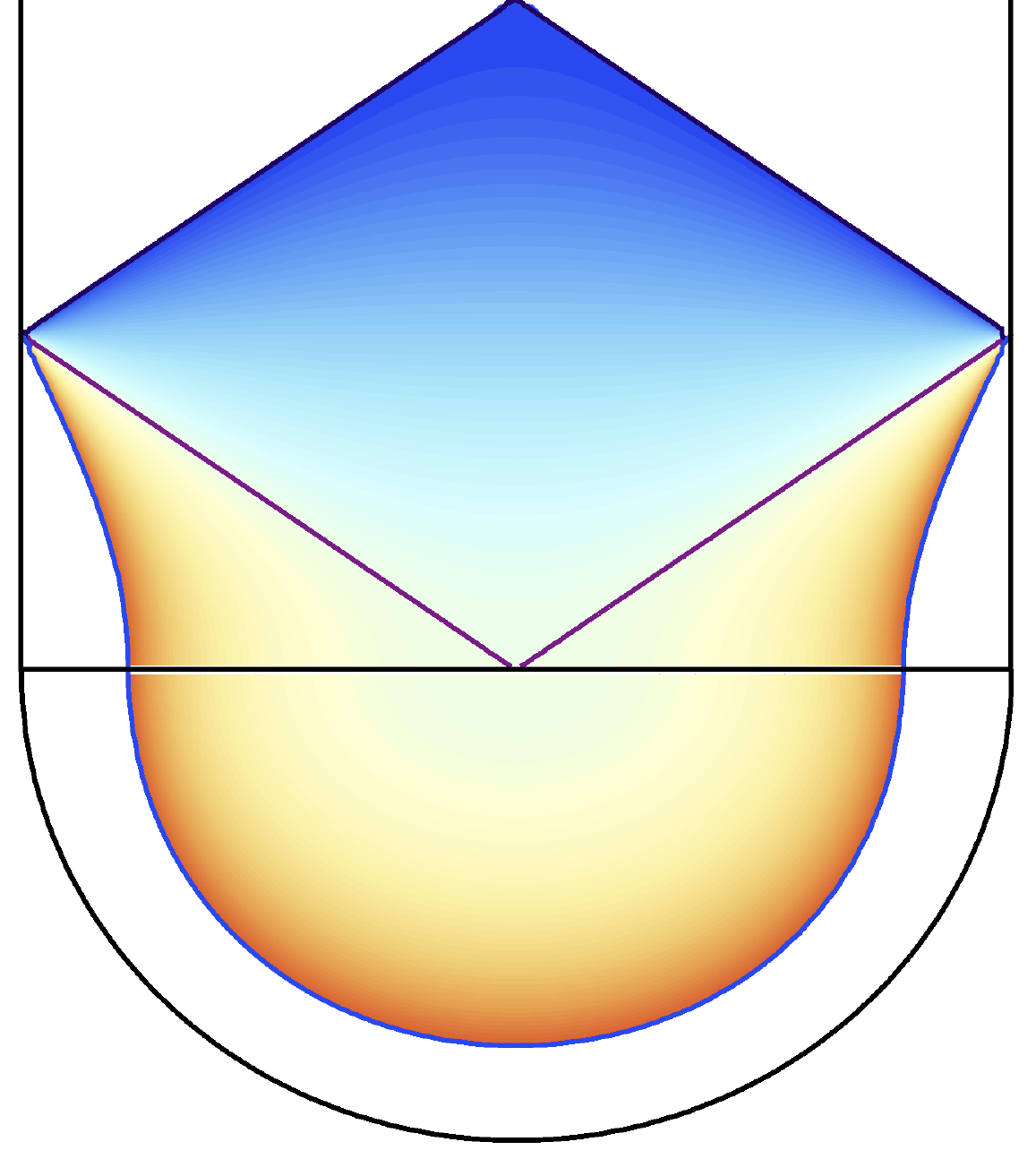}};
\draw (0,-2.8) node {(a)};
\end{tikzpicture}
\hspace{0.2cm}
\begin{tikzpicture}
\node[inner sep=0pt] (russell) at (0,0){\includegraphics[scale=0.39]{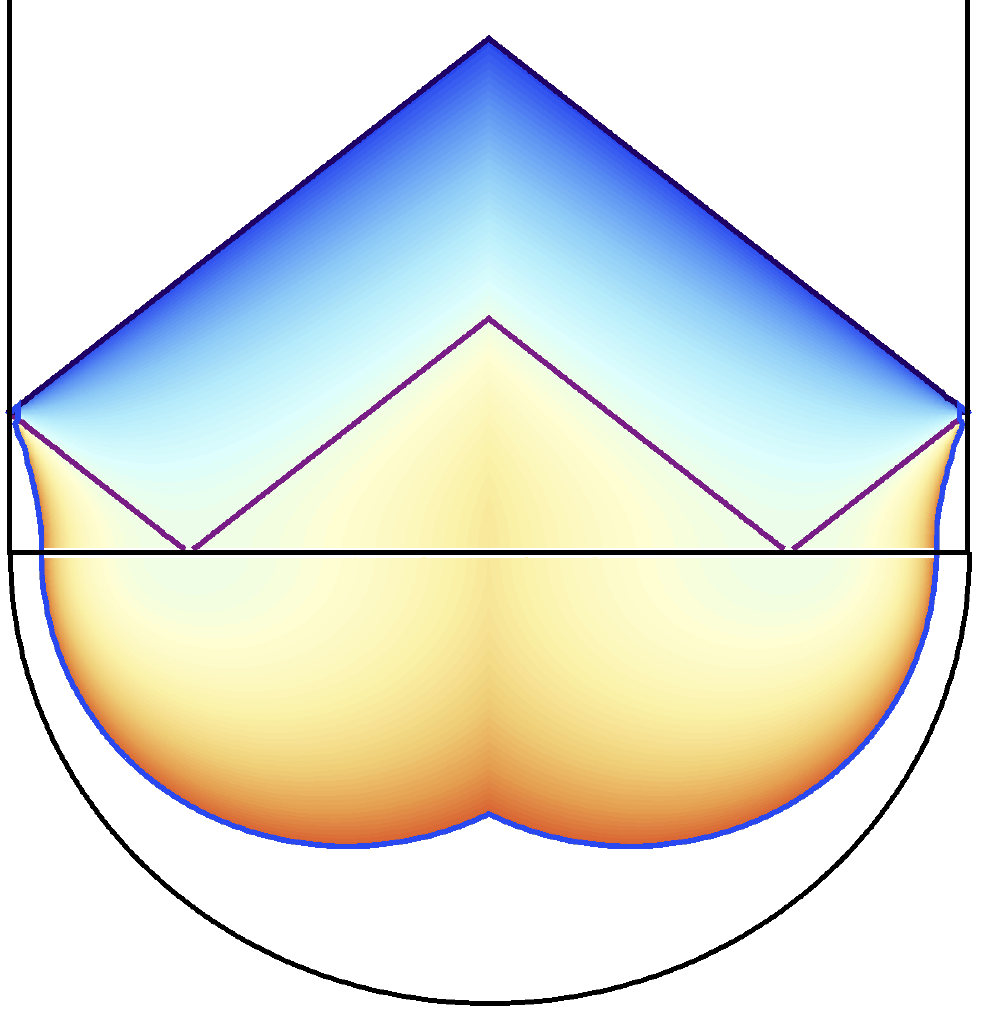}};
\draw[color={rgb:red,10; black,3}, fill={rgb:red,10; black,3}] (0,-1.18) circle (0.08);
\draw[thick][color={rgb:red,10; black,3}] (0,- 1.15) -- (0,1.88);
\draw (0,-2.67) node {(b)};
\end{tikzpicture}
\hspace{0.2cm}
\begin{tikzpicture}
\node[inner sep=0pt] (russell) at (0,0){\includegraphics[scale=0.32]{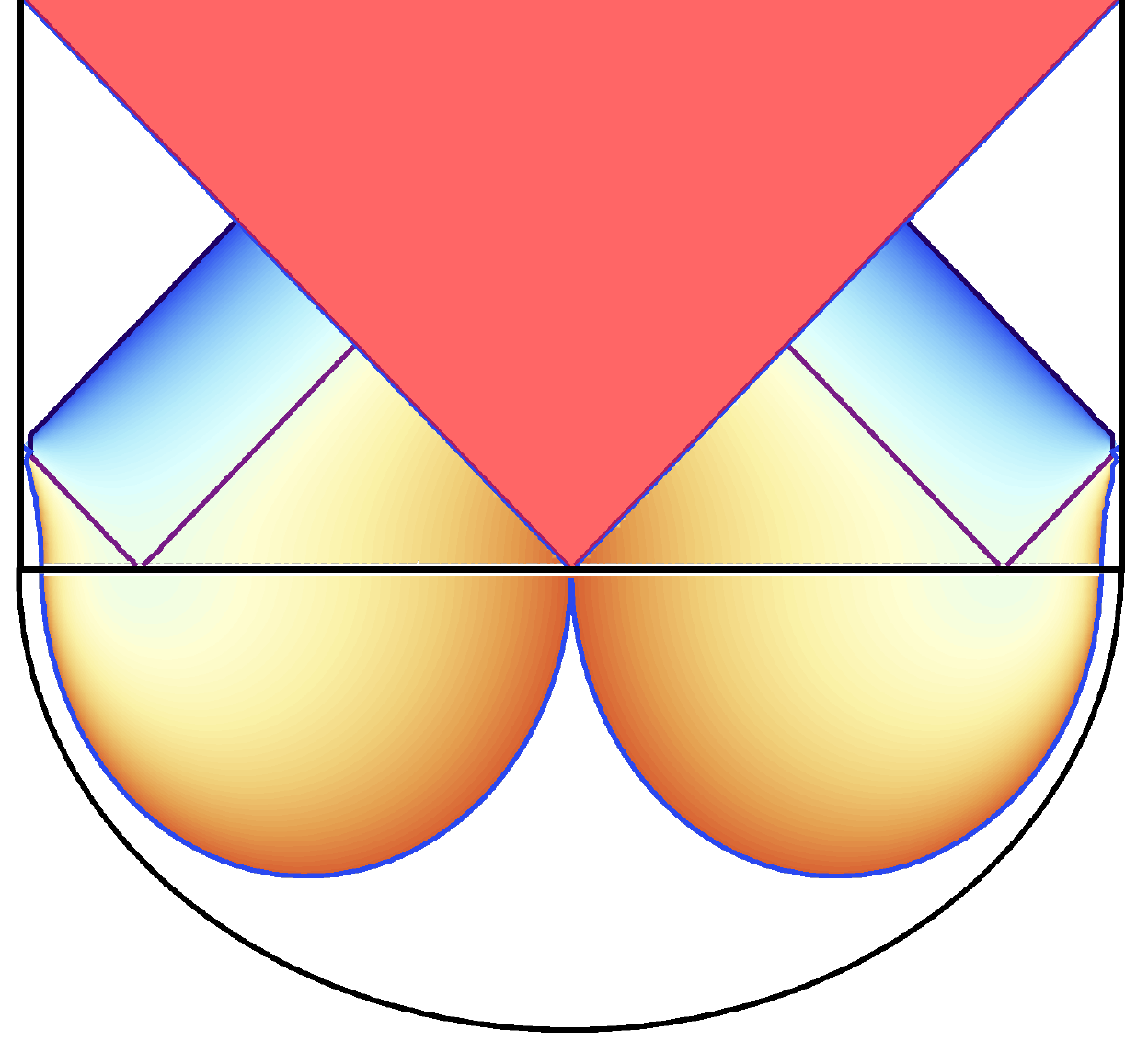}};
\draw[color={rgb:red,10; black,3}, fill={rgb:red,10; black,3}] (0,-0.15) circle (0.08);
\path[fill=white!10]  (-0.01,-0.095) -- (2,2) -- (-2.02,2)-- cycle;
\draw (0,-2.68) node {(c)};
\node[inner sep=0pt, rotate=90] (russell) at (2.8,0){\includegraphics[scale=0.6]{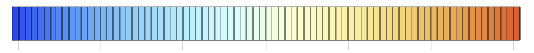}};
\draw (2.7,-2) node {\small $-\Phi_h$};
\draw (2.7,1.85) node {\small $\Phi_b$};
\end{tikzpicture}
\end{center}
\vspace{-0.5cm}
\caption{\small Dilaton profile. The values of the dilaton in arbitrary units go from large positive values of $\Phi$ at the boundary (red end of color spectrum) to large negative values ending at the inner horizon (blue end of color spectrum):  
We show $\ell \sim 0$ (left), $\ell \sim N/\beta J$ (middle) and $\ell \sim N$ (right).}
\label{fig:dprof}
\end{figure}

The dilaton blows up near the asymptotic boundary of AdS. The boundary curve is defined such that $\Phi_{\rm b} = \Phi_r / \epsilon$ for a cut-off $\epsilon$ and a finite renormalized dilaton $\Phi_r$. The Schwarzian captures the limiting dynamics as $\epsilon$ goes to zero. In euclidean space, the dilaton $\Phi$ is smaller than $\Phi_{\rm b}$ everywhere inside the cut-off curve and has a local minimum at each horizon. For the TFD in euclidean space (circle in the Poincare disk) one has concentric circles of constant dilaton. In the continuation to Lorentzian signature across the $t=0$ time slice, the boundary curve splits into two hyperbolas (corresponding to the left and right QM) that hit the boundary of AdS after finite global time. Inside of the lorenzian bulk, the 4d singularity is located where $\Phi + \Phi_0 =0$ so one could imagine taking as a cut-off $\Phi=-\Phi_0$ or to be safe $-\Phi_{\rm b}$. From the 4d perspective the inner horizon is located $-\Phi_h$ and it is well known to be unstable. Therefore we will cut-off the geometry at $-\Phi_h$. We show this situation in panel (a) of figure \ref{fig:dprof}. 

For the PETS with an operator inserted during the euclidean evolution, we need to glue two of locally TFD solutions along the world line of the bulk particle as shown in panel (b). As reviewed in Appendix \ref{app:conv} each side of the circles have a definite $SL(2,\mathbf{R})$ charge. Charge conservation implies that the dilaton is continuous along the bulk brane when the two halves are glued \footnote{To see this simply start from charge conservation $Q_L = Q_R + Q_m$ (these are three component vectors living in embedding space, see Appendix \ref{app:conv} for notation). Then take the inner product with $Y$ (coordinate in embedding space describing EAdS$_2$). Along the worldline of the particle $Q_m \cdot Y=0$ so $\Phi_L = \Phi_R$. }. As expected from the equations of motion, the slope of the dilaton is discontinuous with a jump proportional to the mass $\ell$ of the boundary particle $\nabla \Phi_L - \nabla \Phi_R \sim \ell$. 

The dilaton inside of the cut-off surface stays always below the UV cut-off value fixed by $\Phi_{\rm bdy}$. Then the bulk space between the cut-off surface and the singularity can be trusted. As the mass of the bulk brane is increased the gradient $\nabla \Phi$ grows without bound, $\nabla \Phi \to \infty$ for $\ell \to \infty$. This implies that there is a critical value $\ell_{\rm cr}$ of the mass such that for $\ell > \ell_{\rm cr}$ the low energy approximation that gives JT gravity breaks down. The geometry inside the causal future of the particle position at $t=0$ (V-shaped region in panel (c) of figure \ref{fig:dprof}) describes a strongly coupled region. The answer for what happens inside this region might depend on fine-grained details of the operator and of the SYK dynamics. If the operator is a projection that acts on all fermions such that $\ell \sim N$, it was argued in \cite{Kourkoulou:2017zaj} that the region inside the V-shaped region must be removed. 
 
\subsection{Multiple Insertions}
\vspace{-2mm}

In this subsection we will comment on the generalization of PETS produced by multiple operator insertions. In the leading large $N$ limit and low energy limit of the SYK model, we can then compute the relevant correlation functions using the results of \cite{Mertens:2017mtv}.\footnote{For our purpose, it will be sufficient to focus on channels where bulk propagators do not cross and therefore do not involve the R-matrix of \cite{Mertens:2017mtv}.} 
\begin{figure}
\vspace{-0.6cm}
\begin{center}
\begin{tikzpicture}[scale=1.12]
\draw[thick] (0,0) coordinate (a) arc (30:160:2) coordinate (e); 
\draw[thick] (e) arc (60:300:0.7898) coordinate (f); 
\draw[thick] (f) arc (200:330:2) coordinate (g); 
\draw[thick] (0,0) arc (120:-120:1.15); 
\draw (a)+(0.1,0.4) node {$X_1$};
\draw (d)+(0.1,-0.4) node {$X_2$};
\draw (e)+(-0.15,0.4) node {$X_3$};
\draw (f)+(-0.15,-0.4) node {$X_4$};
\draw[thick,color={rgb:red,10; black,3}] (e) -- (f);
\draw[thick,color={rgb:red,10; black,3}] (a) -- (d);
\draw[fill=black] (-1.732,-1) coordinate (b) circle (0.08);
\draw[fill=black] (0.577,-1) coordinate (c) circle (0.08);
\draw[fill=black] (-4.00632,-1) coordinate (o) circle (0.08);
\draw[dashed] (a) -- (b);
\draw[dashed] (a) -- (c);
\draw[dashed] (d) -- (c);
\draw[dashed] (d) -- (b);
\draw[dashed] (o) -- (e);
\draw[dashed] (o) -- (f);
\draw[dashed] (b) -- (e);
\draw[dashed] (b) -- (f);
\draw (-5.032,-1) node {$\tau_2$};
\draw (1.977,-1) node {$\tau_1$};
\draw (-1.7,1.2) node {$\tau_3$};
\draw (-1.7,-3.25) node {$\tau_4$};
\draw (-1.132,-0.4) node {$\rho_p$};
\draw (0.537,-0.4) node {$\rho_k$};
\draw (-4,-0.5) node {$\rho_q$};
\draw (-1.232,-1.5) node {$\theta_{p1}$};
\draw (-2.182,-1.5) node {$\theta_{p2}$};
\draw (0.537,-1.5) node {$\theta_k$};
\draw (-3.96632,-1.5) node {$\theta_q$};
\draw[thick] (b)+(0.4,0) arc (0:-30:0.4);
\draw[thick] (b)+(0.4,0) arc (0:30:0.4);
\draw[thick] (c)+(-0.25,0) arc (180:240:0.25);
\draw[thick] (c)+(-0.25,0) arc (180:120:0.25);
\draw[thick] (b)+(-0.45,0) arc (180:160:0.45);
\draw[thick] (b)+(-0.45,0) arc (180:200:0.45);
\draw[thick] (o)+(+0.2,0) arc (0:60:0.2);
\draw[thick] (o)+(+0.2,0) arc (0:-60:0.2);
\draw[color={rgb:red,10; black,3}, fill={rgb:red,10; black,3}] (a) circle (0.1);
\draw[color={rgb:red,10; black,3}, fill={rgb:red,10; black,3}] (d) circle (0.1);
\draw[color={rgb:red,10; black,3}, fill={rgb:red,10; black,3}] (e) circle (0.1);
\draw[color={rgb:red,10; black,3}, fill={rgb:red,10; black,3}] (f) circle (0.1);
\end{tikzpicture}
\end{center}
\vspace{-0.35cm}
\caption{\small The boundary curve that maximizes the Schwarzian action with two pairwise operator insertions.}
\label{fig:curve2}
\end{figure}
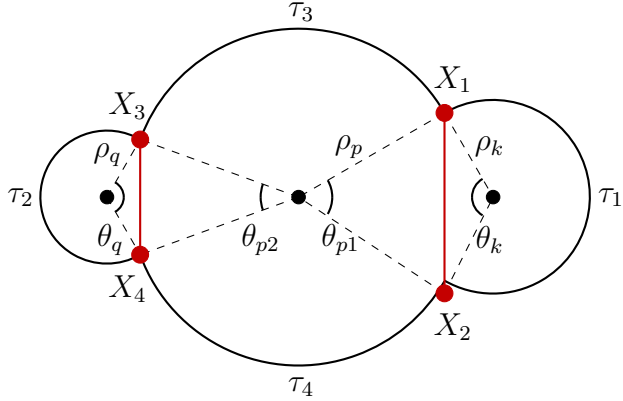

Using the same procedure as described above for the two-point function, we derive that the $2n$-point functions is determined by a semiclassical action which has the form 
\beq
I = \sum_{\rm prop.} \frac{k_i^2}{2C}\tau_i  + \sum_{i,j~{\rm paired}}\left( \ell \log \left(\cos \frac{\theta_i}{2}+ \cos \frac{\theta_j}{2} \right)^2 +\theta_i k_i + \theta_j k_j \right) - \sum_i 2 \pi k_i. 
\eeq 
In the first term the sum is over propagators over times $\tau_i$ with an intermediate state energy $k_i^2/2C$. The second term is a sum over pairs of insertions into a bilocal field. Finally the third term is a sum over momenta $k$ that are different off-shell. From this action one can obtain the saddle point equations and bulk geometry by a gluing procedure similar to the case of two operators.

\subsubsection{Example: Four-Point Function}
\vspace{-1mm}
As a concrete exercise we will apply the ideas above to the four-point function. This shares some general features with the case of an arbitrary number of insertions. We will use the results in section \ref{sec:genmultins}. 

In figure \ref{fig:curve2} we show the geometry backreacted by the two bilocal insertions. We determine the shape of the geometry from the semi-classical expression of the time-ordered four point function\footnote{If the four operators were identical we should sum over all channels. Here we assume that they are only pairwise identical.}
$$\lb \mathcal{O}_{\ell_1}(X_1)\mathcal{O}_{\ell_1}(X_2)\mathcal{O}_{\ell_2}(X_3)\mathcal{O}_{\ell_2}(X_4)\rb.$$
Applying the rules of the previous section the effective action computing this correlator is given by 
\bea
I&=& \frac{k^2}{2C}\tau_1 + \frac{q^2}{2C} \tau_2 + \frac{p^2}{2C}\tau_3 +\frac{p^2}{2C}\tau_4\nn
&&\hspace{-0.6cm}+\ell_1 \log\left( \cos \frac{\theta_k}{2} + \cos \frac{\theta_{p1}}{2} \right)^2+\theta_k k + \theta_{p1} p+\ell_2 \log\left( \cos \frac{\theta_q}{2} + \cos \frac{\theta_{p2}}{2} \right)^2 + \theta_q q + \theta_{p2} p \nn
&& - 2 \pi p - 2 \pi q - 2 \pi k.
\ea
where $\sum_i \tau_i = \beta$. The geometric role of each variable is shown in figure \ref{fig:curve2}. The first line corresponds to a sum over each propagator over a time $\tau_i$. There are four of them contributing, although only three different ones (off-shell) due to a conservation law. In the second line we sum over both pairings. Finally the third line has a sum over channels coming from the density of states. Note that, since off-shell only three momenta differ, $p$ contributes as $2 \pi p$ and not $4\pi p$.   

Since we will use the results in this section later we will write down the saddle-point equations in detail. From varying the momenta $k$, $p$, $q$ we obtain 
\beq
2 \pi - \theta_k = \frac{k \tau_1}{C},~~~2\pi - \theta_q = \frac{q \tau_2}{C},~~~2\pi - \theta_{p1} - \theta_{p2} = \frac{p(\tau_3 + \tau_4)}{C}.
\eeq
From varying the opening angles $\theta$'s we obtain the equations
\bea
&&\frac{k}{\sin \frac{\theta_k}{2}} = \frac{p}{\sin \frac{\theta_{p1}}{2}} = \frac{\ell_1}{\cos \frac{\theta_k}{2} + \cos \frac{\theta_{p1}}{2}},\nn
&&\frac{q}{\sin \frac{\theta_q}{2}} = \frac{p}{\sin \frac{\theta_{p2}}{2}}= \frac{\ell_2}{\cos \frac{\theta_q}{2} + \cos \frac{\theta_{p2}}{2}}.
\ea 
By eliminating the angles $\theta$ can obtain an equation for $p$, $k$ and $q$. 

\section{Entanglement Entropy of PETS}\label{sec:EEPETS}
\vspace{-1mm}
In this section we will combine the results of the previous section (and Appendix \ref{app:conv}) to compute from first principles the entanglement entropy of the partially entangled thermal states using the replica trick. We will begin by reviewing the case of the TFD. Then we will consider operators without backreaction and finally the most general case. The upshot of the calculation will be that the entanglement entropy is determined by the global minimum of the dilaton. This is consistent with the holographic entropy prescription \cite{Ryu:2006bv}.

We will consider a bipartite PETS defined in the introduction. In this section we will study the QM$_{\rm R}$ density matrix $\rho = {\rm Tr}_L |\Psi\rb \lb \Psi |$ after performing a partial trace over the left QM. We will compute the Renyi entropy of these states and from it deduce the entanglement entropy. The Renyi entropy is defined as 
\beq
\mathcal{S}_n = \frac{1}{1-n} \log \frac{{\rm Tr} \rho^n}{({\rm Tr} \rho)^n}.
\eeq
where $n$ indicates the replica index and the limit $n\to1$ gives the entanglement entropy. Another observable with this properties is the modular entropy defined as
\beq\label{renyi}
S_n=- n^2 \frac{\partial}{\partial n} \left[\frac{1}{n} \log {\rm Tr} \rho^n \right].
\eeq
This is a more natural candidate for an entropy associated to the system of $n$ replicas, as explained in \cite{baez}. By using our methods we could in principle compute both. Nevertheless, only the modular entropy $S_n$ has a clear holographic interpretation, as found in \cite{Dong:2016fnf} building upon \cite{Lewkowycz:2013nqa} \footnote{Another advantage is the fact that one can compute $S_n$ without worrying about the normalization of the density matrix. An attempt to divide by $({\rm Tr} \rho)^n$ in equation \eqref{renyi} will give the same $S_n$ after taking the derivative with respect to $n$.}.

For these reasons explained above, in this paper we will focus on $S_n$ which, with slight abuse of terminology, we will still refer to as Renyi entropy. 

As a brief warm up, we will begin by analyzing the TFD state 
\beq
\label{tfd}
| \mbox{\sc TFD} \rangle\nspc \, = \, \begin{tikzpicture}[scale=.52, baseline={([yshift=0cm]current bounding box.center)}]
\draw[thick] (-0.2,0) arc (190:350:1.52);
\draw[fill,black] (-0.2,0.0375) circle (0.05); 
\draw[fill,black] (2.8,0.0375) circle (0.05);
\draw (1.2,-.6) node {\scriptsize\bf $\frac \beta 2$};;
\end{tikzpicture} 
\eeq
Its partition function $Z(\beta)$ is a path integral over thermal circle of length $\beta$, as shown in figure \ref{PETSgeom2}. In the semiclassical limit, large $C$, it is given by 
\bea
\log Z\is \log  \int [df] ~e^{C\int d\tau \{ \tan \frac{1}{2}f(\tau),\tau \} }, \nn
&=& S_0 + \beta E_0 + \frac{2\pi^2 C}{\beta} + \ldots,
\ea
where the dots indicate subleading terms. From the point of view of the Schwarzian theory the extremal zero-point values of entropy and energy $S_0$ and $E_0$ are undetermined but large \footnote{$S_0$ is a zero-point entropy while $S_n$ denotes the $n$-th modular entropy. We hope this will not cause confusion since we will never take the $n\to0$ limit of the modular entropy in this paper.}. In the case of SYK $S_0, E_0 \sim N$ while near extremal corrections are subleading $S-S_0 \sim N/(\beta J)$. The value of the dilaton at the horizon of a black hole geometry is given by $\Phi_h =\frac{2 \pi }{\beta}\Phi_r=G_N \frac{16\pi^2 C}{\beta}$ and fixed by the temperature. 

Using the replica trick, the trace of $\rho^n$ is equivalent to the partition function of a circle of size $n\beta$. This is a simple extension of the result above  
\bea
S_n \is - n^2 \frac{\partial}{\partial n}\left[ \frac{1}{n} \log Z(n\beta)\right]=(1-n\partial_n)\log Z(n\beta)= S_0 + \frac{4\pi^2 C}{n \beta}.
\eea
We can rewrite this result as 
\bea
S_{n} \is S_0 + \frac{\Phi_h(n)}{4 G_N},
\eea
where $\Phi_h(n)$ is the dilaton at the horizon of a black hole of size $n\beta$. This is the minimal value and also lies at the fix point of the replica $\mathbf{Z}_n$ symmetry \footnote{This is not true for the more standard definition $\mathcal{S}_n$ since $\mathcal{S}_n = S_0 + (1+n)2\pi^2 C/n \beta = S_0 + \frac{n+1}{2} \frac{\Phi_h(n)}{4 G_N} $. The right-hand side is not given by $\Phi_h(n)/4 G_N$ unless $n=1$, and we can see even in this simple example the advantage of the modular entropy \eqref{renyi}.}. The entanglement entropy $S = \lim_{n\to 1} S_n = - {\rm Tr} \rho \log \rho$ is given by $S = S_0 + \frac{4\pi^2 C}{\beta}$. Of course since $\rho \sim e^{-\beta H}$ we could have directly guessed this thermodynamic relation between entropy and free energy. This thermodynamic relation will no longer be true for PETS. \footnote{From the 4d perspective, the derivation of the Jackiw-Teitelboim model reduces a near extremal black hole to the near horizon region $AdS_2 \times S^2$ (see for example \cite{Nayak:2018qej}). The dilaton then is the perturbation from extremality of the size of the horizon $A_h= \Phi_0 + \Phi_h$. The entanglement entropy above is therefore the usual Bekenstein entropy of a near extremal black hole since $S_0 = \Phi_0/4G_N$ is related to the extremal dilaton in the same way.}

We can repeat this entropy calculation for the PETS described in the introduction. In particular we will consider 
\bea
\li\spc \Psi\spc \ra & \cong & \ 
\, e^{-\frac 1 2 \beta_\llL H} \, {\cal O}_\ell\,\spc 
e^{-\frac 1 2 \beta_\rrR H} \  \;\; = \;\; \raisebox{1pt}{\begin{tikzpicture}[scale=.52, baseline={([yshift=0cm]current bounding box.center)}]
\draw[thick] (-0.65,-1) arc (245:355:1.6);
\draw[thick] (-3.75,0.3) arc (190:300:1.9);
\draw[fill,black] (1.6,0.25) circle (0.05);
\draw[fill,black] (-3.75,0.3) circle (0.05);
\draw[color={rgb:red,10; black,3}, fill={rgb:red,10; black,3}]   (-.8,-.95) circle (0.14); 
\draw (-2.4,-.54) node {\scriptsize $\frac 1 2 \beta_\llL$};
\draw (.3,-.54) node {\scriptsize $\frac 1 2 \beta_\rrR$};;
\end{tikzpicture}} \
\eea
To simplify some formulas below we will parametrize this state by $\tau=\beta_\rrR/2$ and $\beta = \beta_\llL + \beta_\rrR$ or equivalently $\frac{1}{2} \beta_\llL = \frac{1}{2} \beta - \tau$. For reasons that will be clear below we will focus on $\beta_\rrR \neq \beta_\llL$ or $\tau \neq \beta/4$. We will later generalize this state to multiple insertions in section \ref{sec:genmultins}. 

 The procedure, similarly to the previous computation, is to consider a thermal circle of size $n \beta$ for $n$ replicas, with $2n$ operator insertions. Then the trace of the replicas is given, in terms of $\beta$ and $\tau$, by the following correlator
\bea
{\rm Tr} \rho^n &=& Z_0(n \beta) \lb \mathcal{O}(-\tau) \mathcal{O}(\tau) \mathcal{O}(\beta-\tau) \mathcal{O}(\beta+\tau)  \mathcal{O}(2\beta-\tau) \mathcal{O}(2\beta+\tau)\ldots \rb_{n\beta},\nn
&\equiv& G_n(\tau,\beta),
\ea
where the dots indicate the remaining of the $2n$ operators and the unperturbed TFD partition function we reviewed above is $\log Z_0(n\beta) = S_0 + n \beta E_0 + \frac{2 \pi^2 C}{n \beta}$. In the equation above we added the partition function since, following the notation of \cite{Mertens:2017mtv}, we defined correlators to be normalized to $1$ for $\mathcal{O}=\mathbb{1}$. Then the Renyi entropy we focus on in this paper \eqref{renyi} is given in terms of correlation functions of the Schwarzian theory 
\bea\label{renyiGn}
S_n \is - n^2 \partial_n \left[\frac{1}{n} \log G_n(\tau,\beta)\right].
\eea
The expression in equation \eqref{renyiGn} is naturally divided into the sum of two terms. The logarithm of the correlator always involves $S_0 + n \beta E_0+\ldots$. This gives a contribution of the order $N$ zero-point entropy $S_n = S_0+ \ldots $. The goal will be to compute the leading near-extremal correction $S_n - S_0 \sim C$ contribution to the entropy, when $C$ is large with $\ell/C$ fixed \footnote{As mentioned in section \ref{sec:JTback}, the dimensionless $\ell$ should be compared with a dimensionless ratio such as $2 \pi C/ (\beta_\rrR + \beta_\llL)$. Since we will use units in which $\beta$ is of order one we will simply compare $\ell$ with $C$.}.  

\subsection{Warm-up: Light Operators}
We will begin as a warm-up by analyzing the limit $1\ll \ell \ll C$ \footnote{For $\ell\ll 1$ one can apply entanglement entropy perturbation theory.}. Correlators satisfy large $N$ factorization and the building blocks are given by the semiclassical answer of equation \eqref{eq:sl2r2pt}, $\lb \mathcal{O}(\tau) \mathcal{O}(0)\rb = (\pi/\beta \sin(\pi \tau/\beta))^{2\ell}$, without backreaction. 

Before writing down the general answer let us begin by taking $n=2$. Using factorization, the fact that all four operators are identical and the negligible backreaction gives a simple answer
\beq
{\rm Tr}\rho^2 = e^{S_0 + 2 \beta E_0+\frac{2 \pi^2 C}{2\beta}} \left[ \left( \frac{\pi}{2\beta \sin \frac{\pi \tau}{\beta}} \right)^{4\ell}+ \left( \frac{\pi}{2\beta \cos \frac{\pi \tau}{\beta}} \right)^{4\ell}+ \left( \frac{\pi}{2\beta } \right)^{4\ell} \right],
\eeq
The channel combinatorics makes it hard to extend to arbitrary $n$. But for $1\ll \ell \ll C$ one can simplify this considerably. To be concrete assume first that $0<\tau<\beta/4$. We can rewrite the previous expression in a convenient way as
\bea
{\rm Tr}\rho^2 &=& e^{S_0 + 2 \beta E_0+\frac{2 \pi^2 C}{2\beta}} \left( \frac{\pi}{2\beta \sin \frac{\pi \tau}{\beta}} \right)^{4\ell} \left[1+ \left( \tan \frac{\pi \tau}{\beta} \right)^{4\ell}+ \left( \sin \frac{\pi \tau}{\beta} \right)^{4\ell} \right],\nn
&\approx& e^{S_0 + 2 \beta E_0+\frac{2 \pi^2 C}{2\beta}} \left( \frac{\pi}{2\beta \sin \frac{\pi \tau}{\beta}} \right)^{4\ell}  [1+ \mathcal{O}(e^{-\ell})].
\ea
\begin{figure}[t!]
\begin{center}
\begin{tikzpicture}[scale=1]
\draw[thick, color={rgb:red,10; black,3}, dashed] (-5.9,0.6) arc (38:-38:1);
\draw[thick, color={rgb:red,10; black,3}, dashed] (-5.9+3.8,0.6) arc (180-38:180+38:1);
\draw[thick, color={rgb:red,10; black,3}, dashed] (-4-0.6,1.9) arc (270-38:270+38:1);
\draw[thick, color={rgb:red,10; black,3}, dashed] (-4-0.6,-1.9) arc (90+38:90-38:1);
\draw[dashed] (-6.5,0) -- (-1.5,0);
\draw[dashed] (-4,-2.5) -- (-4,2.5);
\draw[thick] (-2,0) arc (360:0:2);
\draw[fill=black] (-4,0) circle (0.08);
\draw[color={rgb:red,10; black,3}, fill={rgb:red,10; black,3}] (-5.9,0.6) circle (0.08);
\draw[color={rgb:red,10; black,3}, fill={rgb:red,10; black,3}] (-5.9+3.8,0.6) circle (0.08);
\draw[color={rgb:red,10; black,3}, fill={rgb:red,10; black,3}] (-5.9,-0.6) circle (0.08);
\draw[color={rgb:red,10; black,3}, fill={rgb:red,10; black,3}] (-5.9+3.8,-0.6) circle (0.08);
\draw[color={rgb:red,10; black,3}, fill={rgb:red,10; black,3}] (-4-0.6,1.9) circle (0.08);
\draw[color={rgb:red,10; black,3}, fill={rgb:red,10; black,3}] (-4+0.6,1.9) circle (0.08);
\draw[color={rgb:red,10; black,3}, fill={rgb:red,10; black,3}] (-4-0.6,-1.9) circle (0.08);
\draw[color={rgb:red,10; black,3}, fill={rgb:red,10; black,3}] (-4+0.6,-1.9) circle (0.08);
\draw (-6.3,0.3) node {\small $\tau$};
\draw (-4-0.3,2.2) node {\small $\tau$};
\draw (-6.1,1.6) node {\small $\beta-2\tau$};
\end{tikzpicture}
\hspace{1cm}
\begin{tikzpicture}[scale=1, rotate=45]
\draw[thick, color={rgb:red,10; black,3}, dashed] (-5.9,0.6) arc (38:-38:1);
\draw[thick, color={rgb:red,10; black,3}, dashed] (-5.9+3.8,0.6) arc (180-38:180+38:1);
\draw[thick, color={rgb:red,10; black,3}, dashed] (-4-0.6,1.9) arc (270-38:270+38:1);
\draw[thick, color={rgb:red,10; black,3}, dashed] (-4-0.6,-1.9) arc (90+38:90-38:1);
\draw[dashed,rotate=-45] (-6.5+1,-2.84) -- (-1.5+1,-2.84);
\draw[dashed, rotate=-45] (-4+1.17,-2.5-2.8) -- (-4+1.17,2.5-2.8);
\draw[thick] (-2,0) arc (360:0:2);
\draw[fill=black] (-4,0) circle (0.08);
\draw[color={rgb:red,10; black,3}, fill={rgb:red,10; black,3}] (-5.9,0.6) circle (0.08);
\draw[color={rgb:red,10; black,3}, fill={rgb:red,10; black,3}] (-5.9+3.8,0.6) circle (0.08);
\draw[color={rgb:red,10; black,3}, fill={rgb:red,10; black,3}] (-5.9,-0.6) circle (0.08);
\draw[color={rgb:red,10; black,3}, fill={rgb:red,10; black,3}] (-5.9+3.8,-0.6) circle (0.08);
\draw[color={rgb:red,10; black,3}, fill={rgb:red,10; black,3}] (-4-0.6,1.9) circle (0.08);
\draw[color={rgb:red,10; black,3}, fill={rgb:red,10; black,3}] (-4+0.6,1.9) circle (0.08);
\draw[color={rgb:red,10; black,3}, fill={rgb:red,10; black,3}] (-4-0.6,-1.9) circle (0.08);
\draw[color={rgb:red,10; black,3}, fill={rgb:red,10; black,3}] (-4+0.6,-1.9) circle (0.08);
\end{tikzpicture}
\vspace{-0.6cm}
\end{center}
\caption{\small Replica Geometry for $1\ll \ell \ll C$ with $n=4$. We indicate the channel that dominates when $0<\tau<\beta/4$ (left) and $\beta/4<\tau<\beta/2$ (right). The central black dot indicates the $\mathbf{Z}_n$ symmetric horizon with minimal dilaton.}
\label{figsmallell}
\end{figure}
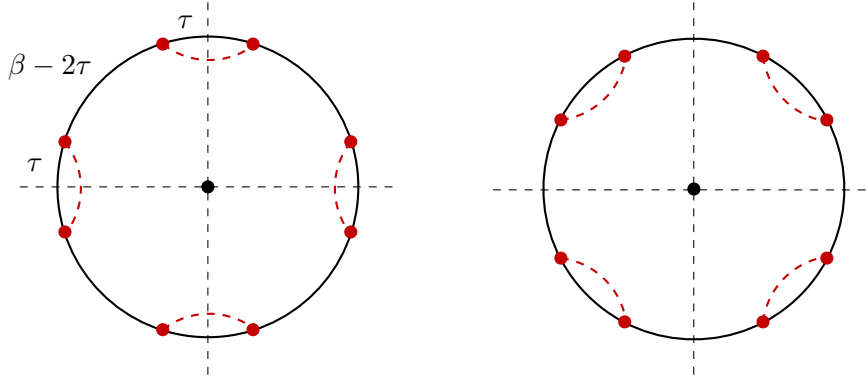
If we assume $\ell \gg 1$ then the second and third term are negligible regardless of $\tau$ (the tangent is smaller than one only for $\tau<\beta/4$). If $\tau>\beta/4$ then the channel contracting operators separated by $\beta-2\tau$ (term with cosine above) dominates. Using this we can run the same argument for arbitrary $n$ and $\tau$. In figure \ref{figsmallell} we show the situation for $n=4$ as an example. The general answer can be written in each case as
\beq
{\rm Tr}\rho^n \approx \begin{cases}
e^{S_0 + n \beta E_0 + \frac{2 \pi^2 C}{n\beta}} \left( \frac{\pi}{n\beta \sin \frac{\pi 2\tau}{n\beta}} \right)^{2n\ell},~~~~0<\tau<\beta/4,\\
e^{S_0 + n \beta E_0+\frac{2 \pi^2 C}{n\beta}} \left( \frac{\pi}{n\beta \sin \frac{\pi (\beta-2\tau)}{n\beta}} \right)^{2n\ell},~~~~\beta/4<\tau<\beta/2.
\end{cases}
\eeq
Note that we are not using a properly normalized density matrix. This is not a problem for computing $S_n$ \eqref{renyi} (although $\mathcal{S}_n$ is sensitive to normalization). Using this result the Renyi entropy is given by 
\beq\label{renyiLCsmall}
S_{n} = S_0 + \frac{4\pi^2 C}{n\beta}+ 2 \ell \left( n - \frac{2 \pi x}{\beta} \cot \frac{2 \pi x}{n\beta} \right),~~~~x={\rm min}\left(\tau,\frac{\beta}{2}-\tau\right).
\eeq
From this expressions taking the limit $n\to1$ is straightforward. The corrections to these expressions are of order $\mathcal{O}(1/\ell)$ and $\mathcal{O}(1/C)$ so that for $1\ll \ell \ll C$ this is well justified. The only subtelty occurs at precisely $\tau=\beta/4$. For this choice there is a phase transition where the Renyi (or entanglement) entropy is continuous but with a jump in the first derivative. Another feature is the symmetry under $\tau \to \beta/2-\tau$ ($\beta_\llL \leftrightarrow \beta_\rrR$) due to the fact that $|\Psi\rb \in \mathcal{H}_L\otimes \mathcal{H}_R$ is a pure state and therefore $S_{\rm L}=S_{\rm R}$.  

\subsection{Heavy Operators}
\vspace{-2mm}
The main issues appearing when attempting to compute $S_n$ is the channel combinatorics and the presence of contractions with crossing legs that imply non-trivial gravitational interactions (the appearance of the R-matrix from \cite{Mertens:2017mtv}). Both these problems can be avoided in the semiclassical limit of large $C$ and large $\ell \sim C$ in a way similar to the previous case. Correlators of an arbitrary number of operators, as reviewed in section \ref{sec:JTback}, are exponential in $C$ meaning that 
 \beq
 G_n(\tau,\beta) = \sum_{{\rm channel~} k} e^{C I_{k}(\tau,\beta, \ell/C)}\approx e^{C I_{\rm max}(\tau,\beta, \ell/C)}[1 + \mathcal{O}(e^{-C})].
 \eeq
Therefore to exponential accuracy in $C$, the correlator is dominated by the channel which minimizes the classical action $I_{\rm k}$ appearing in the exponent. This is similar to the situation in higher dimensions where large $N$ ensures that one picks the saddle-point of minimal action as long as there are no degeneracies. 

\begin{figure}[t!]
\begin{center}
\begin{tikzpicture}[scale=1.05]
\node[inner sep=0pt] (russell) at (-1-0.5,0)
    {\includegraphics[scale=0.3]{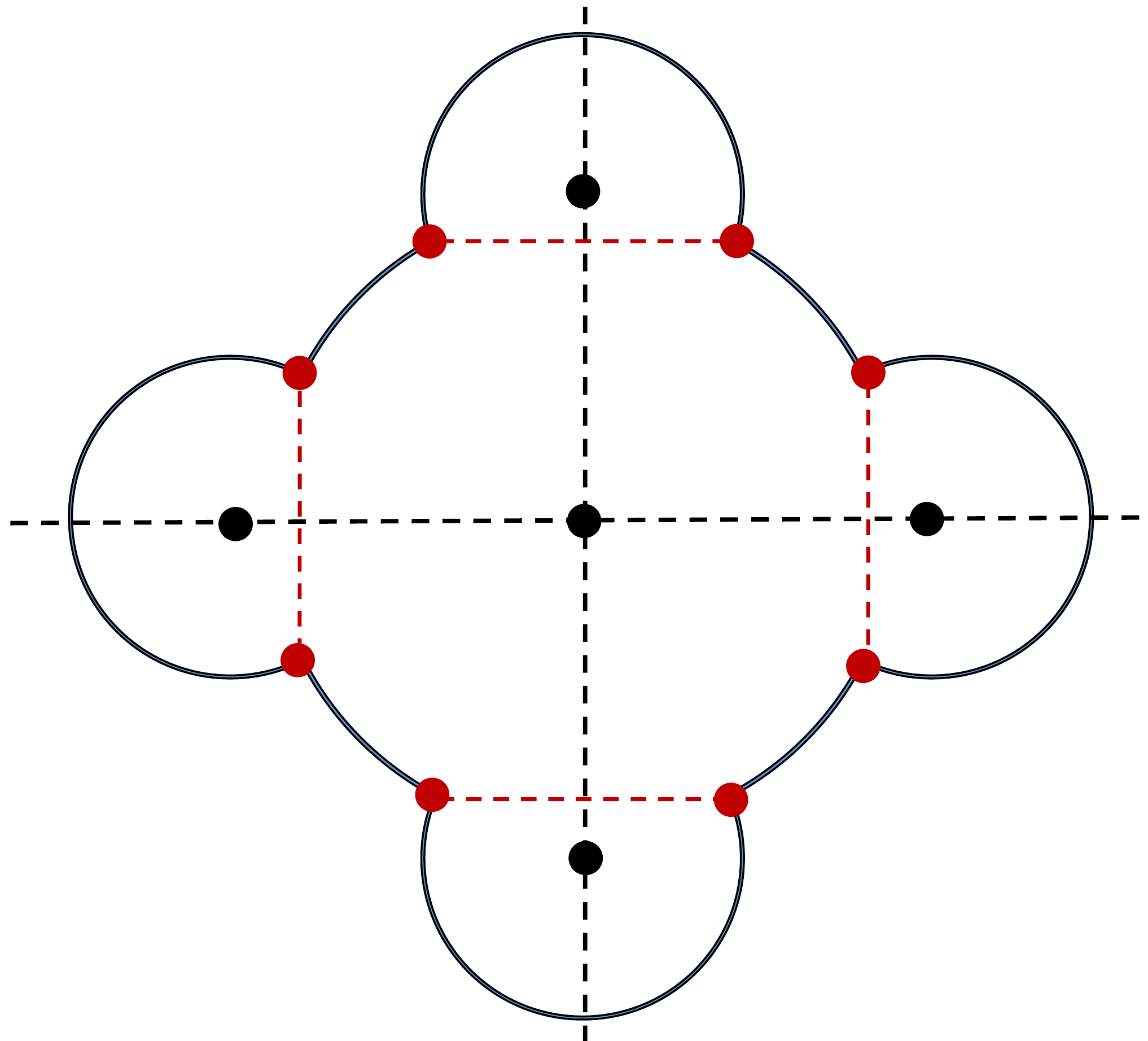}};
    \draw (-4-0.5,0.25) node {\small $k_{1}$};
    \draw (2.1-0.5,-0.25) node {\small $k_{3}$};
    \draw (-0.3-0.5,2.7) node {\small $k_{2}$};
    \draw (-1.5-0.5,-2.7) node {\small $k_{4}$};
    \draw (-2.4-0.5,1.4) node {\small $p$};
    \draw (-2.4-0.5,-1.4) node {\small $p$};
    \draw (1.45-1-0.5,1.4) node {\small $p$};
    \draw (1.45-1-0.5,-1.4) node {\small $p$};
    \draw (-1.1-0.5,-3.5) node {(I) $0<\tau<\beta/4$ ($p<k$)};
\node[inner sep=0pt] (whitehead) at (5.4,0)
    {\includegraphics[scale=0.3]{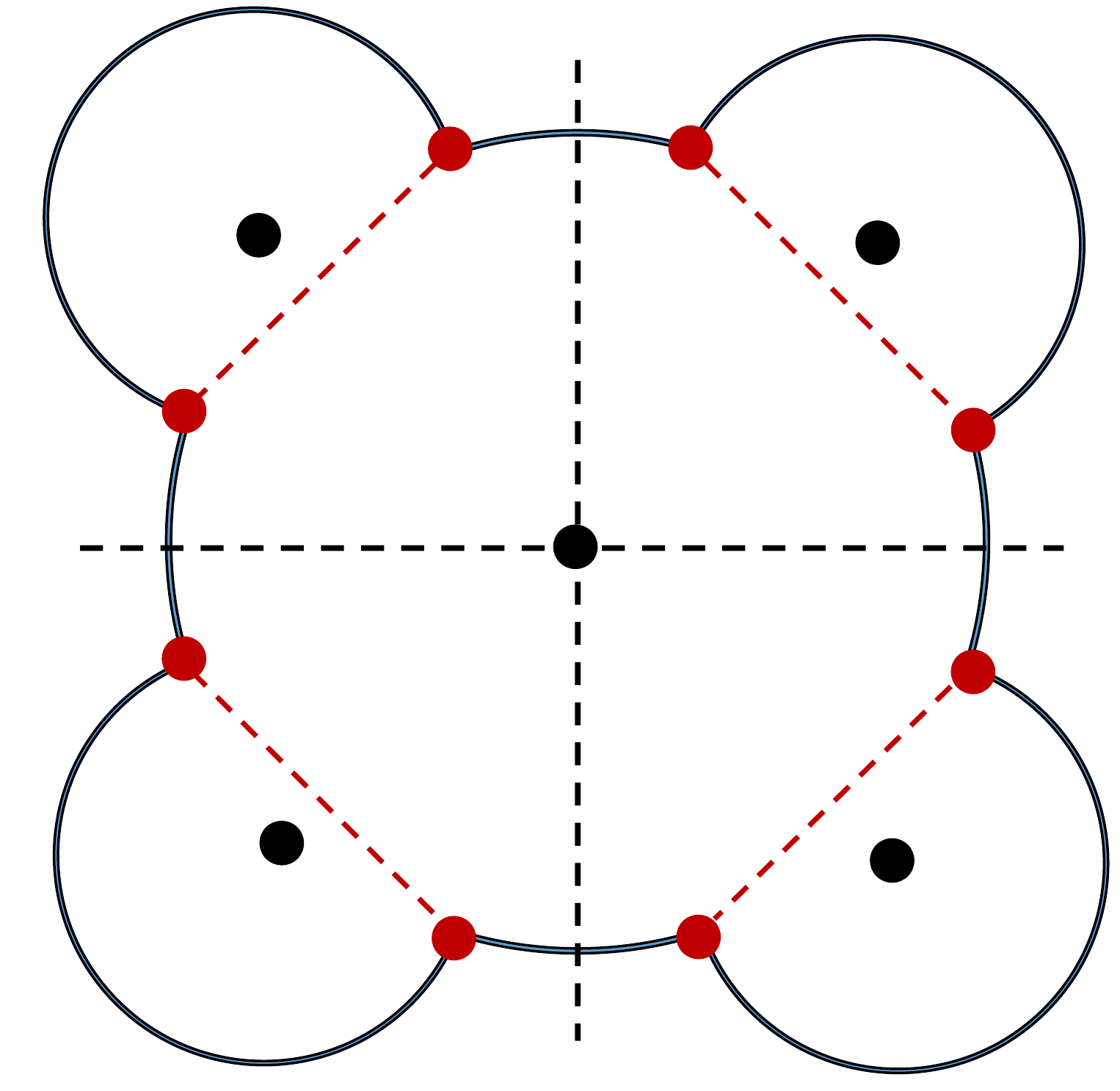}};
    \draw (6-2.5,0.25) node {\small $k$};
    \draw (6+1.5,-0.25) node {\small $k$};
    \draw (6-0.3,2) node {\small $k$};
    \draw (6-0.8,-2) node {\small $k$};
    \draw (6-3,1.5) node {\small $p_{1}$};
    \draw (6+1.9,1.5) node {\small $p_{2}$};
    \draw (6-3,-2) node {\small $p_{4}$};
    \draw (6+1.9,-2) node {\small $p_{3}$};
     \draw (5.5,-3.5) node {(II) $\beta/4<\tau<\beta/2$ ($k<p$)};
    \end{tikzpicture}
\end{center}
\vspace{-0.5cm}
\caption{\small Geometry generated by the insertion of heavy operators (red dots) for $n=4$ (replicas are separated by dashed lines). Case I: Four local dilaton local minima at the horizons (black dots) with $\Phi_i \propto k_{i}(n)\equiv k(n)$ and at the central, (global) minimum $\mathbf{Z}_n$ symmetric, horizon with $\Phi_{\rm min} \propto p(n)<k(n)$. Case II: Four local minima at the horizons (black dots) with $\Phi_i \propto p_{i}(n)=p(n)$ and at the $\mathbf{Z}_n$ symmetric horizon with $\Phi_{\rm min} \propto k(n)<p(n)$.}
\label{fig:cloverdiagram}
\end{figure}

This solves both problems since a single channel dominates and moreover channels with crossing legs never win (the reason is analogous to the statement in Lorenzian time that OTOC cannot be bigger than time ordered ones). For the calculation of the Renyi entropy of the state defined above there are two cases in which different channels dominate (1) $0<\tau<\beta/4$ and (2) $\beta/4<\tau<\beta/2$. 
\vspace{1.5mm}
\begin{center}\textbf{Case I: }$0<\tau<\beta/4$ ($\beta_\rrR<\beta_\llL$)\end{center}
\vspace{.5mm}
We need the correlator of $2n$ operators placed periodically at a distance alternating between $2\tau$ and $\beta-2\tau$. For case I the channel that dominates has a contraction between operators separated a distance $2\tau$. We show this channel in figure \ref{fig:cloverdiagram}, where we define the intermediate channel momenta $k_{i}$ and $p$. 

The configuration has a $\mathbf{Z}_n$ symmetry of permuting the replicas and a $\mathbf{Z}_2$ symmetry of time reversal. This is important for finding the saddle-point of the classical action giving this correlator (momenta running along outer circles $k_{i}$ are different off-shell but coincide on-shell $k_{i} \equiv k$ thanks to the $\mathbf{Z}_n$ symmetry). In general the correlator is given semiclassically as 
\beq
\log G_n(\tau,\beta) = S_0 + n \beta E_0+ 2 \pi p + \sum_{i=1}^n 2 \pi k_{i} +\sum_{i=1}^n \tilde{I}_i(p,k_{i},\ell,\tau,\beta)
\eeq
the explicit formula for the terms $\tilde{I}_i$ and the saddle-point equations can be obtained from the general methods explain in section \ref{sec:JTback}. Using that $k_i \equiv k$ the correlator simplifies to 
\beq
\log G_n(\tau,\beta) =S_0 + n\beta E_0+ 2 \pi p + n \left[2 \pi k + \tilde{I}(p,k,\ell,\tau,\beta)\right]
\eeq
Using this we can compute the Renyi entropy $S_n$ as 
\beq
S_n = - n^2 \partial_n \frac{1}{n} \log Z_n = S_0 -n^2\frac{\partial}{\partial n}  ~\frac{1}{n} \left( 2 \pi p + n \left[2 \pi k +  \tilde{I}(p,k,\ell,\tau,\beta)\right] \right)
\eeq
Now we can see the advantage of this definition of the Renyi entropy. When taking the derivative with respect to $n$ one has the explicit $n$ depence and the implicit dependence through the saddle-point value of $p(n)$ and $k(n)$. When evaluated on the saddle-point solution, the derivative with respect to the implicit dependence on $n$ vanishes exactly. Taking derivatives only to the explicit factors of $n$ simplifies considerably  
\bea
S_{n}^{\rm Renyi} &=& S_0 + 2 \pi p(n)=S_0 + \frac{\Phi_{h}(n)}{4 G_N},\nn
&=&  S_0 + \min_{\small Y\in\hspace{0.7mm}{\rm Bulk}} \frac{\Phi(Y)}{4 G_N},
\ea
where the saddle-point equation defining $p(n)$ and $k(n)$ is given by 
\bea
\frac{ k }{C}2\tau + 2 \arctan \frac{k+p}{\ell} +2 \arctan \frac{k-p}{\ell} &=&2 \pi ,\\
 \frac{p }{C}(\beta-2\tau) + 2 \arctan \frac{k+p}{\ell} - 2 \arctan \frac{k-p}{\ell}&=& \frac{2\pi}{n}.
\ea
$Y$ is the position in embedding space parametrizing the bulk dual to the boundary with $n$ replicas. As we see in figure \ref{fig:cloverdiagram}, there are several local minima of the dilaton (only two are different due to $\mathbf{Z}_n$ symmetry). The Renyi entropy is given by the global minimum which corresponds to the value $\Phi_h(n)$ at the $\mathbf{Z}_n$ symmetric central horizon in the figure. Indeed for $0<\tau<\beta/4$ and any $\ell$ the condition $p(n)<k(n)$ is always satisfied. 

This is consistent with the holographic prescription derived in higher dimensions in \cite{Dong:2016fnf}. The standard Renyi entropy $\mathcal{S}_n$ does not have such a simple formula but it is still computable using the explicit expressions of the semiclassical action. 
 \begin{figure}[t!]
\begin{center}
\begin{tikzpicture}[scale=1.14]
\node[inner sep=0pt] (russell) at (0,0)
    {~\includegraphics[scale=0.35]{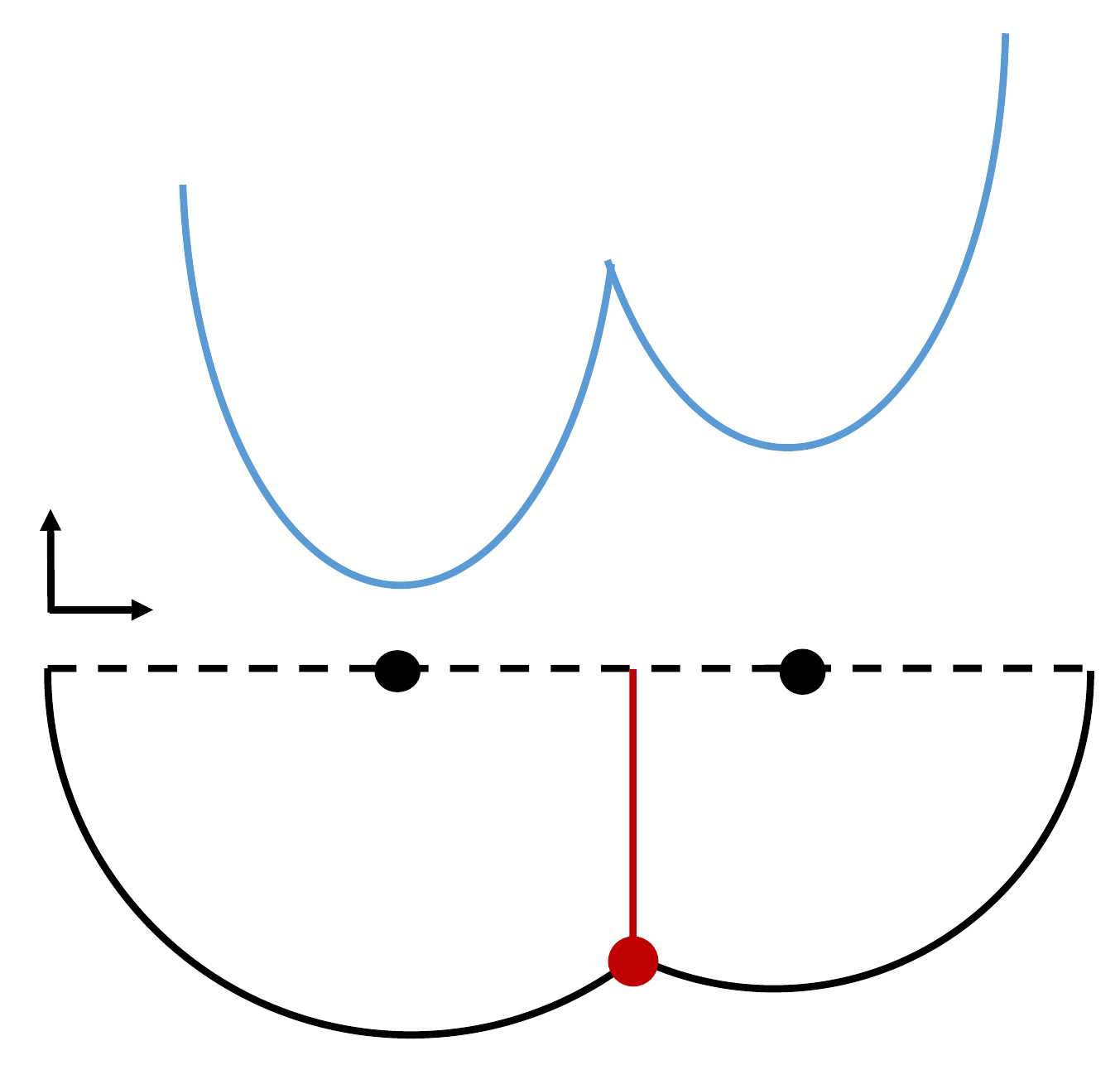}};
    \draw (-2.13,0.05) node {\small $\Phi$};
    \draw (-1.35,-0.3) node {\small $r$};
    \draw (-0.45,-0.9) node {\small $\Phi_L$};
    \draw (0.9,-0.9) node {\small $\Phi_R$};
    \draw (2.55,-0.65) node {\small ${\rm QM}_R$};
    \draw (-2.35,-0.65) node {\small ${\rm QM}_L$};
    \end{tikzpicture}
\end{center}
\vspace{-0.5cm}
\caption{\small Dilaton profile (blue curve) at time $\tau=0$ as a function of the radial direction. Below we show the Euclidean evolution that creates the state with an insertion of $\mathcal{O}_\ell$ (red dot). The profile has two local minima at the horizons $\Phi_{L/R}$ at the left/right horizons. We show the case $\tau<\beta/4$ for which $\Phi_L<\Phi_R$, then the microscopic entanglement entropy is given by $\Phi_L$.}
\label{fig:dilatonprof}
\end{figure}
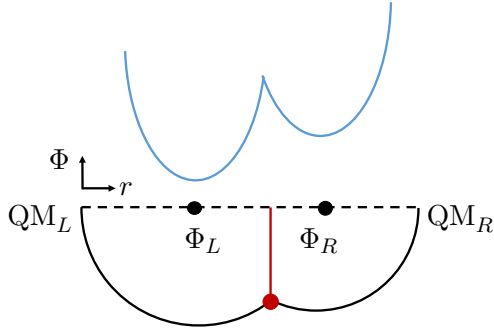

Now we can take the $n\to1$ limit. The local minimal values of the dilaton at the left and right horizons are 
\beq
\frac{\Phi_L}{4 G_N} = 2\pi p(n=1)~~~~\text{and}~~~~\frac{\Phi_R}{4 G_N} = 2 \pi  k(n=1),
\eeq
we show this in terms of the black hole bulk geometry in figure \ref{fig:dilatonprof}. We explained in section \ref{sec:JTback} that one of the horizons might be hidden and would not be part of the geometry for a range of $\ell$. It is easy to see that from the two horizons the one with minimal horizon dilaton is always visible.

\begin{center}\textbf{Case II: }$\beta/4<\tau<\beta/2$ ($\beta_\llL<\beta_\rrR$)\end{center}
\vspace{.5mm}
After case I, deriving the results for case II is straightforward. The channel that dominates now has contractions between nearest neighboring operators separated by $\beta_\llL=\beta-2\tau$. The Renyi entropy is given by the momentum which does not appear with a factor of $n$ in the semiclassical action. For case I this was $p$ and for case II it is $k$ instead (see right panel of figure \ref{fig:cloverdiagram}). This gives
\beq
S_n = S_0 + 2 \pi k(n) =  S_0 + \min_{\small Y\in\hspace{0.7mm}{\rm Bulk}} \frac{\Phi(Y)}{4 G_N}.
\eeq
Then the holographic prescription is still valid for case II. The saddle-point equations are different though and now become
\bea
\frac{ k }{C}2\tau + 2 \arctan \frac{k+p}{\ell} +2 \arctan \frac{k-p}{\ell} &=&\frac{2\pi}{n},\\
 \frac{p }{C}(\beta-2\tau) + 2 \arctan \frac{k+p}{\ell} - 2 \arctan \frac{k-p}{\ell}&=& 2\pi ,
\ea
which coincides with the previous case for $n=1$ but in general might be different. In this case now $k(n)<p(n)$ for any choice. The situation for $n=1$ also gets reversed with respect to the previous case. The prescription of choosing the minimal value of the dilaton still gives the right answer

\subsection{Summary}\label{sec:eepetssummary}
\vspace{-2mm}
Putting everything together we can write the general result valid as long as $\tau \neq \beta/4$ ($\beta_\llL \neq \beta_\rrR$). Looking at figure \ref{fig:dilatonprof} we see that the dilaton profile has two potential local minima given by the horizons value $\Phi_{L}=\Phi(Y_{H_L})$ and $\Phi_R=\Phi(Y_{H_R})$, which are proportional to $p$ and $k$ respectively. For arbitrary $n$ the minima occurs in the $\mathbf{Z}_n$ invariant point which is always given by ${\rm min}(p,k)$. Therefore the entanglement entropy for these partially entangled states labeled by the operator insertion $\ell$ and $\tau$ is given by
\bea
S&=&S_0 + 2 \pi ~{\rm min} (p,k), \nn
&=&S_0 + \min_{\small Y\in\hspace{0.7mm}{\rm Bulk}} \frac{\Phi(Y)}{4 G_N},
\ea
where the choice of $p$ or $k$ is equivalent to finding the minimum between the local minima $\Phi_{L}$ and $\Phi_R$. The dependence of the entanglement entropy on the PETS parameters $\ell$ and $\tau$ is given implicitly by the saddle point equation 
 \bea
 \label{speq1}
\frac{ k }{C}2\tau + 2 \arctan \frac{k+p}{\ell} +2 \arctan \frac{k-p}{\ell} &=&2\pi,\\
 \frac{p }{C}(\beta-2\tau) + 2 \arctan \frac{k+p}{\ell} - 2 \arctan \frac{k-p}{\ell}&=& 2\pi , \label{speq2}
\ea
taken from the previous section for $n=1$. This can be easily rewritten in terms of $\Phi_L$ and $\Phi_R$.
\begin{figure}[t!]
\begin{center}
\begin{tikzpicture}
\node[inner sep=0pt] (russell) at (0,0)
    {\includegraphics[scale=0.37]{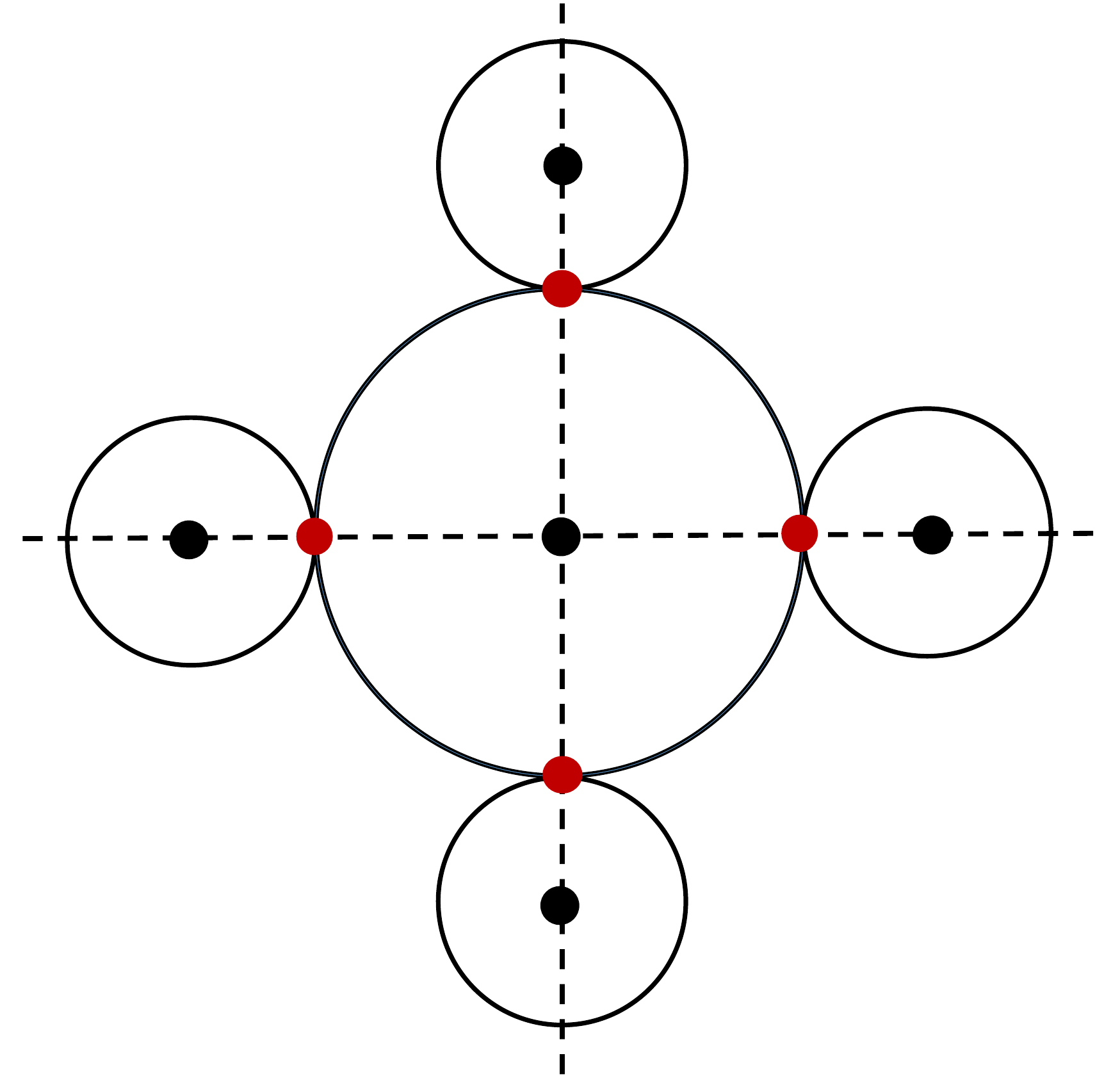}};
    \end{tikzpicture}
\end{center}
\vspace{-0.5cm}
\caption{\small Clover diagram for $\ell/C\gg1$, $\tau<\beta/4$ and $n=4$. Dashed lines separate replicas and the dots indicate the local horizons of each part of the geometry. }
\label{fig:cloverlargeL}
\end{figure}

In general this system of equations needs to be solved numerically. In practice we can derive analytic formulas in two cases. First when $\ell/C \ll1$. This was studied above as a warm-up using a different approach but the same answer can be derived from equations \eqref{speq1} and \eqref{speq2}. This gives (for $\tau<\beta/4$) the approximation
\beq
k(n)\approx  \frac{2 \pi C}{n\beta} + 2\ell\left(n+ \frac{2 \pi \tau}{\beta} \cot\frac{2 \pi \tau}{n\beta} \right) ,~~~p(n) \approx \frac{2 \pi C}{n\beta} + 2\ell\left(n- \frac{2\pi \tau}{\beta} \cot\frac{2\pi \tau}{n\beta} \right).
\eeq
This corresponds to a small perturbation to the TFD value of the dilaton due to the operator backreaction, since the correction is of order $\delta S \sim \ell \sim \mathcal{O}(1)$. We see that since $\tau<\beta/4$ the global dilaton minimum is indeed given by $p(n)$.

 On the other extreme we can take $\ell/C \gg 1$. Then the `clover' diagram describing the backreaction (see figure \ref{fig:cloverlargeL}) gives a graphical representation that gives the approximation (for the $\tau<\beta/4$ case, $\beta_\rrR < \beta_\llL$) 
\beq
k(n) \approx \frac{2 \pi C}{2\tau} = \frac{2 \pi C}{\beta_\rrR} ,~~~p(n) \approx \frac{2\pi C }{n(\beta-2\tau)} = \frac{2 \pi C}{n \beta_\llL}.
\eeq
When $\tau>\beta/4$ the roles of $p$ and $k$ are replaced. In this case the correction is of order $\delta S \sim \mathcal{O}(C)$ due to the semiclassical backreaction. We also see that indeed the global minimum of the dilaton corresponds to $p$ for $\tau<\beta/4$ (and $k$ in the other extreme).  

For general values of $\ell/C$ the parameter $p$ and $k$ interpolate between these limits. As a summary of this discussion the Renyi entropies are given in each limit by the approximations 
\hspace{0.5cm}
\beq
S_n = \begin{cases} 
S_0 + \frac{4 \pi^2 C}{n\beta} + 2\ell\left(n- \frac{\pi x}{\beta} \cot\frac{ \pi x}{n\beta} \right),~~~{\rm for}~~\ell/C \ll 1,\\ 
S_0 + \frac{4 \pi^2 C}{n~ (\beta-x)},~~~{\rm for}~~\ell/C \gg1. 
\end{cases}
\eeq
\hspace{0.5cm}
which is valid for any $\tau$ and again we used $x={\rm min}(2\tau,\beta-2\tau )$. A numerical solution of the saddle point equations shows that $p(n)$ (or $\Phi_L$) and $k(n)$ (or $\Phi_R$) interpolate smoothly and monotonically from the $\ell\to 0$ to the $\ell \to \infty$ limits that we derived above. 

One could ask in what sense are these states partially entangled since the entropy seems to increase with $\ell$. Consider the case such that $p<k$ and therefore $\Phi_L < \Phi_R$. The TFD corresponds to the state of maximal entropy with the constrain of a fixed average energy $E_R$. The energy of the right QM in the PETS can be easily computed to be exactly $E_R = E_0 + k^2/2C$. The state that maximizes the entropy is the thermal one with an inverse temperature chosen such that $\lb H_R\rb = E_0 + k^2/2C$. This can be purified as a TFD living in a tensor product of left- and right QM, with entropy $S_{\rm TFD}=S_0 + 2\pi k$. This should be compared with the actual entropy which is $S_{\rm PETS}=S_0 + 2 \pi p$. Since $p<k$ we see that $S_{\rm PETS} < S_{\rm TFD}=S_{\rm max}$. It is in this sense that our states are partially entangled. 

The entropy $2\pi k$ has another interpretation. In the case that the bulk brane falls behind the right horizon an observer in QM$_R$ cannot notice the state not being thermal, unless one measures complicated observables that are able to see behind the horizon (see figure \ref{PETSgeom22} below, we still take $\Phi_L<\Phi_R$). In the usual statistical mechanical sense, a coarse-grained observer will believe he or she is outside a TFD with temperature associated to the right horizon $\Phi_R/4G_N = 2 \pi k$. The coarse-grained energy will be correct, $E_R= E_0 + k^2/2C$ but the entropy $S_{\rm c.g.}=S_0 + 2\pi k$ will be off with respect to the microscopic one $S=S_0 + 2\pi p$. 

Finally, another measure one can take to characterize the loss of entanglement from a bulk perspective is the decay of left-right correlators at time $t=0$. In the TFD state they are given by 
\beq
\lb \mathcal{O}_L (0) \mathcal{O}_R (0) \rb_{\rm TFD} = (\pi /\beta)^{2\ell_P},
\eeq
where $\ell_P$ is the dimension of the probe operators, not related to the one of the operator insertion that created the PETS. For the PETS we are considering the left-right correlator is difficult to compute. Nevertheless for $\ell_P\gg1$ we can approximate it by a renormalized geodesic distance. This gives 
\beq
\lb \mathcal{O}_L \mathcal{O}_R \rb_{\rm PETS} = \left( \tan \frac{\theta_k}{4} \tan \frac{\theta_p}{4} \right)^{\ell_P} \lb \mathcal{O}_L \mathcal{O}_R \rb_{{\rm TFD},\beta}
\eeq
This prefactor goes from $1$ when $\ell$ is small (since for small backreaction $\theta \approx \pi$). For a large perturbation of the TFD $\ell\gg C$ and $\theta \approx 0$ giving $\lb \mathcal{O}_L \mathcal{O}_R \rb_{\rm PETS} \to 0$. When the left-right correlator becomes smaller than $e^{-S}$ where $S$ is the entanglement-entropy, one can say the two QM are not connected by a smooth semiclassical wormhole (firewall instead?). For $\ell \gg C$ the correlator behaves as $\left( \tan \frac{\theta_k}{4} \tan \frac{\theta_p}{4} \right)^{\ell_P} \approx \left( \frac{k p}{\ell^2}\right)^{\ell_P}$ where $p k\approx \frac{4 \pi^2 C^2}{\beta_\rrR \beta_\llL}$ and therefore the ratio between the PETS correlator and the TFD decays as $\exp(-2 \ell_P \log \frac{\ell}{C})$, controlled by the distance between horizons.

\medskip

\section{Bulk Reconstruction}\label{sec:bulkrec}
\vspace{-1mm}
In this section we will comment about bulk reconstruction of PETS. We will focus on the case in which the operator ${\cal O}_\ell$ has dimension $\ell \sim C \sim N/\beta J$, so that the bulk geometry is the one shown in figure \ref{PETSgeom22}. We will argue in this section that the regions that can be reconstructed from either side are as shown in the right panel of figure \ref{PETSgeom22}. We will give two separate arguments in support of this proposal, one using entanglement wedge reconstruction \cite{Czech:2012bh} and one using a tensor network representation of the bulk state \footnote{In \cite{Mertens:2017mtv} the correlators of the Schwarzian theory were related to local insertions in 2D Liouville between ZZ-branes. It would be interesting to study bulk reconstruction using the insertion of cross-caps as in \cite{Verlinde:2015qfa}. We leave this for future work.}.

\subsection{Entanglement wedge reconstruction}
\vspace{-2mm}
To put this discussion in context, we first take a step back to a pure higher dimensional AdS bulk space-time.
Pick a region $\mathcal{A}$ included in the boundary of AdS. A natural question to ask is to what extent can we reconstruct bulk operators using CFT operators living on this region $\mathcal{A}$. Semiclassically, one can apply the BDHM/HKLL prescription to reconstruct local operators included in the bulk causal wedge $\mathcal{C}_{\mathcal{A}}$ of $\mathcal{A}$ \cite{Banks:1998dd, Hamilton:2006az, Kabat:2011rz}. In contrast, it is believed that from operators in region $\mathcal{A}$ one should be able to reconstruct operators inside the entanglement wedge $\mathcal{E}_{\mathcal{A}}$ of $\mathcal{A}$, defined as the bulk domain of dependence of the region between $\mathcal{A}$ and its extremal RT surface \cite{Czech:2012bh, Headrick:2014cta, Pastawski:2015qua, Almheiri:2014lwa, Heemskerk:2012mn}.
 Since the entanglement wedge can in general be bigger than the causal wedge, $\mathcal{C}_{\mathcal{A}}\subset \mathcal{E}_{\mathcal{A}}$, it is an interesting problem to find natural ways to represent local operators $\phi$ such that $\phi$ lives in the algebra of operators associated to $\mathcal{E}_{\mathcal{A}}$ but not on $\mathcal{C}$.

In \cite{Faulkner:2017vdd} the authors propose a concrete way to construct operators in the entanglement wedge (see also \cite{Jafferis:2015del} and \cite{Almheiri:2017fbd}). Their construction involves defining the zero-mode of a CFT operator $\mathcal{O}$ under modular flow associated to the modular Hamiltonian $K_{\mathcal{A}}$, defined via $\rho_{\mathcal{A}}=e^{-K_{\mathcal{A}}}$ with $\rho_{\mathcal A}$ the density matrix associated to region $\mathcal{A}$. The modular zero-mode is given by (a properly regulated version of)
\beq
\label{modzm}
\mathcal{O}_0 = \int ds ~e^{ iK_{\mathcal{A}} s} \hspace{0.1cm}\mathcal{O} \hspace{0.1cm} e^{- iK_{\mathcal{A}} s}.
\eeq
It is argued in \cite{Faulkner:2017vdd} that this highly non-local CFT operator, that lives in the algebra of operators of inside the region $\mathcal{A}$, is dual to an operator that lives on the RT surface 
\beq
\label{modlocal}
\mathcal{O}_0 = \int_{\rm RT} d\mu(Y_{\rm RT}) ~\phi(Y_{\rm RT}).
\eeq
Here $Y$ labels a point in AdS restricted to the RT surface. In this expression the integral is over the RT surface associated to $\mathcal{A}$, namely the boundary (in the bulk) of the entanglement wedge, and $\phi(Y_{\rm RT})$ is a local bulk operator. The measure $d\mu(Y)$ of the integral over $Y$ is given by a bulk boundary propagator. We will not need its explicit form here. The proposal gives a concrete construction of (non-local) bulk operators outside of the causal wedge, at the edge of the entanglement wedge.
\begin{figure}[t!]
\begin{center}
\begin{tikzpicture}[scale=0.79]
\path[fill=red!10]  (-2.75,0) -- (-5,2.2) -- (2,2.2) -- (0,0) -- (2,-2.2) -- (-5,-2.2) -- cycle ;
\path[fill=green!10]  (0,0) -- (2,2.2) -- (2,-2.2)-- cycle;
\path[fill=blue!10]  (-2.75,0) -- (-5,2.2) -- (-5,-2.2) -- cycle;
\draw[dashed] (-5,-2.2) -- (-.5,2.2);
\draw[dashed] (-5,2.2) -- (-.5,-2.2);
\draw[thick] (-5,-2.2) -- (-5,2.2);
\draw[thick] (2,-2.2) -- (2,2.2);
\draw[thick,color=red!40] (-1.0,-2.2) -- (-1.0,2.2);
\draw[dashed] (-2,2.2) -- (2,-2.2);
\draw[dashed] (2,2.2) -- (-2,-2.2);
\draw[dashed] (-5,0) -- (2,0);
\draw[fill=black] (-2.75,0) circle (0.09);
\draw[thin] (0,0) circle (0.05);
\draw[thick,decoration = {zigzag,segment length = 2mm, amplitude = 0.75mm},decorate] (-5,2.2)--(2,2.2);
\draw[thick,decoration = {zigzag,segment length = 2mm, amplitude = 0.75mm},decorate] (-5,-2.2)--(2,-2.2);
\draw (-2.6,-0.7) node[color=black] {H$_{\rm L}$};
\draw (0.15,-0.7) node[color=black] {H$_{\rm R}$};
\draw (-1.5,-3.5) node[color=black] {(a) Naive: CW};
\end{tikzpicture}
\hspace{2cm}
\begin{tikzpicture}[scale=0.79]
\path[fill=red!10]  (-2.75,0) -- (-5,2.2) -- (-.5,2.2) --cycle ;
\path[fill=red!10]  (-2.75,0) -- (-5,-2.2) -- (-.5,-2.2) --cycle;
\path[fill=green!10]  (-2.75,0) -- (-.5,2.2) -- (2,2.2) -- (2,-2.2) -- (-0.5,-2.2) -- cycle;
\path[fill=blue!10]  (-2.75,0) -- (-5,2.2) -- (-5,-2.2) -- cycle;
\draw[dashed] (-5,-2.2) -- (-.5,2.2);
\draw[dashed] (-5,2.2) -- (-.5,-2.2);
\draw[thick] (-5,-2.2) -- (-5,2.2);
\draw[thick] (2,-2.2) -- (2,2.2);
\draw[thick,color=red!40] (-1.0,-2.2) -- (-1.0,2.2);
\draw[dashed] (-2,2.2) -- (2,-2.2);
\draw[dashed] (2,2.2) -- (-2,-2.2);
\draw[dashed] (-5,0) -- (2,0);
\draw[fill=black] (-2.75,0) circle (0.085);
\draw[thin] (0,0) circle (0.05);
\draw[thick,decoration = {zigzag,segment length = 2mm, amplitude = 0.75mm},decorate] (-5,2.2)--(2,2.2);
\draw[thick,decoration = {zigzag,segment length = 2mm, amplitude = 0.75mm},decorate] (-5,-2.2)--(2,-2.2);
\draw (-2.6,-0.7) node[color=black] {H$_{\rm L}$};
\draw (0.15,-0.7) node[color=black] {H$_{\rm R}$};
\draw (1.1,.26) node [color=blue] {\footnotesize $b$};
\draw (-.6,.22) node [color=blue]{\footnotesize $a$};
\draw (-1.8,.22) node [color=blue]{\footnotesize $c$};
\draw (-4.1,.26) node [color=blue] {\footnotesize $d$};
\draw (-1.5,-3.5) node[color=black] {(b) Correct: EW};
\end{tikzpicture}
\end{center}
\vspace{-0.6cm}
\caption{\small Dual geometry to a PETS created by acting with a heavy operator during euclidean evolution. We indicate the region reconstructed by the left QM by blue and by the right QM by green. The reconstruction of the red region requires both sides. Naively, each side can only see its causal wedge (CW). Instead, we argue below that each side can in principle reconstruct the full entanglement wedge (EW). The right EW includes regions $a, b$ and $c$. }
\label{PETSgeom22}
\end{figure}
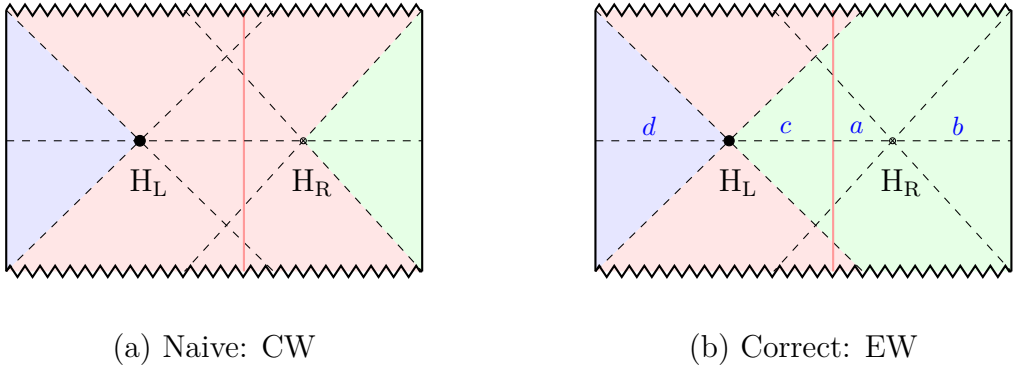

Let us apply this construction to the geometry of figure \ref{PETSgeom22}. We pick parameters such that the left horizon H$_L$ is the horizon with minimal dilaton and 
therefore the extremal surface. According to the result of our previous section, its area fixes the entanglement between left- and the right QM systems. In the panel (a) of figure \ref{PETSgeom22} we show the causal wedges of QM$_L$ (region in red) and QM$_R$ (region in blue). A naive intuition would be that observables in QM$_R$ can only reconstruct operators in the green region and operators in QM$_L$ can only reconstruct operators in the blue region. The red region is outside the causal wedges and naively would require a two-sided reconstruction in terms of operators that act in both QM$_L$ and QM$_R$. 

The construction of \cite{Faulkner:2017vdd} provides an example showing that the naive expectation is not correct. Instead one should consider the full entanglement wedges. The entanglement wedge associated to the left QM coincides with its causal wedge and its shown in blue in figure \ref{PETSgeom22}. In this low dimensional setting, the RT surface becomes a point namely 
\beq
\text{ RT surface} = \text{left horizon H}_{\rm L}.
\eeq
The entanglement wedge of the right QM becomes therefore the green region in panel (b) of figure \ref{PETSgeom22} which includes the interior spatial regions $a$ and $c$. Therefore we propose that this picture is the correct one describing the bulk reconstruction in this PETS. We will motivate this proposal in two ways.

As a first motivation, we can use the construction in \cite{Faulkner:2017vdd} as explained above. Our setup has two advantages. First by construction the Hilbert space factorizes $\mathcal{H} = \mathcal{H}_L \otimes \mathcal{H}_R$. Secondly, the RT surface is a point. Therefore the modular flow zero-mode defined in \eqref{modzm} becomes a local insertion \eqref{modlocal} located at the left horizon. We can formally define the modular Hamiltonian associated to the density matrix of the right QM written down in equation \eqref{rhoom} defined as $\rho_\rrR= e^{-K_\rrR}$ with $\rho_\rrR$ the density matrix of the PETS given in eqn \eqref{rhoom}. Then the suggestion of \cite{Faulkner:2017vdd} implies that, up to normalizations,
\beq
\int ds ~e^{ is\spc K_\rrR} \hspace{0.1cm}\mathcal{O} \hspace{0.1cm} e^{- is \spc K_\rrR } = \phi_{\mathcal{O}}(Y_{{\rm H}_{\rm L}})
\eeq
where $\mathcal{O}$ is an operator living in the right QM.

This is an interesting result for the following reason. A naive observer in the right QM would be led to believe (by doing generic measurements) that she lives in a thermal state and therefore $K_\rrR^{\rm \! \! naive} =\beta_\rrR H_\rrR$. We call this the coarse-grained modular hamiltonian. Then by applying the prescription of \cite{Faulkner:2017vdd} she would end up reconstructing an operator in the boundary of the causal wedge, the right horizon
\beq
\mathcal{O}_0^{\rm naive}=\int ds ~e^{ i s H_\rrR } \hspace{0.1cm}\mathcal{O} \hspace{0.1cm} e^{- i s H_\rrR } = \phi_{\mathcal{O}}(Y_{{\rm H}_{\rm R}}).
\eeq
Of course, upon closer inspection, if the observer is able to do fine-grained measurements and discover that her density matrix is not thermal, correcting for the modular flow will allow her to reconstruct up to the left horizon. Even though it is believed that the entanglement wedge reconstruction gives the right microscopic answer, in practice it may still be extremely hard to reconstruct operators between the two horizons H$_{\rm L}$ and H$_{\rm R}$ using the right QM alone. 

\subsection{Tensor network representation}
\vspace{-2mm}
The above conclusion is supported by the following tensor network argument, first presented in a talk at IAS by Almheiri in \cite{AlmheiriTalk}. In \cite{AlmheiriTalk} it was shown how the QEC property of AdS/CFT \cite{Almheiri:2014lwa}  can be applied to reconstruct the interior of pure SYK black holes. This section highlights and generalizes his approach and points out its equivalence with the QEC procedure for constructing the black hole interior developed in the earlier work \cite{Verlinde:2012cy}.

\begin{figure}
\begin{center}
\begin{tikzpicture}[scale=0.85]
\path[fill=red!10] (0,0) -- (0,0) -- (-2,2) -- (2,2) --cycle ;
\path[fill=red!10] (0,0) -- (0,0) -- (-2,-2) -- (2,-2) --cycle ;
\path[fill=green!10]  (0,0) -- (2,2) -- (2,-2)-- cycle;
\path[fill=blue!10]  (0,0) -- (-2,2) -- (-2,-2) -- cycle;
\draw[dashed] (-2,0) -- (2,0);
\draw[thick] (-2,-2) -- (-2,2);
\draw[thick] (2,-2) -- (2,2); 
\draw[dashed] (-2,2) -- (2,-2);
\draw[dashed] (-2,-2) -- (2,2);
\draw[fill=black] (0,0) circle (0.08);
\draw (1.2,.25) node[color=blue] {\footnotesize $b$};
\draw (-1,.25) node[color=blue] {\footnotesize $d$};
\draw[thick,decoration = {zigzag,segment length = 2mm, amplitude = 0.75mm},decorate] (-2,2)--(2,2);
\draw[thick,decoration = {zigzag,segment length = 2mm, amplitude = 0.75mm},decorate] (-2,-2)--(2,-2);
\end{tikzpicture}
\hspace{3.2cm}
\begin{tikzpicture}[scale=0.85]
\path[fill=green!10] (-4,0) -- (-2,2) -- (0,2) -- (0,-2) -- (-2,-2) --cycle ;
\draw[dashed] (-2,0) -- (0,2);
\draw[dashed] (0,-2) -- (-3,1);
\draw[dashed] (-2,0) -- (-3,-1);
\draw[dashed] (-4,0) -- (0,0);
\draw (-.8,.25) node[color=blue] {\footnotesize $b$};
\draw (-3,.25) node[color=blue] {\footnotesize $a$};
\draw[thick] (0,-2) -- (0,2);
\draw[thick,color={rgb:red,10; black,3}] (-2,-2) -- (-4,0);
\draw[thick,color={rgb:red,10; black,3}] (-4,0) -- (-2,2);
\draw[thin] (-2,0) circle (0.05);
\draw[thick,decoration = {zigzag,segment length = 2mm, amplitude = 0.75mm},decorate] (-2,2)--(0,2);
\draw[thick,decoration = {zigzag,segment length = 2mm, amplitude = 0.75mm},decorate] (-2,-2)--(0,-2);
\draw[color={rgb:red,10; black,3}, fill={rgb:red,10; black,3}] (-4,0) circle (0.12);
\end{tikzpicture}\ \ \ \ \ \  \ \\[9mm]
\begin{tikzpicture}[scale=.44]
\path[fill=blue!10] (0,0) -- (3,0) -- (3,3) -- (0,3) --cycle ;
\path[fill=green!10] (5,0) -- (8,0) -- (8,3) -- (5,3) --cycle ;
\draw[thin] (5,0) -- (8,0) -- (8,3) -- (5,3) --cycle ;
\draw[thin] (0,0) -- (3,0) -- (3,3) -- (0,3) --cycle ;
\draw (6.5,-1) node [color=blue] {\footnotesize $b$};
\draw (1.5,-.95) node [color=blue] {\footnotesize $d$};
\draw (1.5,1.7) node {\large T$^\dag$};
\draw (6.5,1.7) node {\large T};
\draw[thin,blue] (8,.5) -- (9,.5) ;
\draw[thin,blue] (8,1) -- (9,1) ;
\draw[thin,blue] (8,1.5) -- (9,1.5) ;
\draw[thin,blue] (8,2) -- (9,2) ;
\draw[thin,blue] (8,2.5) -- (9,2.5) ;
\draw[thin,blue] (-1,.5) -- (0,.5) ;
\draw[thin,blue] (-1,1) -- (0,1) ;
\draw[thin,blue] (-1,1.5) -- (0,1.5) ;
\draw[thin,blue] (-1,2) -- (0,2) ;
\draw[thin,blue] (-1,2.5) -- (0,2.5) ;
\draw[thin,blue] (3,.6) -- (5,.6) ;
\draw[thin,blue] (3,1.2) -- (5,1.2) ;
\draw[thin,blue] (3,1.8) -- (5,1.8) ;
\draw[thin,blue] (3,2.4) -- (5,2.4) ;
\draw[thick,red] (1.5,-.5) -- (1.5,1) ;
\draw[thick,red] (6.5,-.5) -- (6.5,1) ;
\draw[->,thick,red] (6.5,1) -- (8,1) ;
\draw[->,thick,red] (1.5,1) -- (0,1) ;
\end{tikzpicture}
\hspace{2.2cm}
\begin{tikzpicture}[scale=.44]
\path[fill=red!10] (-1,0) -- (-1,3) -- (-2.3,1.5) --cycle ;
\path[fill=green!10] (0,0) -- (3,0) -- (3,3) -- (0,3) --cycle ;
\path[fill=green!10] (5,0) -- (8,0) -- (8,3) -- (5,3) --cycle ;
\draw[thin] (-1,0) -- (-1,3) -- (-2.3,1.5) --cycle ;
\draw[thin] (0,0) -- (3,0) -- (3,3) -- (0,3) --cycle ;
\draw[thin] (5,0) -- (8,0) -- (8,3) -- (5,3) --cycle ;
\draw (6.5,-.98) node [color=blue] {\footnotesize $b$};
\draw (1.5,-.95) node [color=blue] {\footnotesize $a$};
\draw (1.5,1.7) node {\large T$^\dag$};
\draw (6.5,1.7) node {\large T};
\draw (-1.5,1.65) node {\large P};
\draw[thin,blue] (8,.5) -- (9,.5) ;
\draw[thin,blue] (8,1) -- (9,1) ;
\draw[thin,blue] (8,1.5) -- (9,1.5) ;
\draw[thin,blue] (8,2) -- (9,2) ;
\draw[thin,blue] (8,2.5) -- (9,2.5) ;
\draw[thin,blue] (-1,.5) -- (0,.5) ;
\draw[thin,blue] (-1,1) -- (0,1) ;
\draw[thin,blue] (-1,1.5) -- (0,1.5) ;
\draw[thin,blue] (-1,2) -- (0,2) ;
\draw[thin,blue] (-1,2.5) -- (0,2.5) ;
\draw[thin,blue] (3,.6) -- (5,.6) ;
\draw[thin,blue] (3,1.2) -- (5,1.2) ;
\draw[thin,blue] (3,1.8) -- (5,1.8) ;
\draw[thin,blue] (3,2.4) -- (5,2.4) ;
\draw[thick,red] (1.5,-.5) -- (1.5,1) ;
\draw[thick,red] (6.5,-.5) -- (6.5,.5) ;
\draw[->,thick,red] (6.5,.5) -- (8,.5) ;
\draw[->,thick,red] (1.5,1) -- (8,1) ;
\end{tikzpicture}
\vspace{-3mm}
\end{center}
\caption{\small Tensor network representation of the one-sided reconstruction of interior and exterior operators for the thermo-field double state (left) and the thermal pure state (right). The reconstruction map of the bulk operators is indicated by the corresponding red arrows. Figure taken from \cite{AlmheiriTalk}.}
\label{doublet}
\end{figure}
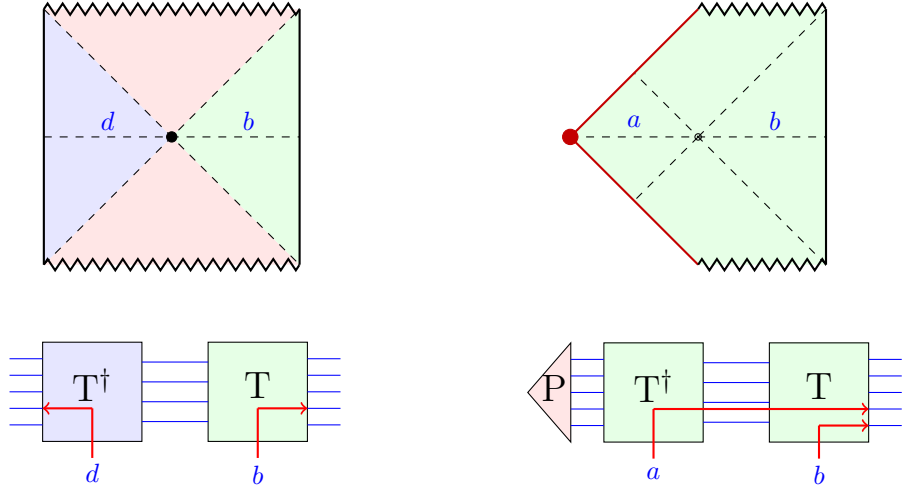

Figure \ref{doublet} shows a tensor network representation for the bulk reconstruction map for the thermo-field double state (left) and the thermal pure state (right) \cite{AlmheiriTalk}. Let us first explain the former. We assume that the left and right CFT Hilbert space can be factorized into the tensor product of a (visible) bulk QFT Hilbert space ${\cal H}^{\rm qft}_{\llL,\rrR}$ and (hidden) horizon Hilbert space ${\cal H}^{\rm hor}_{\llL,\rrR}$
\bea
{\cal H}_{\llL}^{\rm cft} = {\cal H}^{\rm qft}_{\llL} \otimes {\cal H}^{\rm hor}_{\llL}, \qquad \qquad 
{\cal H}_{\rrR}^{\rm cft} = {\cal H}^{\rm qft}_{\rrR} \otimes {\cal H}^{\rm hor}_{\rrR}
\eea
Each tensor $T$ denotes the embedding of the tensor product into the respective CFT Hilbert space. The left- and right horizon Hilbert space is assumed to be in unique maximally entangled state between the two sides, and is therefore represented by the lines connecting the two tensors \cite{AlmheiriTalk}. For the TFD state, each bulk QFT Hilbert space is reconstructed in terms of the corresponding CFT. This reconstruction map is indicated by the red arrows. Hence each side can only reconstruct its causal wedge. In this sense, the thermo-field double state has a firewall: a one-sided infalling observer (that can only use one-sided observables) cannot pass the horizon unscathed. This conclusion follows from the AMPS argument: the one-sided states are thermal mixed states, and do not encode the local entanglement that is required to ensure smoothness of the horizon.

The situation is different for the thermal pure state. Let us write the thermal pure state as 
\bea
|\Psi \rb_\rrR \is {}_{\llL}\!\smpc \lb\spc \sss\spc | \mbox{\sc TFD}\rb_\rrR
\eea
The tensor network for the TFD state is the same as before, but it is now capped off on the left with a projection onto the left state $\lb \spc \sss\spc |$, indicated by the triangle \cite{AlmheiriTalk}. In the space-time diagram the projection is indicated by the `end-of-the-world particle, that cuts off the left asymptotic region \cite{Kourkoulou:2017zaj}. Since there is now only one CFT, the bulk reconstruction has to proceed towards the right. Concretely, the above figure indicates that a bulk QFT operator $A$ inside the black hole region $a$ with matrix elements 
with $A_{nm} =  \lb n|A|m\rb $ acts on the CFT Hilbert space as (c.f. \cite{Verlinde:2012cy} and Appendix \ref{appQECBHI})
\bea
\label{intone}
{\bf A} \is    \sum_{m,n} \spc A_{nm} \,  {\bf P}\spc {\bf T}^\dag\spc |n  \rb \lb m| \, {\bf T}\spc {\bf P}
\eea
where ${\bf P}$ denotes the projection onto the state $|\spc \sss \spc \rb$. This tensor network is a schematic representation of the state-dependent reconstruction map of \cite{Papadodimas:2013wnh}, or equivalently, of the general construction of the interior operators of \cite{Verlinde:2012cy} based on the application of quantum error correction technology. The latter construction is summarized in Appendix \ref{appQECBHI}, and also works for partially mixed states. For the thermal pure state, there is no quantum information theoretic obstruction to reconstruct the black hole interior.

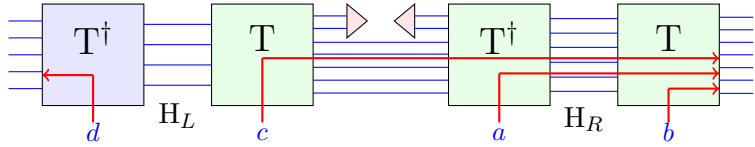
\begin{figure}
\begin{center}\medskip
\begin{tikzpicture}[scale=.45]
\path[fill=red!10] (-1,2) -- (-1,3) -- (-1.6,2.5) --cycle ;
\path[fill=green!10] (0,0) -- (3,0) -- (3,3) -- (0,3) --cycle ;
\path[fill=green!10] (5,0) -- (8,0) -- (8,3) -- (5,3) --cycle ;
\draw[thin] (-1,2) -- (-1,3) -- (-1.6,2.5) --cycle ;
\draw[thin] (0,0) -- (3,0) -- (3,3) -- (0,3) --cycle ;
\draw[thin] (5,0) -- (8,0) -- (8,3) -- (5,3) --cycle ;
\draw (-8,-.3) node {\small H$_L$};
\draw (4,-.4) node {\small H$_R$};
\draw (6.5,-.79) node[color=blue] {\footnotesize $b$};
\draw (1.5,-.84) node[color=blue] {\footnotesize $a$};
\draw (-10.5,-.79) node[color=blue] {\footnotesize $d$};
\draw (-5.5,-.84) node[color=blue] {\footnotesize $c$};
\draw (1.5,2) node {\large T$^\dag$};
\draw (6.5,2) node {\large T};
\draw[thin,blue] (8,.36) -- (9,.36) ;
\draw[thin,blue] (8,.75) -- (9,.75) ;
\draw[thin,blue] (8,1.12) -- (9,1.12) ;
\draw[thin,blue] (8,1.5) -- (9,1.5) ;
\draw[thin,blue] (8,1.87) -- (9,1.87) ;
\draw[thin,blue] (8,2.24) -- (9,2.24) ;
\draw[thin,blue] (8,2.61) -- (9,2.61) ;
\draw[thin,blue] (0,.37) -- (-4,.37) ;
\draw[thin,blue] (0,.75) -- (-4,.75) ;
\draw[thin,blue] (0,1.13) -- (-4,1.13) ;
\draw[thin,blue] (0,1.52) -- (-4,1.52) ;
\draw[thin,blue] (0,1.87) -- (-4,1.87) ;
\draw[thin,blue] (0,2.28) -- (-1,2.28) ;
\draw[thin,blue] (0,2.65) -- (-1,2.65) ;
\draw[thin,blue] (-11,.5) -- (-13,.5) ;
\draw[thin,blue] (-11,1) -- (-13,1) ;
\draw[thin,blue] (-11,1.5) -- (-13,1.5) ;
\draw[thin,blue] (-11,2) -- (-13,2) ;
\draw[thin,blue] (-11,2.5) -- (-13,2.5) ;
\draw[thin,blue] (3,.42) -- (5,.42) ;
\draw[thin,blue] (3,.84) -- (5,.84) ;
\draw[thin,blue] (3,1.28) -- (5,1.28) ;
\draw[thin,blue] (3,1.72) -- (5,1.72) ;
\draw[thin,blue] (3,2.14) -- (5,2.14) ;
\draw[thin,blue] (3,2.57) -- (5,2.57) ;
\draw[thick,red] (1.5,-.5) -- (1.5,.95) ;
\draw[thick,red] (6.5,-.5) -- (6.5,.5) ;
\draw[->,thick,red] (6.5,.5) -- (8,.5) ;
\draw[->,thick,red] (1.5,.95) -- (8,.95) ;
\path[fill=red!10] (-3,2) -- (-3,3) -- (-2.4,2.5) --cycle ;
\path[fill=green!10] (-4,0) -- (-7,0) -- (-7,3) -- (-4,3) --cycle ;
\path[fill=blue!10] (-9,0) -- (-12,0) -- (-12,3) -- (-9,3) --cycle ;
\draw[thin] (-3,2) -- (-3,3) -- (-2.4,2.5) --cycle ;
\draw[thin] (-4,0) -- (-7,0) -- (-7,3) -- (-4,3) --cycle ;
\draw[thin] (-9,0) -- (-12,0) -- (-12,3) -- (-9,3) --cycle ;
\draw[thin,blue] (-7,.6) -- (-9,.6) ;
\draw[thin,blue] (-7,1.2) -- (-9,1.2) ;
\draw[thin,blue] (-7,1.8) -- (-9,1.8) ;
\draw[thin,blue] (-7,2.4) -- (-9,2.4) ;
\draw[thin,blue] (-3,2.28) -- (-4,2.28) ;
\draw[thin,blue] (-3,2.65) -- (-4,2.65) ;
\draw[thick,red] (-5.5,-.5) -- (-5.5,1.4) ;
\draw[thick,red] (-10.5,-.5) -- (-10.5,.9) ;
\draw[->,thick,red] (-10.5,.9) -- (-12,.9) ;
\draw[->,thick,red] (-5.5,1.4) -- (8,1.4) ;
\draw (-10.5,2) node {\large T$^\dag$};
\draw (-5.5,2) node {\large T};
\end{tikzpicture}
\end{center}
\caption{\small Tensor network representation of the partially entangled thermal state.}
\label{petst}
\end{figure}

Finally, we turn to the tensor network representation of the partially entangled thermal states shown in figure \ref{petst}. It is useful to think about PETS as a local operator ${\cal O}_\ell$ sandwiched between two thermal field double states with temperature $\beta_\llL$ and $\beta_\rrR$. Since each TFD state is a tensor product state, this leaves a (partially) entangled state. Each TFD state is represented by a pair of tensors $T$ and $T^\dag$. The operator ${\cal O}$, viewed as an element of the tensor product of two QM Hilbert spaces, is a partially entangled state -- it is partially transmitting (entangled) and partially reflecting (product of pure). In the above tensor network, this is indicated by the partial projections, depicted by the red triangles.

The number of lines between the successive tensors in figure \ref{petst} indicates the amount of entanglement across the corresponding interface. As indicated, the left horizon H$_L$ supports the minimal amount of entanglement, and thus forms the information bottleneck between the left- and right CFT. Hence the left horizon is the bifurcation between the left- and right entanglement wedge. The reconstruction of the bulk QFT modes in each region proceeds as indicated by the right arrows. The rule is that the arrow points in the direction of the nearest interface with the largest number of lines, since this is the direction that dominates entropically: the bulk modes are entangled with the largest nearby Hilbert space. This entropic argument underscores the entanglement wedge reconstruction proposal. 

The QEC reconstruction procedure of \cite{Verlinde:2012cy} directly applies to region $a$, and with minor modification, to region $c$. The density matrix of the right system
\bea
\rho_\rrR \is e^{-\frac 1 2 \beta_\rrR H} \spc \mathcal{O}_\ell \spc e^{- \beta_\llL H} \spc \mathcal{O}_\ell \spc e^{-\frac 1 2 \beta_\rrR H}
\eea
is only partially mixed: its von Neumann entropy is strictly smaller than the thermal entropy. The operator insertions in effect restrict $\rho$ to lie within a certain code subspace of the total Hilbert space. This enables the QEC reconstruction of the interior operators. The density matrix of the left system is maximally mixed, and the QEC procedure does not work in this case. The left entanglement wedge is equal to the left causal wedge, the outside region to the left of the horizon.

\bigskip

\section{Generalizations}\label{sec:genmultins}
\vspace{-2mm}   

In this section we discuss two generalizations of partially entangled states. In the first subsection, we introduce a coarse-graining by including an incoherent sum over different operators of the same scale dimension, all inserted at the same euclidean time instant. Then we briefly discuss the case of two different operators insertions at different euclidean times.

\subsection{Coarse graining and tripartite entanglement}\label{sec:cg}
\vspace{-1mm}

Looking at figure \ref{fig:dilatonprof} we have learned that the entanglement entropy is fixed by (the extremal entropy $S_0$ plus) the one associated to the smaller horizon ($\Phi_\llL$ in the figure). But we could ask the following question. Which physical quantity is associated to the other horizon where the dilaton attains a local minimum $\Phi_\rrR > \Phi_\llL$ in the cases for which both horizons are part of the geometry? If both describe the microscopic von Neumann entropy of the corresponding QM system, the total combined state can no longer be in a pure state. This observation makes it natural to wonder if one should also associate an entropy with the operator insertion itself, and consider the PETS as a tripartite state.

Specifically, instead of the density matrix $\rho =  e^{-\frac 1 2 \beta_\rrR H} \spc \mathcal{O}\spc e^{- \beta_\llL H} \spc \mathcal{O} \spc e^{-\frac 1 2 \beta_\rrR H}$ of the right QM system, we study instead consider the following class of mixed states
\bea
\label{rhoom} 
\rho \is \sum_{i=1}^{K} e^{-\frac 1 2 \beta_\rrR H} \spc \mathcal{O}_i\spc e^{- \beta_\llL H} \spc \mathcal{O}_i \spc e^{-\frac 1 2 \beta_\rrR H}
\;\; = \; \;  \sum_{i=1}^{K}
\raisebox{4pt}{\begin{tikzpicture}[scale=.48, baseline={([yshift=0cm]current bounding box.center)}]
\draw[thick] (-0.65,-1.05) arc (245:350:1.6);
\draw[thick] (-0.65,1.85) arc (115:10:1.6);
\draw[thick] (-.95,1.85) arc (50:310:1.9);
\draw[fill,black] (1.6,0.1) circle (0.05);
\draw[fill,black] (1.6,0.7) circle (0.05);
\draw[dotted,thick,red] (-0.8,-0.7) -- (-.8,1.6);
\draw[color={rgb:red,10; black,3}, fill={rgb:red,10; black,3}]   (-.8,1.75) circle (0.14); 
\draw[color={rgb:red,10; black,3}, fill={rgb:red,10; black,3}]   (-.8,-.95) circle (0.14); 
\draw (-.8,-1.7) node {\footnotesize\bf $i$};
\draw (-.8,2.5) node {\footnotesize\bf $i$};
\draw (-3.5,.45) node {\footnotesize\bf $\beta_\llL$};
\draw (.3,1.4) node {\scriptsize $\frac 1 2 \beta_\rrR$};
\draw (.3,-.65) node {\scriptsize $\frac 1 2 \beta_\rrR$};
\end{tikzpicture}}
\eea
 where we sum over $K\gg1$ operators with dimensions $\ell_i \approx \ell$ constant. Since we assume that $\ell \sim N/\beta J \gg 1$, the scaling dimension is large and it is natural to expect a correspondingly large degeneracy $K$ of operators with dimension close to $\ell$.
The density matrix \eqref{rhoom} does not correspond to tracing out the left QM in a pure state in $\mathcal{H}_R \otimes \mathcal{H}_L$. In particular, the entropy is not the same for the right-QM or left-QM, $S_L \neq S_R$. One can add a Hilbert space associated to the operator $\mathcal{H}_{\rm op}$, with one basis element for each value of the index $i=1,\ldots, K$. Then the state can be purified in the tensor product $\mathcal{H}_R \otimes \mathcal{H}_L\otimes \mathcal{H}_{\rm op}$. We will comment below on this tripartite structure of entanglement.

This generalization has a few motivations. First, taking $K$ large makes it more straightforward to decide which contraction channel dominates in the correlator involved in the computation of the $n$th Renyi entropy. Take the entropy for the right CFT. In the large $K$ limit one can see that the channel in the left panel of figure \ref{fig:cloverdiagram} dominates. This is true independently of $\beta_\llL$ and $\beta_\rrR$ as long as $K$ is taken to be sufficiently large (but not larger than $N$). In this way we obtain 
\bea
S_R \is S_0 + 2 \pi k= S_0 + \frac{\Phi_R}{4 G_N}.
\eea
Here $\Phi_\rrR$ is a local minimum of the dilaton, but not necessarily the global minimum. On the other hand, repeating this analysis for the left QM gives
\bea
S_L \is S_0 + 2 \pi p = S_0 +\frac{\Phi_L}{4 G_N}
\eea
 since now the opposite channel dominates. This should be contrasted with the entropy of pure PETS in which case the entanglement entropy in both cases is equal to the minimum between $S_L$ and $S_R$. 
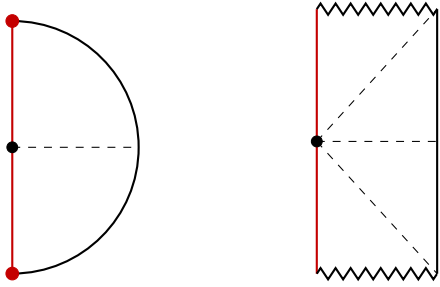
\begin{figure}[t!]
\begin{center}
\begin{tikzpicture}[scale=0.84]
\draw[dashed] (0,-2) -- (2,-2);
\draw[thick,color={rgb:red,10; black,3}] (0,-4) -- (0,0);
\draw[thick] (0,0) arc (90:-90:2);
\draw[color={rgb:red,10; black,3}, fill={rgb:red,10; black,3}] (0,-4) circle (0.1);
\draw[color={rgb:red,10; black,3}, fill={rgb:red,10; black,3}] (0,0) circle (0.1);
\draw[fill=black] (0,-2) circle (0.085);
\end{tikzpicture}
\hspace{2cm}
\begin{tikzpicture}[scale=0.8]
\draw[thick] (2,-2.2) -- (2,2.2);
\draw[thick,color={rgb:red,10; black,3}] (0,-2.2) -- (0,2.2);
\draw[dashed] (0,0) -- (2,2.2);
\draw[dashed] (0,0) -- (2,-2.2);
\draw[dashed] (0,0) -- (2,0);
\draw[fill=black] (0,0) circle (0.09);
\draw[thin] (0,0) circle (0.05);
\draw[thick,decoration = {zigzag,segment length = 2mm, amplitude = 0.75mm},decorate] (0,2.2)--(2,2.2);
\draw[thick,decoration = {zigzag,segment length = 2mm, amplitude = 0.75mm},decorate] (0,-2.2)--(2,-2.2);
\end{tikzpicture}
\end{center}
\caption{\small The euclidean and lorentzian space-time dual to the extremal partially entangled state \eqref{ostate} with $\beta_\llL \to \infty$ and $\ell = k_\rrR$. }
\label{PETSgeom3}
\end{figure}

In the type of states discussed in this section we have computed $S_L$ and $S_R$. What is the maximal value of $S_{LR}$, the entropy of the left- and right system combined? Since the tripartite state living in the enlarged Hilbert space $\mathcal{H}_R \otimes \mathcal{H}_L\otimes \mathcal{H}_{\rm op}$ is pure, the entropy of the reduced density matrix on $\mathcal{H}_R \otimes \mathcal{H}_L$ is equal to the entropy of the density matrix on $\mathcal{H}_{\rm op}$, which in turn is bounded by the degeneracy of operators with scaling dimensions in the neighborhood of $\ell$. We will call the log of this level density the spectral entropy $S_\ell$. We will now argue in favor of the following inequality and equality
\bea\label{eq:propSell}
S_{LR} \leq S_\ell \qquad {\rm with}   \qquad  S_\ell \, = \, 2 \pi \ell.
\eea

To motivate this proposal, consider a PETS with a fixed $\beta_\rrR$ and take the limit $\beta_\llL \to \infty$. In this limit $k_L \to 0$ and the left QM gets frozen to its ground state \footnote{This is a slightly subtle argument since the SYK model has a large number of approximate ground states of order $e^{S_0} \sim e^N$. In any case we assume the dynamics of the left QM to freeze to one of its ground states.}. From the discussion in section \ref{sec:JTback} it is clear that if one takes $k_R = \ell$ then the end of the world brane sits on top of the right horizon as shown in figure \eqref{PETSgeom3}. In this special situation, it is natural to associate an entropy to the object right at the horizon, the massive bulk particle, an entropy that is equal to the RT entropy. This leads to the formula $S_{\rm \ell} = 2\pi k_R = 2 \pi \ell$. A similar argument that motivates this proposal can be made using the results of \cite{Takayanagi:2011zk}. 

We conjecture that the relation in equation \eqref{eq:propSell} is also true for general values of $\beta_\rrR$ and $\beta_\llL$. 
It would be indeed interesting to verify this conjecture, that (in the regime of large scaling dimension $\ell \simeq N/\beta J$) the SYK model has a universal level density of operators given by \eqref{eq:propSell}, from a microscopic viewpoint.

Note that this assignment is consistent with subadditivity and the Araki-Lieb inequality 
\bea
| S_L - S_R| \leq S_{LR} \leq S_L + S_R. & & 
\eea
Subadditivity is satisfied since $S_L + S_R \sim S_0$ which is trivially larger than $2\pi \ell$. One can check that strong subadditivity is also satisfied.

Using the quantities $S_L$, $S_R$ and setting $S_{LR} = S_\ell = 2\pi \ell$, we can write the length of the throat (the distance between left- and right horizon) in terms of entropies
\bea
\cosh D_H \is \frac{S_{LR}^2 + \tilde{S}_L^2+\tilde{S}_R^2}{2 \tilde{S}_L\tilde{S}_R},
\eea
where we defined $\tilde{S} = S- S_0$. 
Similarly we can write the distance between the trajectory of the bulk particle and the right horizon as 
\bea
\sinh D_2 \is  \frac{S_{LR}^2 + \tilde{S}_L^2-\tilde{S}_R^2}{2 {S}_{\ell}\tilde{S}_R} 
\eea
Requiring the $D_2>0$ implies that $S_{LR}$ is always larger than the geometric mean of $S_L - S_R$ and $\tilde{S}_L + \tilde{S}_R$. This implies the Araki-Lieb inequality, which only becomes an equality in the case that $\tilde{S}_L = 2\pi k_L = 0$ and $S_R = 2\pi k_R = 2\pi \ell = S_\ell$.

As a final comment one can study bulk reconstruction in for these states, as in the previous section. Then if $\Phi_L<\Phi_R$ the right QM can only reconstruct its causal wedge (outside of the right horizon), which coincides with its entanglement wedge for these states. If one wants to reconstruct up to the left horizon using the right QM one needs to add the knowledge of the degrees of freedom creating the state, associated to the bulk brane. For the system $\mathcal{H}_R \otimes \mathcal{H}_{\rm op}$ the entanglement wedge reaches the left horizon past the right interior, just like for the pure PETS considered above.

\begin{figure}[t!]
\begin{center}
\begin{tikzpicture}[scale=1.1]
\draw[thick] (-3,0) coordinate (A) arc (180:300:0.7898) coordinate (B); 
\draw[thick] (2,0) coordinate (C) arc (0:-117:0.7898) coordinate (D); 
\draw[thick] (B) arc (230:311:2); 
\draw[dashed] (-3,0) -- (2,0);
\draw[thick,dotted,color={rgb:red,10; black,3}] (B) -- (-1.9,0);
\draw[thick,dotted,color={rgb:red,10; black,3}] (D) -- (0.9,0);
\draw[thin] (-2.5,0) coordinate (Ha) circle (0.05);
\draw[fill=black] (-0.5,0) coordinate (Hb) circle (0.08);
\draw[thin] (1.4,0) coordinate (Hc) circle (0.05);
\draw[color={rgb:red,10; black,3}, fill={rgb:red,10; black,3}] (B) circle (0.1);
\draw[color={rgb:red,10; black,3}, fill={rgb:red,10; black,3}] (D) circle (0.1);
\draw (1.7,-.89) node {$\tau_1$};
\draw (-0.5,-1.38) node {$\tau_3$};
\draw (-2.8,-.88) node {$\tau_2$};
\draw (Ha)+(0,0.4) node {\small $\Phi_h \propto q$};
\draw (Hb)+(0,0.4) node {\small $\Phi_h \propto p$};
\draw (Hc)+(0,0.4) node {\small $\Phi_h \propto k$};
\end{tikzpicture}
\hspace{1.8cm}
\begin{tikzpicture}[scale=0.78]
\path[fill=red!10]  (-2.75,0) -- (-5,2.2) -- (2,2.2) -- (0,0) -- (2,-2.2) -- (-5,-2.2) -- cycle ;
\path[fill=green!10]  (-1.5,0) -- (.5,2.2) -- (2,2.2) -- (2,-2.2) -- (.5,-2.2) -- cycle;
\path[fill=blue!10]  (-1.5,0) -- (-3.5,2.2)  -- (-5,2.2) -- (-5,-2.2) -- (-3.5,-2.2) -- cycle;
\draw[dashed] (-5,-2.2) -- (-.5,2.2);
\draw[dashed] (-5,2.2) -- (-.5,-2.2);
\draw[thick] (-5,-2.2) -- (-5,2.2);
\draw[thick] (2,-2.2) -- (2,2.2);
\draw[thick,color=red!90] (-0.4,-2.2) -- (-0.4,2.2);
\draw[thick,color=red!90] (-2.2,-2.2) -- (-2.2,2.2);
\draw[dashed] (-2,2.2) -- (2,-2.2);
\draw[dashed] (-3.5,2.2) -- (0.5,-2.2);
\draw[dashed] (0.5,2.2) -- (-3.5,-2.2);
\draw[dashed] (2,2.2) -- (-2,-2.2);
\draw[dashed] (-5,0) -- (2,0);
\draw[thin] (-2.75,0) circle (0.06);
\draw[thin] (0,0) circle (0.06);
\draw[fill=black] (-1.5,0) circle (0.1);
\draw[thick,decoration = {zigzag,segment length = 2mm, amplitude = 0.75mm},decorate] (-5,2.2)--(2,2.2);
\draw[thick,decoration = {zigzag,segment length = 2mm, amplitude = 0.75mm},decorate] (-5,-2.2)--(2,-2.2);
\end{tikzpicture}
\end{center}
\vspace{-0.4cm}
\caption{\small Backreaction by two operator insertions (red dots) with three local horizons (black circles and black dot). Assuming that $p < k, q$, the middle horizon is the extremal RT surface that separates the left- and right entanglement wedges, indicated by the blue and green regions on the right.}
\label{fig:twoinsertionsPETS}
\end{figure}
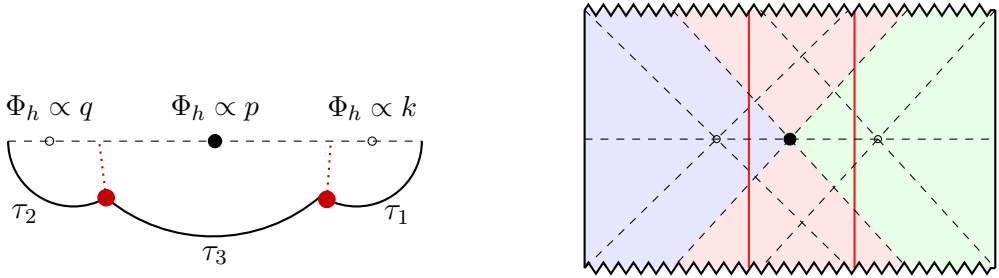

\subsection{Mulitple operators}

\vspace{-1mm}

In this section we will generalize the previous analysis of PETS to multiple insertions. For simplicity we will begin with two insertions. Generalization to more operators is straight-forward. 

We show the Euclidean part that produces the state in figure \ref{fig:twoinsertionsPETS}. To create this state we insert two operators of dimensions $\ell_1$ at $\tau_1$ and $\ell_2$ at $\tau_1+\tau_3$ and we define $\tau_2$ such that $\tau_1 + \tau_2 + \tau_3=\beta/2$. To each propagator for time $\tau_1$, $\tau_2$ or $\tau_3$ we associate a momentum (horizon dilaton) $\tau_1 \to k$, $\tau_2 \to q$ and $\tau_3 \to p$ as shown in figure \ref{fig:twoinsertionsPETS}.

\begin{figure}[t!]
\begin{center}
\begin{tikzpicture}[scale=0.34]
\path[fill=green!20] (0,0) -- (5,0) -- (5,5) -- (0,5) --cycle ;
\path[fill=red!20] (5,5) -- (5,0) -- (10,0) -- (10,10) --cycle ;
\path[fill=blue!20] (5,5) -- (0,5) -- (0,10) -- (10,10) --cycle ;
\fill (5,5) circle (3pt) node[above] {$T$};
\fill (5,0) circle (1pt) node[below] {\footnotesize ${\beta}/{4}$};
\fill (0,5) circle (1pt) node[left] {\footnotesize ${\beta}/{4}$};
\fill (10,0) circle (1pt) node[below] {\footnotesize ${\beta}/{2}$};
\fill (0,10) circle (1pt) node[left] {\footnotesize ${\beta}/{2}$};
\fill (0,0) circle (3pt) node[left] {$O$};
\draw (2.5,2.5) node[color=blue] {$p$};
\draw (7.5,2.5) node[color=blue] { $k$};
\draw (2.5,7.5) node[color=blue] { $q$};
\draw[->,thick,black] (0,0) -- (12,0) node[pos=1,below] {$\tau_1$};
\draw[->,thick,black] (0,0) -- (0,12) node[pos=1,left] {$\tau_2$};
\end{tikzpicture}
\hspace{2.4cm}
\begin{tikzpicture}[scale=0.34]
\path[fill=green!20] (0,0) -- (5,0) -- (3.33,3.33) -- (0,5) --cycle ;
\path[fill=red!20] (3.33,3.33) -- (5,0) -- (10,0) -- (10,10) --cycle ;
\path[fill=blue!20] (3.33,3.33) -- (0,5) -- (0,10) -- (10,10) --cycle ;
\fill (3.33,3.33) circle (3pt) node[above] {$T$};
\fill (5,0) circle (1pt) node[below] {\footnotesize $\;\; {\beta}/{4}$};
\fill (0,5) circle (1pt) node[left] {\footnotesize ${\beta}/{4}$};
\fill (10,0) circle (1pt) node[below] {\footnotesize ${\beta}/{2}$};
\fill (0,10) circle (1pt) node[left] {\footnotesize ${\beta}/{2}$};
\fill (3.33,0) circle (1pt) node[below] {\footnotesize ${\beta}/{6}\;\; $};
\fill (0,3.33) circle (1pt) node[left] {\footnotesize ${\beta}/{6}$};
\fill (0,0) circle (3pt) node[left] {$O$};
\draw (1.5,2.5) node[color=blue] {$p$};
\draw (7.5,2.5) node[color=blue] {$k$};
\draw (2.5,7.5) node[color=blue] {$q$};
\draw[dotted,black]  (3.33,0) -- (3.33,3.33);
\draw[dotted,black]  (0,3.33) -- (3.33,3.33);
\draw[dotted,black]  (5,0) -- (5,5);
\draw[dotted,black]  (0,5) -- (5,5);
\draw[->,thick,black] (0,0) -- (12,0) node[pos=1,below] {$\tau_1$};
\draw[->,thick,black] (0,0) -- (0,12) node[pos=1,left] {$\tau_2$};
\end{tikzpicture}
\end{center}
\vspace{-0.8cm}
\caption{\small Left: Phase diagram near $\ell=0$. $T$ denotes the `tri-critical' point. Right: Phase diagram near $\ell=\infty$. $T$ denotes the `tri-critical' point. We see that it moves towards the origin.}
\label{fig:phasediagram}
\end{figure}
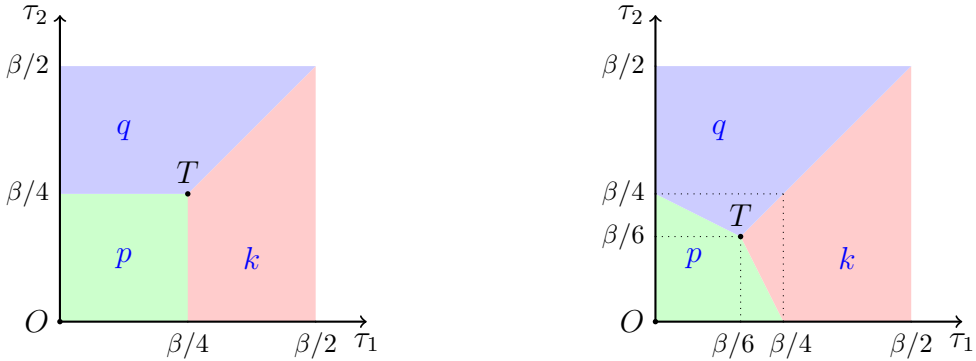

These dilaton values $p$, $k$, $q$ are fixed by the saddle point equations describing the backreaction of JT gravity as explained in section \ref{sec:JTback}. Repeating the analysis of the previous sections, or equivalently applying the holographic prescription, we can compute the entanglement entropy of this state by 
\beq
S = S_0 + 2 \pi ~ {\rm min} (p,q,k).
\eeq
Depending on the choice of parameters different choices of horizon dilaton dominates. For the case of a single insertion this choice was simply determined by whether $\tau<\beta/4$ or $\tau>\beta/4$. In the two operator case we show a phase diagram as a function of time insertions $\tau_1$ vs $\tau_2$ in figure \ref{fig:phasediagram}. This has an interesting behavior in terms of the `tricritical' point that divides the three different regions. The location of the tricritical point can be found in terms of a transcendental equation derived from the saddle point relations, which should be solved numerically. In figure \ref{fig:phasediagram} we show the cases $\ell \to 0$ (left panel) and $\ell \to \infty$ (right panel) which can be analytically found. The phase diagram for intermediate $\ell$ interpolates between these extreme cases.

We can also consider `multi'-partite states for which one averages over the microscopic choice of operators possible such that their dimensions are approximately $\ell_1$ and $\ell_2$. Following section \ref{sec:cg} we imagine having a large number $K$ of operators with dimensions $\ell_i \approx \ell$ for $i=1,\ldots, K$. In a large $K$ limit this controls the factorization channel that dominates the Renyi entropy calculation and gives 
\bea
S_L \! \is \! S_0 + 2 \pi q,\qquad \qquad \ 
S_R \, =\, S_0 + 2 \pi k.
\eea
We can compare these quantities with the entanglement entropy by looking at figure \ref{fig:phasediagram}. 

Another interesting feature of this kind of composite PETS is the following. Let us choose parameters such that the global minimum of the dilaton is located at the middle horizon, so that the entanglement entropy equals $S= S_0 + 2 \pi p$. We choose $\ell$ large enough such that the two bulk particles are hidden behind the left and right horizons. Then, as opposed to figure \ref{PETSgeom22}, the extremal RT surface is outside the right and left causal wedges and has no overlap with it. This situation is shown on the right in figure \ref{fig:twoinsertionsPETS}. Still, following the discussion in section \ref{sec:bulkrec}, left- or right observers with sufficient detailed understanding of the microscopic wave function would be able to perform a one-sided bulk reconstruction of their full entanglement wedge, indicated by the green and blue regions in figure~\ref{fig:twoinsertionsPETS}. 

\section{Concluding Remarks}
\vspace{-2mm}

Motivated by the geometric approach of \cite{Kourkoulou:2017zaj} to pure states in the SYK model, we have studied the holographic dual of a general class of partially entangled thermal states (PETS) specified by the insertion of the single scaling operators into the euclidean time evolution that creates the thermo-field double state. We studied the bulk dual of a PETS 
in the low energy approximation of the SYK model described by the Schwarzian theory.
We argued that the partially entangled thermal states describe a composite black hole with two horizons, separated by an expanded interior region with a massive bulk particle.
We computed the entanglement entropy of these states and compared with the usual holographic RT prescription. We argued, both from an entanglement wedge and a tensor network perspective, that a one-sided reconstruction can be extended into the interior geometry of the black hole.

It would be interesting to generalize this setup to higher dimensions, maybe using SYK-like models such as \cite{Gu:2016oyy, Turiaci:2017zwd, Murugan:2017eto, Berkooz:2017efq} or 2D generalizations of the Schwarzian action \cite{Turiaci:2016cvo}. The bulk membrane dual to the PETS might have a more interesting structure in these cases. Another interesting application would be to study, even within AdS$_2$, how to apply the GJW teleportation protocol \cite{GJW} in this context. Since entanglement is a resource for these kinds of operations, it should be harder to make the PETS wormhole traversable. 

One important point which we leave for future work is to study partially entangled states in the regime where the entanglement is a finite fraction of $N$. Semi-classically, these states look like factorized black hole geometry as indicated in panel c) of figure \ref{fig:dprof}. The physics of this transition depends on the microscopic SYK dynamics.
This looks like a hard problem, but may be tractable using dynamical mean field theory or via numerical methods. 

Another interesting modification of the thermo-field double state is obtained by considering the insertion of a topological interface. A topological interface in a 2D CFT is defined by considering a boundary state in the tensor product of two identical CFTs and then using the folding trick \cite{Oshikawa:1996dj}\cite{Bachas:2001vj} to reorient it such that the reflection from left-movers into right-movers is replaced by a transmission from CFT${}_1$ to CFT${}_2$ (see e.g. \cite{Brehm:2015plf}). Since the resulting interface is topological (commutes with the Virasoro algebra), inserting it into the euclidean path integral of the TFD state does not lead to any (localizable) gravitational backreaction. In particular, the effective temperature on both sides will always be the same. Hence the quantum numbers that specify the topological interface should be considered as non-trivial potential quantum numbers of the state associated with the ER bridge of a two-sided black hole geometry. 

\bigskip

\begin{center}
{\bf Acknowledgements}
\end{center}

\vspace{-2mm}

We want to thank Ahmed Almheiri, Netta Engelhardt, Himanshu Khanchandani, Aitor Lewkowycz, Juan Maldacena, Thomas Mertens, Xiaoliang Qi, Douglas Stanford, Edward Witten and Zhenbin Yang for useful discussions and comments. H.T.L is supported by a Croucher Scholarship for Doctoral Study and a Centennial Fellowship from Princeton University.
The research of H.V. is supported by NSF grant PHY-1620059.

\begin{appendix}
\section{A Complete Basis of Partially Entangled States}\label{app:SYKPETS}

In this Appendix we discuss a general class of partially entangled thermal states in SYK whose one-sided correlation functions coincide with their thermal expectation value while the two-side correlation functions and entanglement entropy can be different.

Consider $4N$ Majorana variables $\psi^i$ spanning a $2^{2N}$ dimensonal Hilbert space.
Introduce the basis of $2^{2N}$ states $|\sss\rangle$ defined by
\bea
\label{sdef2}
\(\psi^{2k-1}-is_k\psi^{2k}\)|\spc \sss \spc \rangle\! \is \! 0  \ \Leftrightarrow \ S_k|\spc \sss \spc \rangle\equiv 2i\psi^{2k-1}\psi^{2k}|\spc \sss \spc \rangle\, =\, s_k|\spc \sss \spc \rangle\,.
\eea
We partition the $4N$ Majorana fermions into two groups of $2N$ Majorana fermions $\{\psi_{L,R}\}$, from which, two sub-Hilbert space $\mathcal{H}_{L,R}$ of dimension $2^{N}$ can be built. We consider a class of states
\bea
\label{petsonetwo}
|\Psi\rangle\is |\sss;  \beta_\llL, \beta_\rrR \rangle=e^{-\frac 1 2 \beta_\llL H_\llL}\otimes e^{-\frac 1 2 \beta_\rrR H_\rrR}|\spc \sss \spc \rangle\,,
\eea
where $H_{\llL,\rrR}$ are Hamiltonian of the same form acting on $\mathcal{H}_{\llL,\rrR}$ respectively. By choosing different partitions, we can obtain a class of states.
 with different amount of entanglement between ${\cal H}_\llL$ and ${\cal H}_\rrR$. 
The thermo-field double is the state for which everything is transmitted and for which all $s_k = 1$. There are $2^{N}$ other states with the same amount of entanglement as the TFD state. 
For generic states \eqref{petsonetwo}, $K$ fermion pairs are reflected back and $N-K$ fermion pairs are transmitted from the left to the right system. 
These states are all partially entangled thermal states with entanglement entropy between zero (product states) and the thermal entropy (TFD type states).

We denote the operator that flips the sign of $\psi_k$ by $\sigma_k$. Note that $\sigma_kH_{\llL,\rrR}\sigma_k^{-1}=H_{\llL,\rrR}$ after averaging over the random SYK couplings. 
The inner products of these PETS do not depend on the partition 
\bea
\langle \Psi|\Psi\rangle \! \is\! 2^{-2N}\sum_k\langle \spc \sss\spc |\sigma_k^{-1} e^{-\frac 1 2 \beta_\llL H_\llL-\frac 1 2 \beta_\rrR H_\rrR} \sigma_k \sigma_k^{-1}e^{-\frac 1 2 \beta_\llL H_\llL-\frac 1 2 \beta_\rrR H_\rrR}\sigma_k|\spc \sss\spc \rangle \nonumber\\[2mm]
\! \is \! 
\text{Tr}[e^{-\beta_\llL H_\llL}\otimes e^{-\beta_\rrR H_\rrR}] \, = \, 
Z(\beta_L)Z(\beta_\rrR)\,.
\eea
The one-sided two-point correlators are
\bea
G_{\text{diag}}(\tau_1,\tau_2)\!\! &\equiv& \! \! \frac{\langle \Psi|\psi^i(\tau_1)\psi^i(\tau_2)|\Psi\rangle}{\langle \Psi|\Psi\rangle}=G_{\beta_{L,R}}(\tau_1-\tau_2),\quad \text{ if $\psi^i\in\{\psi_{\llL,\rrR}\}$}\,,
\eea
which is the same as the thermal expectation value of the temperature associated to the subsystem and does not depend on the details of the partition. The temperature can be different in general. We can also compute off-diagonal two-point functions. Only the combination $s_k\psi^{2k-1}\psi^{2k}$ has nontrivial expectation value at leading order. Using that it is flip invariant, we compute 
\bea
G_{\text{off}}(\tau_1,\tau_2)\!\! &\! \equiv\! & \!\! s_k\frac{\langle \Psi|\psi^{2k-1}(\tau_1)\psi^{2k}(\tau_2)|\Psi\rangle}{\langle\Psi|\Psi\rangle}
\, = \, \textstyle -2iG_{\beta_L}(\tau_1\!+\!\smpc \frac 1 2 \beta_\llL )G_{\beta_R}(\tau_2\!+\!\smpc \frac 1 2 \beta_\rrR)\,.~~~
\eea
If $\psi^{2k-1},\psi^{2k}$ belong to the same partition, the correlation function can be interpreted as one-sided off-diagonal correlation function. If they belong to different partition, the correlation function can be interpreted as a two-side correlation function and $\psi^{2k-1}$ is the counterpart of $\psi^{2k}$ at the other side.

\section{Bulk Kinematics and Dynamics}\label{app:conv}
\subsection{Kinematics}
We will summarize the coordinate systems we use to describe AdS$_2$. It is useful to work in embedding space $Y=(Y^{-1},Y^0,Y^{1})$. The metric is $ds^2=\eta_{AB} dY^A dY^B$ with the inner product $Y_1 \cdot Y_2 = \eta_{AB} Y_1^AY_2^B$. Then by restricting to $Y\cdot Y=-1$ we obtain Euclidean AdS$_2$ if $\eta={\rm diag}(-1,1,1)$ or Lorenzian AdS$_2$ if $\eta={\rm diag}(-1,-1,1)$. For definiteness we will focus in Euclidean signature. 

A first set of convenient coordinates are $(\rho, \tau)$ such that 
\beq
Y=\pm (\cosh \rho, \sinh \rho \sin \tau ,\sinh \rho \cos \tau)
\eeq
In these coordinates the metric of AdS$_2$ is 
\beq\label{rhometric}
ds^2=d\rho^2 + \sinh^2 \rho ~d\tau^2.
\eeq
This covers the Rindler patch when analytically continued to Lorenzian signature $\tau \to i t_R$. Another choice of coordinates parametrizes the hyperboloid as 
\beq
Y =\pm \left(\frac{1+x^2+y^2}{1-x^2-y^2} ,\frac{2x}{1-x^2-y^2},\frac{2 y}{1-x^2-y^2} \right)
\eeq 
This gives polar coordinate for the plane $(x,y)=(r \cos \theta, r \sin \theta)$. The metric is 
\beq
ds^2 = \frac{dx^2 + dy^2}{1-x^2-y^2}
\eeq
Then AdS$_2$ is mapped to the Poincare disk $x^2+y^2<1$. To compare with the Rindler parametrization one can take $\theta=\tau$ and $r\equiv\sqrt{x^2+y^2}=\tanh \rho/2$.

Finally one can define coordinates giving the Poincare patch of AdS as 
\beq
Y=\pm \left(\frac{1+t^2+z^2}{2z},\frac{t}{z},\frac{1-t^2-z^2}{2z}\right)
\eeq
this gives the metric $ds^2=z^{-2}(dt^2+dz^2)$. 

In any coordinate system, geodesic distance between two points can be computed as $\cosh D_{12} = - Y_1 \cdot Y_2$. This can be rewritten in any of the coordinates above. Geodesics in this geometry are parametrized by a ray $X$ in embedding space and is given by $Y$ such that 
\beq
X\cdot Y=0.
\eeq

\subsection{Dynamics}

\vspace{-1mm}

We summarize the classical solutions of JT gravity in terms of embedding coordinates following and using the notation of \cite{Maldacena:2016upp}. It is then straightforward to translate results to any coordinate system described in the previous section according to convenience. Since the effective coupling is proportional to the combination $\Phi_b/G_N$ involving the boundary dilaton we will take a dilaton normalization such that $8 \pi G_N =1$ without loss of generality (since in 2D $G_N$ is dimensionless). 

The dilaton in regions without matter behaves as \cite{Almheiri:2014cka}\cite{Maldacena:2016upp}\cite{Engelsoy:2016xyb}
\beq
\Phi (Y)= Z\cdot Y,~~~~{\rm for}~Y^2=-1
\eeq
where $Z$ is an arbitrary vector in embedding space. Natural boundary conditions for JT gravity giving boundary gravitons described by the Schwarzian action fixes metric and dilaton. Then the boundary is described by $Z \cdot Y = \Phi_b=\Phi_r/\epsilon$, where $\epsilon$ denotes the cut-off. This describes a circle (set of points at fixed geodesic distance from a center) in AdS$_2$. The center of this circle coincides to the horizon, where the value of the dilaton is minimal. It is easy to find this location as 
\beq
Y_h = (-Z\cdot Z)^{-1/2} Z,~~~~\Phi_h = (-Z \cdot Z)^{1/2}.
\eeq
To summarize, the sourceless solution is fixed by a three-component vector in embedding space $\mathbb{R}^{2,1}$. Its direction fixed the location of the horizon and its magnitude fixes the horizon dilaton. Moreover for a fixed $Z$ the boundary curve has length $\beta/\epsilon$ (where $\epsilon$ denotes the cut-off) with inverse temperature $\beta$ related to the magnitude of $Z$ (or equivalently $\Phi_h$) as $\Phi_h = \frac{2\pi}{\beta} \Phi_r$. In this units, the Bekenstein-Hawking entropy of this geometry is $ S= 2 \pi \Phi_h$. Finally, the ADM energy is $E= \Phi_h^2/(2\Phi_r)$. By adding a topological term to the action $\int d^2x ~\Phi_0 R$, with $\Phi_0$ constant, one can account for a possible zero-point entropy $S_0$.

These boundary trajectories of constant $Z\cdot Y$ correspond to circles in the Poincare disk coordinates $(x,y)$. Therefore it is natural to draw the backreacted boundary in the Poincare disk coordinates such as figure \ref{fig:curve1} or \ref{fig:curve2} (nevertheless these coordinates distort the size and location of the origin with respect to the flat $(x,y)$ plane).

 From the 2D Liouville perspective of the Schwarzian theory \cite{Mertens:2017mtv} (see \cite{Mertens:2018fds} for more details) the horizon dilaton $\Phi_h$ corresponds to the momentum $k=\Phi_h$ associated to a primary state of Liouville with energy $E=k^2/(2C)$, with $C=\Phi_r$ in units with $8 \pi G_N=1$. This is a natural variable to label intermediate states.

Using these identifications and coordinates defined above it is a straightforward exercise to get the equations and relations presented in section \ref{sec:JTback}.
 
The $SL(2,R)$ charge of a given solution is also fixed by the vector $Z$ as $Q=2Z$ \cite{Maldacena:2016upp}. Then one can interpret the boundary trajectory as a particle in a magnetic field \cite{Kitaev16}. More importantly this allows to add matter in a straightforward way. Within the JT approximation of free matter bulk particles propagate along geodesics $Q_{\rm m} \cdot Y =0$ with the space-like vector $Q_{\rm m}$ giving the $SL(2,R)$ charge of the particle, normalized by the mass square $Q_m^2 = \mu^2$. Then one can glue bulk solutions labeled by $Z=Q_L/2$ and $Z=Q_R/2$ along the particle geodesic with the singlet constrain 
\beq
Q_L + Q_R + Q_m =0.
\eeq 
This charge conservation constrain has the nice property of making the dilaton $\Phi$ continuous along the matter geodesic. But the dilaton slope jumps proportional to its mass 
\beq
\frac{\partial \Phi}{\partial s}\Big|_{L}-\frac{\partial \Phi}{\partial s}\Big|_{R} = 2 \mu,
\eeq
in a particle's rest frame where $Q_{\rm matter} = (0,0,\mu)$. In this notation, the mass of the particle $\mu= \ell$ (for large $\ell$), the dimension of the dual operator. $s$ is a geodesic length in the direction perpendicular to the particle's geodesic. This is consistent with the equations of motion that come from varying the metric which relates the matter stress tensor with the second derivative of the dilaton.

\section{QEC and the Black Hole Interior}\label{appQECBHI}
\vspace{-2mm}

In this Appendix, we briefly summarize the QEC procedure for reconstructing the black hole interior. Consider a holographic large $N$ CFT with a weakly coupled bulk dual. At leading order in $N$, the bulk QFT Hilbert space is a free field Fock space spanned by orthonormal basis states $
 {}_b\la n\ri m \ra_b 
 = \delta_{nm}\,.$ Now consider a given CFT state that corresponds to a semi-classical black hole geometry in the bulk. Due to the Hawking effect, the bulk QFT state represents a thermal mixed state with a non-zero particle density. Following \cite{Verlinde:2012cy}, we represents the embedding of the bulk low energy QFT Hilbert space into the CFT Hilbert space via the following random tensor representation 
 \bea
\label{kdeco}
\li \Psi \ra = \sum_n   {\bf T}_n\spc  \li \Psi_0 \ra \spc \li n \ra_b, \qquad \qquad 
\sum_n {\bf T}_n^\dag {\bf T}_n = \mathbb{1}\, .
\eea
The ${\bf T}_n$ are the AdS/CFT analogues of the Kraus operators employed in \cite{Verlinde:2012cy};
the second relation is the standard unitarity condition. The initial state $\li \Psi_0 \ra$ can be thought of as describing the horizon state.
The ${\bf T}_n$ are assumed to be state independent: they do not depend on $\li \Psi_0 \ra$. 
 
$\li \Psi \ra$ looks thermal from the bulk QFT perspective. This gives useful statistical information
\bea
\label{therm}
\la \Psi_0| {\bf T}^\dag_m {\bf T}_n \li \Psi_0\ra \is w_n \delta_{mn}\, 
\eea
where $w_n  = \frac{1}{Z}e^{-\beta E_n}$ with $\sum_n w_n = 1$ denote the Boltzmann weights.
Let ${\bf P}$ denote the projection operator onto the code subspace ${\cal H}_{\rm code}$. 
Basic statistical reasoning shows that the matrix element of ${\bf T}^\dag_m {\bf T}_n$ between any pair of typical basis states $\li \bar i \ra, \li \bar j \ra \in {\cal H}_{\rm code}$ takes the form
\bea
\label{diag}
\la\, \bar{i}\, \ri {\bf T}^\dag{\!\!}_m {\bf T}_n\li \spc \bar  j\spc \ra   \is w_n\delta_{mn}\;  \delta_{\bar i \bar j}
\eea
The operators $ {\bf T}^\dag_m$ and ${\bf T}_n$ are large complex random matrices,
whose product yields a sum of many terms with different phases. This sum averages out to zero, except when there is constructive interference. 
Property (\ref{diag}) holds with exponential accuracy $e^{S_{\rm code} - S_{\rm BH}} \, = \,  \frac {\dim{H}_{\rm code}} {\dim{\cal H}_{\rm CFT}}$. Note that ${\bf T}_n$ map a bigger to a smaller Hilbert space, and are therefore non-invertible on ${\cal H}_{\rm CFT}$. When restricted to the code subspace ${\cal H}_{\rm code}$, however, it becomes effectively invertible. 

In analogy with quantum error correcting codes, we now define the recovery operators 
\bea
\label{recover}
{\bf R}_ n  = \, {\bf P} \, \frac{{\bf T}_n^\dag}{\sqrt{w_n\!\!}\;\,} \qquad ; \qquad
{\bf R}_n^\dag  = \, \frac{{\bf T}_n}{\sqrt{w_n\!\!}\;\,} \; {\bf P}\, .
\eea
 Using \eqref{diag},
we deduce that (up to exponentially small corrections) the recovery operator ${\bf R}_n$ acts on the state $|\Psi\rangle$ via $
{\bf R}_m \li \Psi \ra = \sqrt{w_m\!\!}\;\, \li \Psi_0 \ra \spc \li m \ra_b$,
provided that $|\Psi_0\rangle \in {\cal H}_{\rm code}$. So by acting with the ${\bf R}_n$'s, one can recover the quantum information contained in the original state 
$|\Psi_0\rangle$.

Any general outside operator $B$, given by some polynomial in the $\calO^\dag$ and $\calO$ oscillators, can be characterized by its matrix elements $B_{mn} =
{}_b\la m | {\bf B} \, \li n \ra_b $.
Following \cite{Verlinde:2012cy}, we define interior operators ${\bf A}$ as follows
\bea
\label{intone}
{\bf A} \is    \sum_{m,n} \spc A_{mn}\,  {\bf R}^\dag{\!\!}_n\spc {\bf R}_m 
\eea
This definition 
generalizes the mirror operator definition in \cite{Papadodimas:2013wnh} to arbitrary non-maximally mixed states. 
Note that these operators 
act linearly on all of ${\cal H}_{\rm code}$. We now compute
\bea
\la \Psi \ri {\bf A} {\bf B} \li \Psi \ra \is \sum_{m,n, p,q}\; \frac{A_{nm}B_{pq} }{\sqrt{w_m w_n \!\!}\; }\; \;\la \Psi_0 \ri \spc {\bf T}^\dag_p \, {\bf T}_n\, {\bf P}\, {\bf T}^\dag_m\spc {\bf T}_q \li \Psi_0 \ra\\[3mm]
\is \sum_{m,n}\; \sqrt{w_m w_n \!\!}\; \; A_{nm} B_{nm} \;\; + \;\; {\rm small} \ {\rm  error}
\eea
with an exponentially small error of order $e^{S_{\rm code} - S_{\rm BH}}$. We see that the mixed expectation values of the exterior and interior operators matches with those derived from low energy effective field theory in a black hole geometry with a smooth horizon. This reconstruction breaks down for very high point correlation functions and when $S_{\rm code}$ approaches $S_{\rm BH}$.

\end{appendix}

\begingroup\raggedright\endgroup

\end{document}